# Terahertz field effect in a two-dimensional semiconductor $MoS_2$

Tomoki Hiraoka[1*], Sandra Nestler[2], Wentao Zhang[1], Simon Rossel[1], Hassan A. Hafez[1], Savio Fabretti[1], Heike Schlörb[2], Andy Thomas[2,3], and Dmitry Turchinovich[1*]

[1]Fakultät für Physik, Universität Bielefeld, 33615 Bielefeld, Germany
[2]Leibniz-Institut für Festkörper- und Werkstoffforschung, Helmholtzstraße 20, 01069 Dresden, Germany
[3]Institut für Festkörper- und Materialphysik, Technische Universität Dresden, Haeckelstraße 3, 01069 Dresden, Germany

Layered two-dimensional (2D) materials, with their atomic-scale thickness and tunable electronic, optical, and mechanical properties, open many promising pathways to significantly advance modern electronics. The field effect caused by a strong electric field, typically of MV/cm level, applied perpendicular to the material layers, is a highly effective method for controlling these properties. Field effect allows the regulation of the electron flow in transistor channels, improves the photodetector efficiency and spectral range, and facilitates the exploration of novel exotic quantum phenomena in 2D materials. However, existing approaches to induce the field effect in 2D materials utilize circuit-based electrical gating methods fundamentally limited to microwave response rates. Device-compatible ultrafast, sub-picosecond control needed for modern technology and basic science applications still remains a challenge. In this study, we demonstrate such an ultrafast field effect in atomically thin $MoS_2$, an archetypal 2D semiconductor, embedded in a hybrid 3D-2D terahertz nanoantenna structure. This nanoantenna efficiently converts an incident terahertz electric field into the vertical ultrafast gating field in $MoS_2$ while simultaneously enhancing it to the required MV/cm level. We observe the terahertz field effect optically as coherent terahertz-induced Stark shift of characteristic exciton resonances in $MoS_2$. Our results enable novel developments in technology and the fundamental science of 2D materials, where the terahertz field effect is crucial.

*Corresponding authors: t.hiraoka1023@physik.uni-bielefeld.de, dmtu@physik.uni-bielefeld.de

## Introduction

Two-dimensional (2D) layered materials, with their atomic-scale thickness and remarkable electronic, optical, and mechanical properties, offer significant potential across a broad range of scientific and technological fields, from classical and quantum computing to photodetection, energy harvesting, spintronics, memristor technology, and the discovery of new quantum phases [1-8]. The primary advantage of layered 2D materials is the broad tunability of their key physical properties, which stem from microscopic interactions within and between the material layers [9,10], as well as between the material and its environment [11-13].

One very effective way to control these intra- and interlayer interactions, and hence the physical properties of 2D materials, is via field effect caused by a strong electric field perpendicular to the material layers [14]. Field effect enables precise control of conductivity in field-effect transistor channels, as well as of the efficiency and detection range of 2D photodetectors, the generation of ultrahigh-frequency signals, and other processes critical to modern technology [15-18]. Moreover, it is crucial for exploration of the fundamental physics of 2D materials, allowing control over bandgaps [14,19-21] and excitonic states [22-28], insulator-metal transition [16,20], structural phase transitions [29], quantum phase transition in moiré systems [8] and more. Due to their atomic thickness, achieving significant control in 2D materials requires very strong electric fields, typically of the order of MV/cm [16-17,20-28].

Technologically, the field effect is realized by electrostatic gating. So far, gating of 2D materials has been limited to static DC-based [16-29] or microwave-based [30-32] methods, thus lacking the technologically-relevant ultrafast, sub-picosecond control capabilities. The key challenge here lies in the practical way of applying the terahertz (THz) gating electric signal of MV/cm strength to the 2D material in the direction perpendicular to its layers. Here, we demonstrate a pronounced THz field effect in a model 2D semiconductor $MoS_2$ [9,10,16,21-23,25,28]. For this, we introduce a hybrid 3D-2D nanoantenna structure that receives the broadband THz electromagnetic signal from the free space and converts it into a vertical gating field in the $MoS_2$, simultaneously enhancing it to a MV/cm level. Using time-resolved optical measurements, we observe the field effect as a coherent THz-induced Stark shift of characteristic exciton resonances in $MoS_2$. Therefore, our approach enables efficient ultrafast 2D (opto)electronic technology controlled by the THz fields, as well as novel experiments in fundamental investigations of 2D materials, where THz field effect is essential.

## A hybrid 3D-2D nanoantenna for THz gating

The design of our hybrid 3D-2D nanoantenna for THz gating of 2D materials is presented in Fig. 1. The antenna consists of two gold electrodes, top and bottom, vertically separated by an $Al_2O_3$ dielectric spacer layer. As shown in Fig. 1a,b, the electrodes are horizontally displaced such that they only overlap in the middle section of the antenna, which has lateral dimensions of 10 μm x 10 μm. In the top view projection, our antenna has a bowtie dipole shape, which enables efficient incoupling of broadband free-space THz fields to its electrodes and strong local field enhancement in the sub-wavelength antenna gap [33,34]. The entire antenna structure is deposited on a glass substrate. The 150-nm-thick $Al_2O_3$ dielectric spacer layer was grown by atomic-layer deposition [35,36] between the electrode planes in order to prevent the dielectric breakdown in the antenna in strong THz fields. Fig. 1c shows the microscope optical image of our hybrid 3D-2D nanoantenna. Several such devices were fabricated in two batches (below referred to as α and β - batches) and tested in this study. For the details see Methods Section 1,2, and the Performance Variation Among Fabricated Devices section below.

The application of a THz electromagnetic field $F_{x,\mathrm{in}}$, vertically illuminating the entire antenna structure and polarized along its electrodes, leads to a THz-field-driven current in the top and bottom electrodes. Due to their vertical separation and only partial horizontal overlap, this creates an electric polarization in the middle section of the antenna in the vertical direction. If a 2D material is placed in this region with layers parallel to the electrodes, it will experience the vertical gating THz field $F_z$, as shown in Fig. 1d. Besides conversion of the horizontally polarized incident THz field into the vertical gating field, our antenna also provides a significant field enhancement in the vertical direction. Finite difference time domain (FDTD) simulations of our antenna allow for the calculation of the gating THz field $F_z$ from the experimentally measured incident THz field $F_{x,\mathrm{in}}$ at the antenna gap position. This resulted in a calculated field enhancement factor of the antenna of $|F_z/F_{x,\mathrm{in}}| =$ 13.3, as shown in Fig. 1e. This field enhancement factor resulted in a maximum gating THz field $|F_z|$ in excess of 1 MV/cm applied to a $MoS_2$ sample, an estimate confirmed in our measurements. We note, however, that our experiments do not allow for the direct measurement of $|F_z/F_{x,\mathrm{in}}|$. Therefore, we present all the relevant THz-field dependent parameters in this work as a function of incident THz field $F_{x,\mathrm{in}}$, a directly measured quantity. The details of the calculation of the gating field $F_z$ from the measured incident THz field $F_{x,\mathrm{in}}$ can be found in Supplementary Section A, while the experimental estimate of the gating field strength is presented in the Discussion section below.

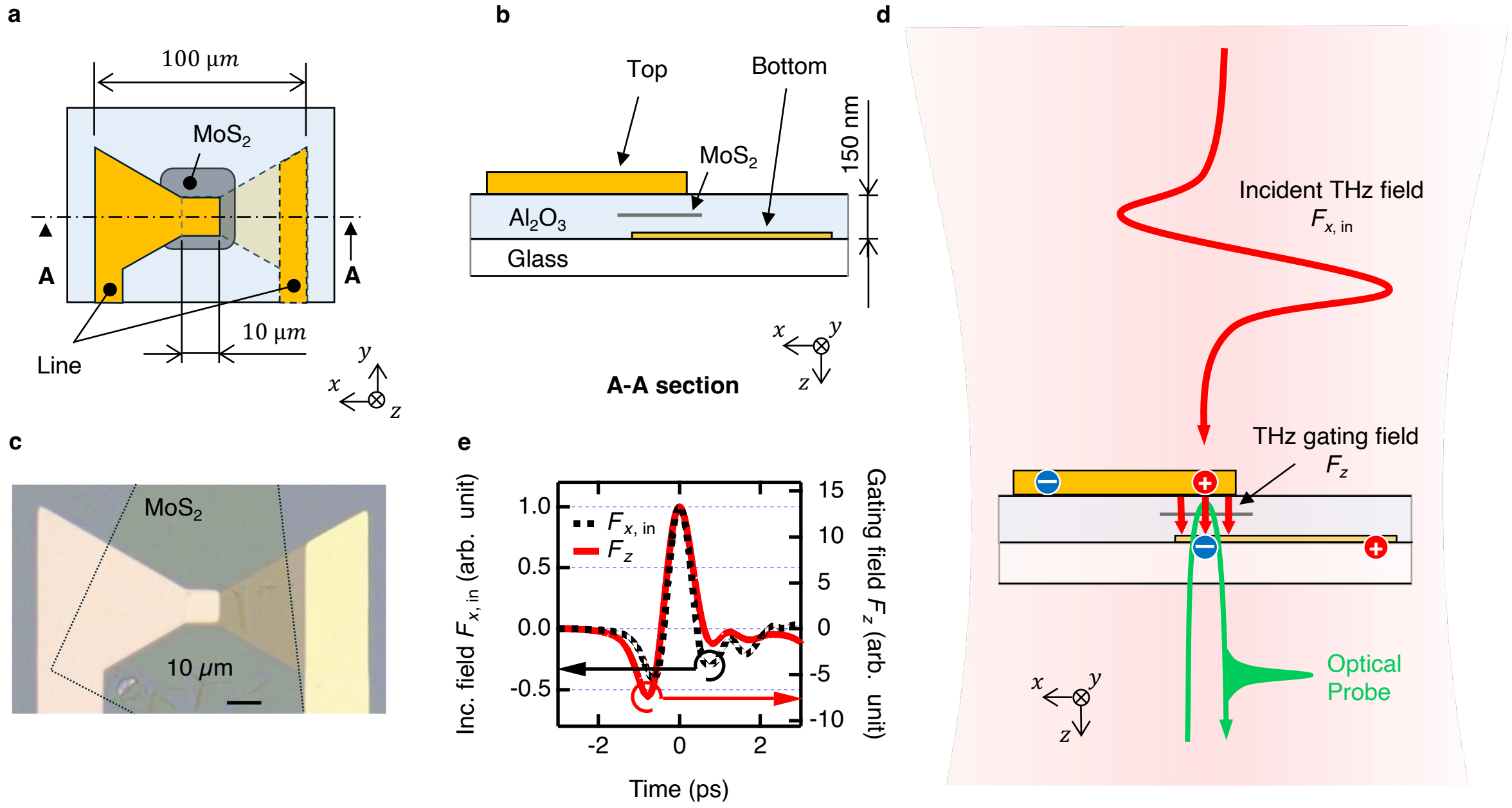

**Figure 1.** Hybrid 3D-2D THz nanoantenna. **a**, Schematic top view of the device around the bowtie antenna part. **b**, Schematic side view of the device (A-A section of **a**.) **c**, Microscope image of a fabricated device. **d**. Schematic of the THz pump-optical probe experiment. **e**, Simulated field enhancement. Black dashed line: measured incident THz field $F_{x,\mathrm{in}}$. Red solid line: gating THz field in the antenna gap $F_z$, calculated numerically from the measured incident THz field $F_{x,\mathrm{in}}$. The calculated field enhancement factor is $|F_z/F_{x,\mathrm{in}}| = 13.3$. Both vertical axes in **e** are on the same arb. unit scale. The traces $F_{x,\mathrm{in}}$ and $F_z$ are horizontally offset to show the peaks of both signals at 0 ps.

Several single-crystalline $MoS_2$ flakes, prepared by mechanical exfoliation, each containing 3-4 layers, were studied in this work. In our experiments, we used optical probing to demonstrate the THz gating effect on the $MoS_2$ flakes, which are positioned horizontally in the middle field-enhancement section of the antenna, as shown in Figs. 1a,b,d,e. In order to ensure optical access to the $MoS_2$, the bottom electrode of our hybrid 3D-2D nanoantenna was made 8 nm thin and, therefore, optically semitransparent [37], whereas the top electrode thickness was 25 nm and 50 nm for the devices in $\alpha$- and $\beta$- batches, respectively. As shown in Fig. 1d, the optical probe pulse enters the structure through the bottom electrode, travels through the spacer region interacting with the $MoS_2$ flake, gets back-reflected off the thick top electrode, and is then directed to the optical detection. We note that, in such an arrangement, the relatively low thickness of the bottom gold electrode of 8 nm was chosen to achieve both optical transparency as required for the optical probing and reasonably high conductance [38] as required for efficient electrical gating. In a situation where no optical access to the 2D material is required, using thicker gold for both electrodes will lead to a higher conductance and, hence, to an even larger field enhancement factor $|F_z/F_{x,\mathrm{in}}|$. The design of our antenna enables efficient incoupling of incident free-space THz signal with the bandwidth of at least 0.1 – 2.5 THz and its conversion into the field-enhanced THz gating field without any loss of bandwidth (see Supplementary Section A).

**THz pump-optical probe measurements**

As a probing mechanism for the THz gating effect, we used time-resolved optical spectroscopy of characteristic exciton resonances in $MoS_2$. For this, a THz pump-optical probe (TPOP) experiment [39] was set up, as schematically shown in Fig. 1d. The pump THz field was generated via tilted pulse front optical rectification of a 2 mJ, 800 nm laser pulse with 100 fs duration in a lithium niobate crystal [40,41]. This produced a broadband single-cycle pulse of the THz field with a frequency range of 0.2 – 2.5 THz and a center frequency of 0.4 THz, propagating into the free space. The THz beam was then focused and directed at our antenna at normal incidence, leading to the maximum THz field strength of $|F_{x,\mathrm{in}}| = 104$ kV/cm, as measured at the position of the antenna structure using free-space electrooptic sampling [42]. In order to prevent the potential breakdown of the studied devices, we made sure that the field-enhanced gating THz signal $|F_z|$ does not exceed the estimated values of 1.2 – 1.5 MV/cm by properly attenuating the incident THz field. For the gating field estimate, see the Discussion section. The incident THz field was polarized along the bowtie antenna structure, as shown in Figs. 1a-d. The diameter of the THz focus spot ($1/e$ width for the field) is estimated as 1 mm at the center frequency of 0.4 THz, which is much larger than the antenna size. A broadband optical probe pulse was produced via supercontinuum broadening of another 800 nm, 100 fs laser pulse with 3 μJ energy in a sapphire slab of 4 mm thickness [43]. The generated optical supercontinuum signal was then passed through a bandpass filter, resulting in an optical probe pulse with 900 fJ energy and covering the photon energy range of ca. 1.7 – 2.3 eV as required for our measurements. The probe pulse chirp correction is addressed in Methods Section 4 and Supplementary Section C.

The THz pump field and the optical probe pulse were directed at the antenna from the side of the top and bottom electrodes, respectively, as shown schematically in Fig. 1d. The optical probing was performed in reflection in a confocal configuration using the 10x magnification microscope objective, where the optical probe was focused on the field-enhancement region of the antenna, subject to the gating field $F_z$. For the analysis of the optical probe reflectance of our structure, an imaging CCD spectrometer was used to simultaneously record the spectra of the incident and reflected optical probe pulses. As a result, we were able to measure the calibrated broadband optical reflectance spectra of the field-enhancement region of our antenna as a function of the time delay between the incident THz pump field and the broadband optical probe pulse. All our experiments were performed at room temperature. See the Methods Section 3 for the details of our experimental setup.

At first, we characterized the response of a reference structure, a hybrid 3D-2D nanoantenna *without* the $MoS_2$ sample, to the incident THz field. The results are shown in Figs. 2a,c,f. A measured static optical reflectance spectrum $R_{\mathrm{st}}$ of the reference structure without applied THz field is shown in Fig. 2a. This spectrum has a simple convex shape originating from the reflectance of gold electrodes and multi-reflection in the antenna gap, as confirmed by our optical simulations (see Supplementary Section B1). Fig. 2c shows the temporal evolution of the differential optical reflectance spectrum $\Delta R/R$ of the reference antenna subject to the incident THz field $F_{x,\mathrm{in}}$ with a peak amplitude of 104 kV/cm, shown in Fig. 2e. Given the field enhancement factor of $|F_z/F_{x,\mathrm{in}}| = 13.3$, this corresponds to an estimated peak gating field in the gap region of the antenna of $F_z$ = 1.4 MV/cm. The THz field leads to a small differential reflectance modulation of the reference antenna structure of $|\Delta R/R|$ < 1%, spectrally spread over the entire range of probe photon energy of 1.75 - 2.15 eV. This modulation is several picoseconds long and does not show any temporal structure present in the incident THz waveform, which is shown in Fig. 2e. We attribute this response to the long-living THz-induced modulation of the refractive index of the spacer layer and/or of the reflectance of the thin gold electrode in the field-enhancement region of the antenna. See Supplementary Section B2 for details. In Fig. 2f, an absolute optical reflectance spectrum $R$ of a reference antenna structure is shown as a function of the THz pump-optical probe delay. It does not demonstrate any significant temporal modulation, and its overall shape resembles that of the static optical reflectance spectrum of the reference antenna without the THz field, shown in Fig. 1a.

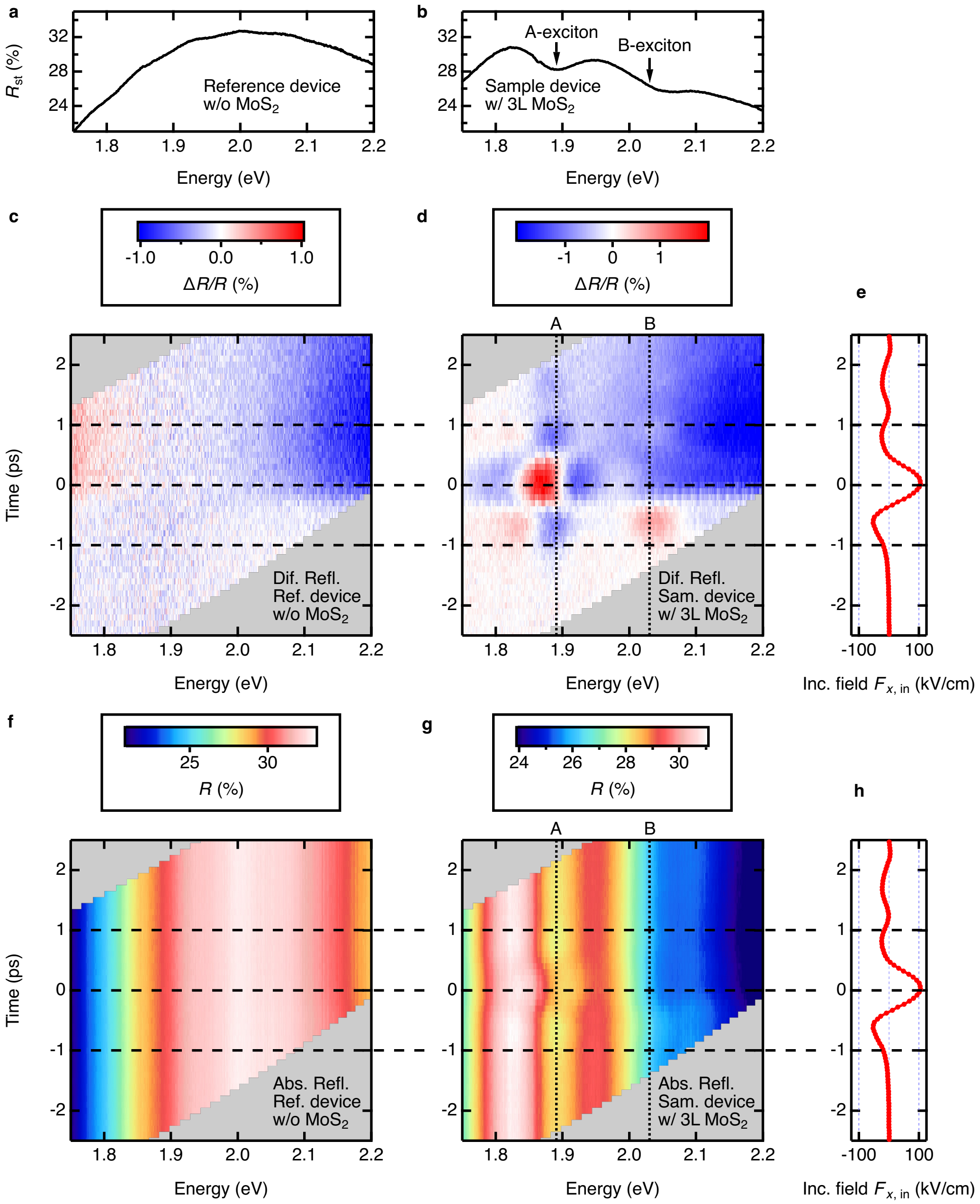

**Figure 2.** THz pump-optical probe (TPOP) response of the reference (without $MoS_2$) and sample (with 3L $MoS_2$) hybrid 3D-2D THz nanoantennas from the production batch α. **a-b**, Static absolute optical reflectance spectra $R_{st}$ of the antenna without $MoS_2$ (**a**) and with 3L $MoS_2$ (**b**) measured in the absence of the THz field. A- and B-exciton resonances in $MoS_2$ are marked with arrows. **c-h**, Transient differential reflectance spectra $\Delta R/R$ (**c-d**) and absolute reflectance spectra $R$ (**f-g**), respectively, observed in a TPOP measurement with incident THz field $F_{x,in}$ shown in **e,h.** Here, (**c,f**) are the results of an antenna without $MoS_2$ and (**d,g**) are that of an antenna with 3L $MoS_2$. Vertical dotted lines in **d,e** indicate the spectral positions of A- and B- excitons. Horizontal dashed lines in **c-h** are guide-to-the-eye to provide the relative timing of a TPOP response in antennas with respect to the incident THz field.

After the reference antenna without the $MoS_2$ sample was characterized, we proceeded with the measurements on an actual device containing the $MoS_2$. Figs. 2b,d,g show the static optical reflectance spectrum $R_{st}$ without the THz field, the time-dependent differential reflectance spectrum $\Delta R/R$, and the time-dependent absolute reflectance spectrum $R$, respectively, of one of our manufactured devices - an antenna containing the 3 layer (3L) flake of $MoS_2$ in its field-enhancement region. The static optical reflectance spectrum of the antenna with 3L $MoS_2$ (Fig. 2b) now shows two pronounced dips around the probe photon energies of 1.89 eV and 2.03 eV. These spectral dips correspond to optical absorption due to well-known A- and B-exciton transitions in $MoS_2$ [9]. When this device was subject to the incident THz field $F_{x,\mathrm{in}}$ with peak amplitude of 104 kV/cm (see Figs. 2 e,h), corresponding to a maximum estimated gating field of $|F_z|$ = 1.4 MV/cm, a very clear spectral and temporal modulation was observed in the differential reflectance spectrum $\Delta R/R$ of the antenna loaded with $MoS_2$ (Fig. 2d). Spectrally, these THz-induced modulations are concentrated around the A- and B-exciton transitions. Temporally, these modulations resemble the structure of the incident THz field, shown in Fig. 2e. The time evolution of the absolute optical reflectance spectrum $R$ of the antenna with $MoS_2$, shown in Fig. 2g, shows modulation of spectral positions of both A- and B-exciton lines, time-correlated with the oscillation of the incident THz field $F_{x,\mathrm{in}}$, shown in Fig. 2h. We have therefore observed a clear effect of THz gating of the $MoS_2$ sample as spectral and temporal modulation of its characteristic A- and B-exciton transitions, an effect also confirmed by our optical simulations (see Supplementary Section B). When the antenna was rotated by 90° with respect to the THz field polarization to avoid the incoupling of the incident THz field into the antenna structure, no measurable modulation of optical probe reflectivity was observed in our devices. In the following, we will perform a detailed analysis of these THz-induced optical modulations of $MoS_2$.

We note that the time evolution of the incident THz field $F_{x,\mathrm{in}}$ and the time evolution of the optical reflectance spectra $R$ and $\Delta R/R$ were measured separately, and the precise relative timing offset between them could not be established experimentally. Therefore, in the data presentation, we chose the timing such that the maximal modulation of optical reflectance of $MoS_2$ would correspond to the strongest THz field in the $F_{x,\mathrm{in}}$ waveform positioned at the time delay of 0 ps. The details of the time offset and the parameters used in our data fitting can be found in Supplementary Section D. We have also attempted to characterize the optical response of our devices under static, DC-bias conditions. However, all attempts to recreate the gating fields of the order of MV/cm in the gap region of our antenna by applying the necessary static voltage to the contact pads of the electrodes led to permanent damage of the devices via dielectric breakdown and unintentional shortcuts, as described in Supplementary Section E. This observation points to yet another advantage of using very fast oscillating THz fields for efficient gating of 2D materials as being non-destructive.

**Quantifying the effect of THz gating in $MoS_2$**

In order to quantify the effect of the THz gating on the excitonic transitions in $MoS_2$ and to extract the THz field-dependent parameters of A- and B-exciton resonances, we have performed the fits of the time-dependent absolute reflectance spectra $R$ of the THz-gated antenna shown in Fig. 2g. The reflectance spectra of the antenna with 3L $MoS_2$ sample could be well described by a third-order polynomial describing the background convex curve, and the two Lorentzian dips representing A- and B-exciton resonances. The details of the fitting are described in Methods Section 5.

Fig. 3a is an image plot of the fitted transient reflectance spectra, and Fig. 3b is the incident THz field $F_{x,\mathrm{in}}$ measured at the position of the antenna. Measured and fitted optical reflectance spectra in the vicinity of A- and B-exciton resonances, corresponding to selected pump-probe delays are shown in Figs. 3c,d, respectively. These pump-probe delays of 0 ps and -0.6 ps correspond to the incident field strength of $F_{x,\mathrm{in}}$ = 104 kV/cm and $F_{x,\mathrm{in}}$ = -52 kV/cm, or to estimated gating fields of $F_z$ = 1.4 MV/cm and $F_z$ = -0.69 MV/cm, respectively. The dashed vertical lines in Fig. 3c,d mark the spectral positions of A- and B- exciton resonances without the THz field. Note that the presented optical reflectance spectra are distorted due to the presence of a spectrally broad convex background. This leads to an apparent shift of spectral minima with respect to the true spectral positions of the exciton resonances, which, however, can be reliably determined from the Lorentzian parts of the fit functions, as presented below. We note that in our experiments, the absolute polarity of the electric field within the incident THz waveform $F_{x,\mathrm{in}}$ could not be determined. Therefore, in the context of our discussion, we define the electric field of the strongest half-cycle in $F_{x,\mathrm{in}}$, presented in Figs. 1-4 as "positive".

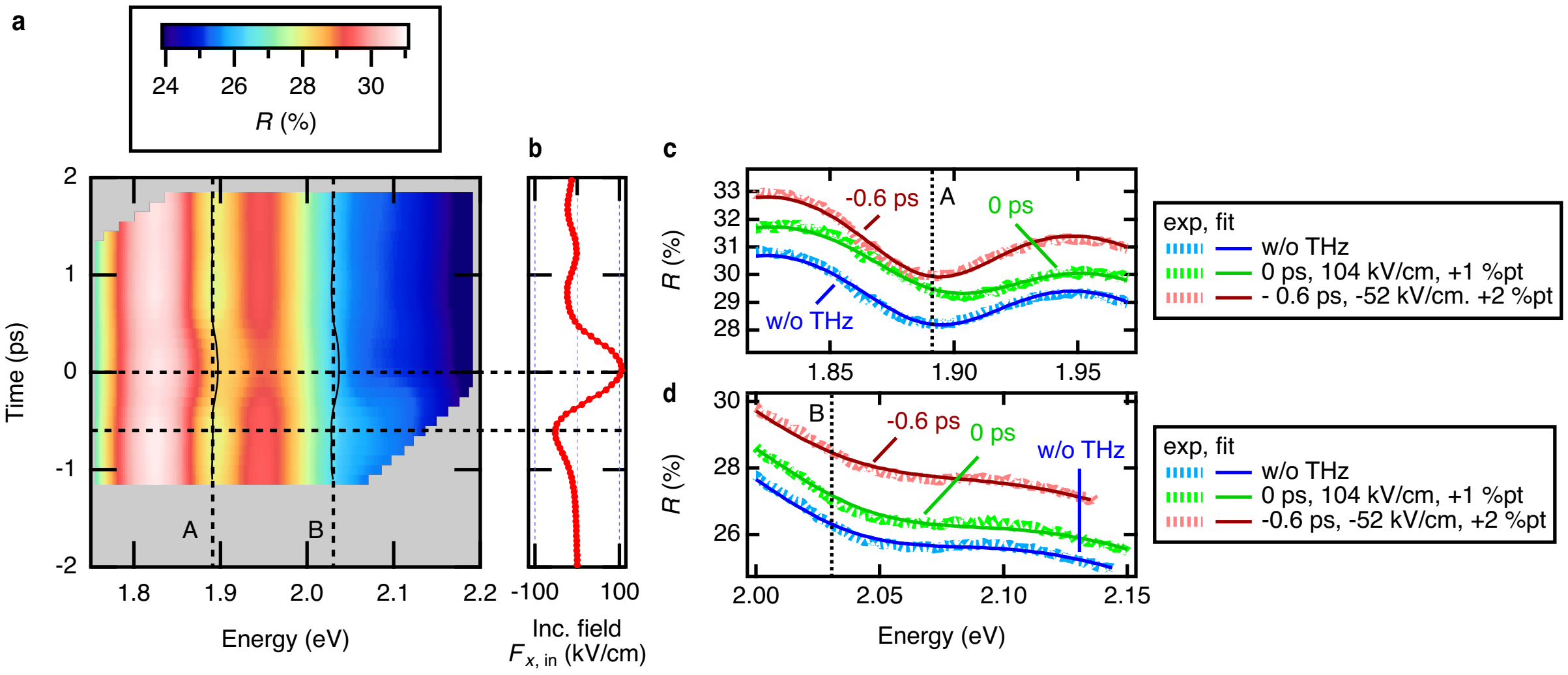

**Figure 3.** The fitting of transient absolute reflectance spectra $R$ of a hybrid 3D-2D THz nanoantenna with 3L $MoS_2$ from the production batch α. **a**, Transient reflectance spectra obtained via fitting of the experimental data. Two vertical dotted lines show the position of A- and B-exciton resonances without the presence of the THz field. Two solid curved lines show the temporal change of position of A- and B-exciton resonances. **b**, incident THz field $F_{x,\text{in}}$. Horizontal dotted lines in **a** and **b** point to time delays of 0 ps and -0.6 ps. **c-d**, Absolute reflectance spectra $R$ in the vicinity of A-exciton (**c**) and B-exciton (**d**) resonances without THz pump (blue), under THz pump of $F_{x,\text{in}} = 104$ kV/cm corresponding to 0 ps time delay (light green, vertically offset by +1 %pt for clarity), and under THz pump of $F_{x,\text{in}} =$ -52 kV/cm corresponding to -0.6 ps time delay (dark red, vertically offset by +2 %pt for clarity). Dotted lines are experimental data, and solid lines are fitting curves. The vertical dotted lines in **c** and **d** show the position of A- and B-exciton resonances, respectively, without the presence of the THz field.

In Figs. 4a,b, we present the dependency of the spectral positions of A- and B- exciton resonances, extracted from the fitting described above, on the THz pump-optical probe delay. The incident THz waveform $F_{x,\text{in}}$ is presented in Fig. 4c as a reference. One observes a clear spectral modulation of the transition energy of both A- and B-excitons, which is time-coherent with the dynamics of the incident THz field. The strongest effect is the blueshift of both resonances at the strongest "positive" half-cycle of the THz field oscillation, and both resonances also show a weaker redshift when the THz field changes its sign. This is the demonstration of the coherent nature of the THz gating field effect on the exciton resonances in $MoS_2$, realizing the ultrafast, sub-picosecond control of the fundamental properties of a 2D material via THz gating.

The maximum THz-driven modulation of resonant energy for both A- and B-exciton was of the order of 6-7 meV. We have established a dependency of the spectral shift of exciton resonances on the applied THz field, using the correlation between the field strength, including its sign, and the time delay within the incident THz waveform $F_{x,\text{in}}$. This correlation is illustrated by color coding in Figs. 4c-e. The THz-field dependency of the spectral positions of A- and B-excitons is presented in Figs. 4d,e, respectively. The strongly dominant blueshift at the "positive" THz gating field and much weaker redshift at the "negative" THz fields are observed for both exciton resonances. We have also tested the response of our devices to the THz gating field with the reversed polarity. For this, we rotated the antenna by 180° about the propagation axis of the incident THz beam (see Fig. 5a). The generally bipolar modulation of the exciton transition energies with applied gating THz fields of opposite signs and the observations related to the reversal of polarity of the THz gating field, which are detailed in the following section, point to an existence of permanent symmetry breaking in the $z$-direction in our structure comprising a hybrid 3D-2D nanoantenna and a $MoS_2$ flake. The possible origin of this symmetry breaking will be addressed below in the Discussion section.

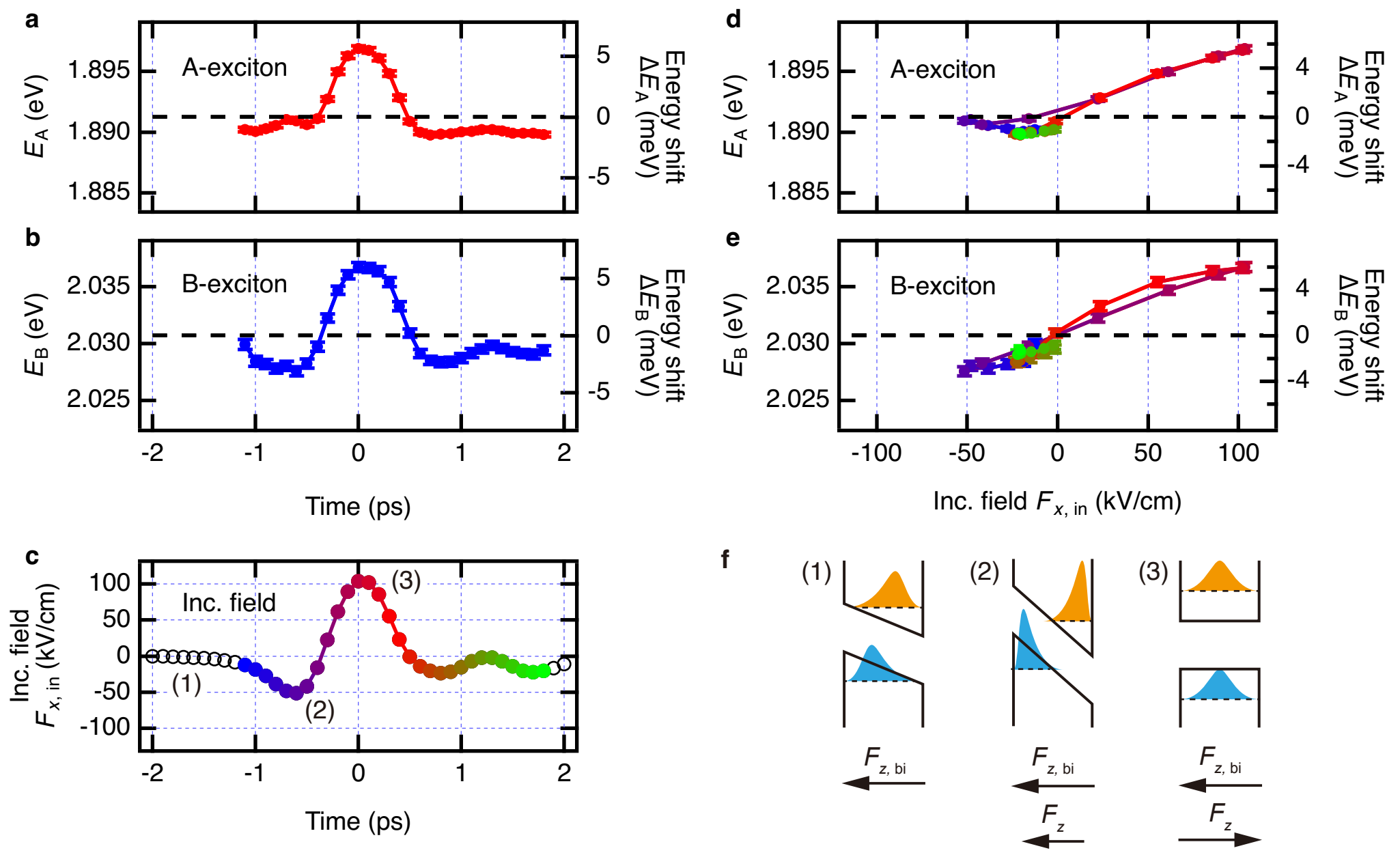

**Figure 4.** Dependency of exciton resonance energy on THz pump-optical probe (TPOP) time delay and on the THz field strength for the hybrid 3D-2D THz nanoantenna with 3L $MoS_2$ from the production batch α. **a-b**, Dependency of the A- and B-exciton resonance energies on the TPOP time delay. Horizontal dashed lines show the resonance positions without the presence of the THz field. **c**, Incident THz field $F_{x,\text{in}}$ in this measurement. Labels (1)-(3) indicate the timing corresponding to the schematic figures (1)-(3) in **f**. **d-e**, Dependency of the resonance energy for A- and B-excitons on the THz field strength. This dependency was established via the time correlation between the time-dependent resonance energies in **a,b** and the time-dependent instantaneous THz field in **c**. This time correlation is illustrated via color coding in **c,d,** and **e**. Error bars in **a,b,d,** and **e** show the standard deviation of the exciton resonance energies as resulting from the corresponding fits. **f,** Illustration of a quantum-confined Stark effect in a 2D semiconductor induced by an out-of-plane THz field $F_z$ under the presence of a built-in field $F_{z,\text{bi}}$. Sub-figures (1)-(3) correspond to the timing labels (1)-(3) in **c**. The solid lines illustrate the quantum-confinement potential, the dashed lines depict the energy levels of the electron and hole, and the areas filled with orange and blue colors depict the squared wave functions of the electron and hole, respectively.

**Performance variation among fabricated devices**

As mentioned above, several devices with varying numbers of $MoS_2$ layers were fabricated and tested in this work. A total of two device batches were produced, termed here α- and β-batches. Besides the difference in the top electrode thickness (25 nm and 50 nm for the devices in $\alpha$- and $\beta$- batches, respectively), another difference between the batches was the annealing condition during the fabrication, as described in Methods Section 1. Below, we present the variation in performance among the fabricated structures, crucial for the explanation of the symmetry breaking within the devices.

In Figs. 5a-f, the measured exciton spectral position shifts for three selected devices are shown: (I) A device with 3L $MoS_2$ from α-batch, (II) a device with 3L $MoS_2$ from β-batch, and (III) a device with 4L $MoS_2$ from α-batch. These measurements include rotation of the antenna structure by 180° in the incident THz beam, thereby inverting the polarity of the THz gating field on $MoS_2$ (see Fig. 5a). Each individual measurement in Figs. 5 b-g is presented in a different color. For the measurement performed with the antenna rotated by 180°, the sign of the incident THz field in the horizontal axes in Figs. 5a-e was inverted. The results shown in Figs. 2-4 and discussed previously are related to the device (I). The individual measurement data for each device can be found in Supplementary Section F and Supplementary Figure Files.

Comparing the devices, we found that the bipolar THz gating can lead to a predominant blueshift (Figs. 5b,c), both blue- and redshifts (Figs. 5d,e), or a predominant redshift (Figs. 5f,g) of the exciton resonances, with both A- and B-exciton resonances following the same shift trend. The possible origin of this behavior is addressed in the Discussion section below.

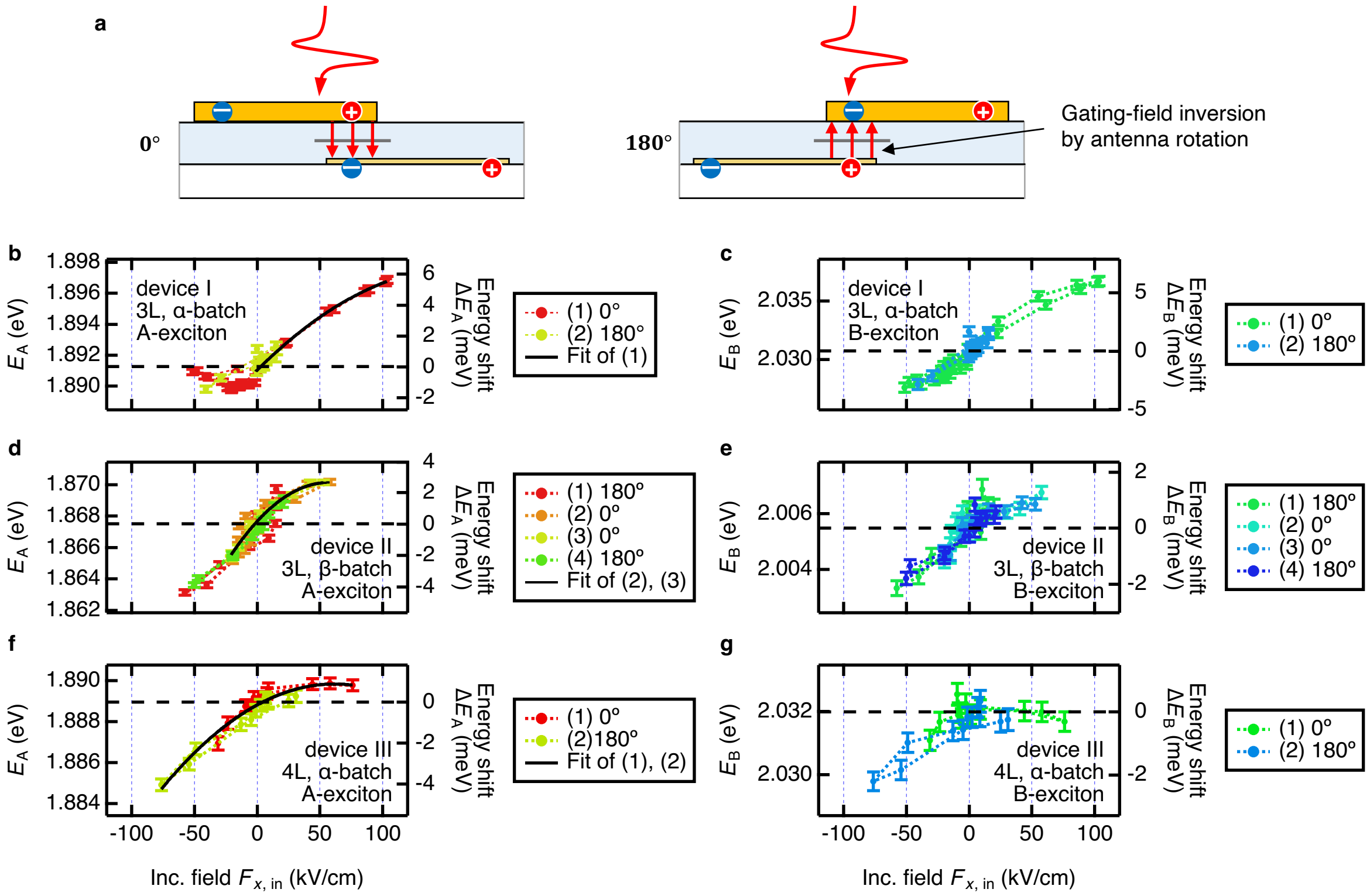

**Figure 5. a**, Illustration of gating field inversion by rotating the antenna by 180° in the incident THz beam. **b-g,** Dependency of the resonance energies for A-exciton (**b,d,f**) and B-exciton (**c,e,g**) on the strength of the incident THz field $F_{x,\text{in}}$, for several tested devices: **b,c** – device I, the antenna with 3L $MoS_2$ from production batch α, **d,e** – device II, the antenna with 3L $MoS_2$ from production batch β, **f,g** device III, the antenna with 4L $MoS_2$ from production batch α. These measurements included inversion of the gating field by rotation of antenna by 180°, as shown in **a**. Solid lines in **b,d,f**: fit of the parabolic part of the dependency of A-exciton resonance energy on the incident THz field $F_{x,\text{in}}$ with a quadratic equation describing the perturbative-regime quadratic quantum confined Stark effect, and the polarizability of the A-exciton established in [22]. These fits allow one to estimate the strength of the built-in field in the tested devices, as well as the THz field enhancement factors of the antennas and the peak THz gating field strength.

## Discussion

As described above, we have demonstrated an efficient THz gating in $MoS_2$ embedded into a hybrid 3D-2D THz nanoantenna. The THz gating resulted in the pronounced shift of transition energy of A- and B-excitons in $MoS_2$, which is time-coherent with the incident broadband THz field. In addition, the polarity of the observed energy shifts varied among the studied devices and included predominant blueshift, bipolar shift, and predominant redshift of the exciton transition in response to the bipolar-oscillating THz gating field (see Figs. 5b-g).

In a 2D semiconductor without broken symmetry in out-of-plane direction, an application of an electric field of *any* polarity along the same direction will lead to an instantaneous redshift of a ground state optical transition – the well-known quantum-confined Stark effect (QCSE). In the perturbative regime, the magnitude of this Stark shift has a quadratic dependency on the electric field strength [44,45]. Let us assume that this electric field in a semiconductor pre-exists, and we refer to it as a built-in field. An application of an *additional* electric field along the same direction, such as, e.g., the THz gating field, depending on its polarity, will lead to either a weakening or to an enhancement of the pre-existing QCSE induced by a built-in field. This will, in turn, result in either the blueshift or the redshift of the optical transition energy, respectively. The direction and strength of the observed shift of the optical transition energy will thus depend on the polarity and strength of the pre-existing built-in field and that of the applied external gating field, as illustrtated schematically in Fig. 4f.

The static field-induced QCSE is by now well studied in $MoS_2$ [22-25] and other 2D semiconductors [26,27]. Here, we compare our THz gating-induced energy shifts of A- and B-exciton transitions with that from existing literature describing the DC field-induced Stark shifts in $MoS_2$. In Ref. [22], Klein et al. studied the QCSE in mono- and few-layer $MoS_2$ films sandwiched between the two enclosing oxide layers, a geometry similar to the design of our hybrid 3D-2D nanoantenna (Fig. 1b). Excitonic Stark shifts of the order of 5-10 meV were observed in response to applied static gating fields of MV/cm scale, which is in line with our observations. It was also found in work [22] that the amount of the Stark shift in single- and multi-layer $MoS_2$ does not depend on the number of layers in the sample but only on the total applied field strength, which is also in agreement with our results. From this, it was concluded that the observed exciton transitions belong to intra-layer excitons in $MoS_2$. Further, it was found that several studied samples had an unintentional built-in electric field in the *z*-direction. The origin of this built-in field can be explained by the presence of the positively charged trap states near the $MoS_2$/oxide interface, such as border traps originating from oxygen vacancies within the oxide layer or interface traps originating from impurities/defects of $MoS_2$ [46-48]. If the concentration of these positive traps is different for the two opposing $MoS_2$/oxide interfaces in the structure, e.g., due to intentional or unintentional difference in layer growth conditions, a potential difference, and hence a net built-in electric field in $MoS_2$ in the *z*-direction will be created, as was indeed demonstrated in [22,46-48]. These uncompensated charged interface traps thus lead to the pre-existing QCSE in $MoS_2$/oxide structures in the *z*-direction, resulting in a generally bipolar shift of excitonic transition energy in response to a bipolar external gating field, which is in full agreement with our observations. For some samples in the literature, a deviation of the Stark shift from its initially quadratic field dependency, followed by its strong saturation with increasing gating field in one certain direction was also reported in [22], again in agreement with our observations presented in Figs. 5b,d. In work [22], this saturation was explained by the charge transfer between $MoS_2$ and the charged trap states at the $MoS_2/Al_2O_3$ interface.

In a later study, Roch et al. [23] similarly observed the QCSE in a $MoS_2$ sample sandwiched between the two hexagonal boron nitride (hBN) layers and subject to MV/cm – scale DC bias fields. The hBN, in contrast to commonly used oxides such as $Al_2O_3$ or $SiO_2$, famously features a very low density of defects and charge trap states. Yet, the observed Stark shifts in [23] were reported to be significantly smaller than those observed by Klein et al. [22] and by us, both using the $MoS_2$/oxide arrangement. This demonstrates the enhancement of exciton polarizability in a $MoS_2$/oxide arrangement as compared to the $MoS_2$/hBN one, once again highlighting the role of the environment in the physics of 2D materials [11-13].

Based on the above, we conclude that our observations—namely, the generally bipolar shift of exciton transition energy under bipolar THz gating and the saturation of the quadratic QCSE at stronger gating fields—are consistent with the results of DC-Stark shift experiments presented in [22,23]. Our findings can be thus explained by the THz-induced QCSE in $MoS_2$ in the presence of a symmetry-breaking built-in electric field applied in the *z*-direction. This built-in field originates from an unintentional imbalance in the concentration of positive trap states at the opposing $MoS_2$/oxide interfaces within the structures, creating a potential difference between the two interfaces. Therefore, the observed variations in device performance can be attributed to different strengths of this built-in field, resulting from variation in interface trap concentrations among the devices. This is likely caused by the variation in fabrication conditions, most notably the device annealing process (see Methods Section 1).

With this physical picture in mind, we use the known dependency of the Stark shift on the applied static bias field in order to experimentally calibrate the THz gating field $F_z$ in our experiments. In the work [22], using $MoS_2$/oxide arrangement and observing the Stark shift similar to ours, the A-exciton energy was described by the quadratic QCSE in the wide range of applied fields, and the polarizability of A-exciton was determined as $\mu = (0.58 \pm 0.25) \times 10^{-8}\,\mathrm{Dm\,V^{-1}}$. Using this value, we fit the parabolic parts of the observed field-dependent A-exciton transition energy $E_\mathrm{A}(F_{x,\mathrm{in}})$ from Figs. in 5b,d,f with the quadratic equation $E_\mathrm{A}(F_{x,\mathrm{in}}) = E_{\mathrm{A},0} - \mu(kF_{x,\mathrm{in}} - F_{z,\mathrm{bi}})^2$, where $E_{\mathrm{A},0}$ is the A-exciton transition energy in $MoS_2$ in the absence of QCSE, $F_{x,\mathrm{in}}$ is the experimentally measured incident THz electric field, $k = |F_z/F_{x,\mathrm{in}}|$ is the field enhancement factor of our antenna, and $F_{z,\mathrm{bi}}$ is the pre-existing built-in field in the $MoS_2$ discussed above. For details, see Methods Section 6. The resulting fits are shown as black solid lines in Figs. 5b,d,f, yielding the following parameters. The exciton transition energy in the absence of QCSE was found to be $E_{\mathrm{A},0} = 1.8977 \pm 0.0008$ eV, $1.8701 \pm 0.0004$ eV, and $1.8898 \pm 0.0001$ eV for devices I, II, and III, respectively, which is in good agreement with Refs. [9,10,22,23]. The sample-to-sample variation in $E_{\mathrm{A},0}$ of about 15 meV is related to the difference in number of layers [9,10] and to dielectric disorder in the samples [12]. The built-in field in our antennas was found to be $F_{z,\mathrm{bi}}$ = $2.4 \pm 0.1$ MV/cm, $1.5 \pm 0.1$ MV/cm, and $0.94 \pm 0.07$ MV/cm for devices I, II, and III, respectively, which is consistent with the values reported for similar $MoS_2$/oxide structures in Ref. [22]. We also determined the field enhancement factors of $k = 15 \pm 2, 25 \pm 4,$ and $15 \pm 1$ for devices I, II, and III, respectively. These experimental values are of the same order as our theoretical estimate of $k = |F_z/F_{x,\mathrm{in}}|$ = 13.3. While the experimental and the calculated field enhancement factors are in good agreement for devices I and III, the experimental value exceeds the theoretical prediction by almost a factor of two for device II. This scattering among the experimentally measured field enhancement factors is most likely caused by the variation of antenna fabrication conditions. The fact that our experimental values always exceed the theoretical ones can originate from the conservatively underestimated value of gold electrode conductivity used in our FDTD simulations, in turn leading to a smaller computed THz-driven current in the top and bottom electrodes and hence to a weaker conversion into the THz gating field than measured experimentally.

As already mentioned, in order to prevent potential damage to our devices, in our experiments, we kept the maximum gating field at a level not exceeding $|F_z|$ = 1.2 – 1.5 MV/cm by monitoring the observed Stark shifts and using the field calibration as described above. For this, the incident THz field was attenuated in order not to exceed this maximum value of $|F_z|$ when necessary.

We note that apart from the THz-induced QCSE as described above, several alternative physical mechanisms could potentially also lead to the modification of optical absorption at the exciton resonances. However, none of these mechanisms would be fully compatible with our observations in the same way as the QCSE. Free-carrier generation and carrier heating in the $MoS_2$ via a strong THz field can be eliminated since these effects will be insensitive to the polarization of the driving field, in contradiction to our observations shown in Fig. 5. THz-field-driven doping-level change in the $MoS_2$, which could occur if a $MoS_2$ flake were in good electrical contact with either electrode in the field-enhancement region, is also not likely. In this case, such an electrical spacer breakdown or current leak would have to occur in all our measured devices without exception. This is highly improbable, especially since all our devices feature well-insulating 75 nm-thick ALD-grown $Al_2O_3$ spacer layers.

**Conclusions and outlook**

We have demonstrated a pronounced THz field effect in a 2D semiconductor $MoS_2$ using a specially designed hybrid 3D-2D nanoantenna. It converts an incident broadband THz electromagnetic field into a vertical gating field within the material, simultaneously amplifying it by more than one order of magnitude to MV/cm level, while maintaining its full bandwidth of at least 0.1 – 2.5 THz. The THz field effect was observed via the time-resolved optical probing of characteristic exciton resonances in $MoS_2$, and the specific gating mechanism in this case was the quantum-confined Stark effect, which is time-coherent with the driving THz field.

Our approach to THz gating in 2D materials can be readily applied to numerous advances in nanotechnology and fundamental science. Besides the THz-driven QCSE described above, the ultrafast control of the doping level in 2D materials embedded in a hybrid 3D-2D nanoantenna is feasible. Here, the THz-injected carrier densities could become as high as $10^{13}$ cm$^{-2}$ in case of direct contact between the 2D material and one of the electrodes (see Supplementary Section B4), thus enabling f.ex., the direct THz-rate modulation of 2D transistors [2,15,16] via ultrafast control of the doping level in the channel. We also note that the carrier density of the order of $10^{13}$ cm$^{-2}$ is sufficient even to induce a phase transition in a material such as $MoS_2$ [16]. In case of direct contact between a 2D material and both electrodes, an ultrafast out-of-plane current in the 2D material can be induced, f.ex., enabling the ultrafast control of 2D memristors [7]. Other possible applications of our hybrid 3D-2D nanoantenna structures could be sub-picosecond activation of 2D photodetectors and optical modulators [4,17], memristor networks [49], direct THz control over the phase transitions including quantum phases in moiré systems [8,16,20,29], ultrafast manipulation of neutral and charged excitons in 2D semiconductors [28], direct electric field activation of THz layer-breathing phonons in van der Waals heterostructures [50,51], and many more.

Our THz gating method has significant potential for further development. The nanoantenna geometry and structure can be further optimized for more efficient resonant or non-resonant incoupling and conversion of the input control signals, also from other spectral ranges such as mid-infrared. It can also be further optimized to allow the non-destructive application of static bias fields of MV/cm scale, thereby realizing an active control over the polarity and depth of the THz field effect in 2D materials. The dielectric spacer layers enclosing the 2D material can be engineered in order to deterministically create the built-in bias fields within the structures via control of the interface trap states. Further, the electronic polarizability in a 2D material, and hence its basic response to the gating fields in the antenna, can be controlled via the suitable choice of dielectric layer materials such as high defect density $Al_2O_3$ or $SiO_2$, or low defect density hBN.

In summary, our results pave the way for novel technology and fundamental science of 2D materials where field effects induced on a sub-picosecond time scale are essential.

## Author contributions

T.H. and D.T. conceived and coordinated the project. T.H., S.N., S.F., H.S., and A.T. designed, fabricated and performed basic characterization of the hybrid 2D-3D THz nanoantenna. T.H., W.Z. and H.H. designed and built the measurement infrastructure. T.H. conducted the optical measurements and analyzed the data together with D.T. S.R. performed optical simulations. S.F., H.S., A.T., and D.T. supervised the project. T.H. and D.T. wrote the paper. All co-authors discussed the results and commented on the manuscript.

## Funding

We acknowledge the financial support from the European Union's Horizon Europe (HORIZON-MSCA-2021-PF-01, Project number 101060427 - UCoCo), Horizon 2020 research and innovation program (EXTREME-IR - Grant Agreement No. 964735), Deutsche Forschungsgemeinschaft (DFG) within Project No. 468501411-SPP2314 INTEGRATECH and Project No. 518575758 HIGHSPINTERA, Bundesministerium für Bildung und Forschung (BMBF) within Project No. 05K2022 PBA Tera-EXPOSE, and Bielefelder Nachwuchsfond.

## Acknowledgement

We are grateful to Dr. Mohammed Nouh, Dr. Arslan Usman, Dr. Hilary Masenda, and Prof. Martin Koch for useful instructions regarding the 2D materials fabrication. We are also grateful to Dr. Shinya Takahashi, Dr. Kohei Nagai, Dr. Satoshi Kusaba, Dr. Kento Uchida, Prof. Takashi Arikawa, and Prof. Koichiro Tanaka for fruitful discussions throughout the entire project, and assisting with the characterization of the test devices using their real-time THz near-field microscope. We are grateful to Prof. Heejae Kim, Prof. Ryusuke Matsunaga, Prof. Ikufumi Katayama, and members of Ultrafast Science research unit at Bielefeld University for fruitful discussions and support throughout this project.

## Data availability

All the data supporting this study and its findings are available within the article and Supplementary Information. Additional supporting data for this study are available from the corresponding authors upon request. Source data are provided with this paper.

## Ethics declarations

**Competing interests:** Authors declare no competing interests.

## References

[1] Akinwande, D. *et al.* Graphene and two-dimensional materials for silicon technology. *Nature* **573**, 507–518 (2019).
[2] Fiori, G. *et al.* Electronics based on two-dimensional materials. *Nat. Nanotechnol.* **9**, 768–779 (2014).
[3] Liu, X. & Hersam, M. 2D materials for quantum information science. *Nat. Rev. Mater.* **4**, 669–684 (2019).
[4] Wang, Q. H., Kalantar-Zadeh, K., Kis, A., Coleman, J. N. & Strano, M. S. Electronics and optoelectronics of two-dimensional transition metal dichalcogenides. *Nat. Nanotechnol.* **7**, 699–712 (2012).
[5] Jariwala, D., Davoyan, A. R., Wong, J. & Atwater, H. A. Van der Waals Materials for Atomically-Thin Photovoltaics: Promise and Outlook. *ACS Photonics* **4**, 2962–2970 (2017).
[6] Ahn, E. C. 2D materials for spintronic devices. *Npj 2D Mater. Appl.* **4**, 1–14 (2020).
[7] Ge, R. *et al.* Atomristor: Nonvolatile resistance switching in atomic sheets of transition metal dichalcogenides. *Nano Lett.* **18**, 434–441 (2018).

[8] Andrei, E. Y. *et al.* The marvels of moiré materials. *Nat. Rev. Mater.* **6**, 201–206 (2021).
[9] Splendiani, A. *et al.* Emerging photoluminescence in monolayer $MoS_2$. *Nano Lett.* **10**, 1271–1275 (2010).
[10] Mak, K. F., Lee, C., Hone, J., Shan, J. & Heinz, T. F. Atomically thin $MoS_2$: a new direct-gap semiconductor. *Phys. Rev. Lett.* **105**, 136805 (2010).
[11] Novoselov, K. S., Mishchenko, A., Carvalho, A. & Castro Neto, A. H. 2D materials and van der Waals heterostructures. *Science* **353**, aac9439 (2016).
[12] Raja, A. *et al.* Dielectric disorder in two-dimensional materials. *Nat. Nanotechnol.* **14**, 832–837 (2019).
[13] Dai, Z., Liu, L. & Zhang, Z. Strain engineering of 2D materials: Issues and opportunities at the interface. *Adv. Mater.* **31**, e1805417 (2019).
[14] Roldán, R. & Castellanos-Gomez, A. A new bandgap tuning knob. *Nat. Photonics* **11**, 407–409 (2017).
[15] Novoselov, K. S. *et al.* Electric field effect in atomically thin carbon films. *Science* **306**, 666–669 (2004).
[16] Radisavljevic, B. & Kis, A. Mobility engineering and a metal-insulator transition in monolayer $MoS_2$. *Nat. Mater.* **12**, 815–820 (2013).
[17] Xia, F., Mueller, T., Lin, Y.-M., Valdes-Garcia, A. & Avouris, P. Ultrafast graphene photodetector. *Nat. Nanotechnol.* **4**, 839–843 (2009).
[18] Kovalev, S. *et al.* Electrical tunability of terahertz nonlinearity in graphene. *Sci Adv* **7**, (2021).
[19] Chaves, A. *et al.* Bandgap engineering of two-dimensional semiconductor materials. *npj 2D Materials and Applications* **4**, 1–21 (2020).
[20] Kim, J. *et al.* Observation of tunable band gap and anisotropic Dirac semimetal state in black phosphorus. *Science* **349**, 723–726 (2015).
[21] Chu, T., Ilatikhameneh, H., Klimeck, G., Rahman, R. & Chen, Z. Electrically Tunable Bandgaps in Bilayer $MoS_2$. *Nano Lett.* **15**, 8000–8007 (2015).
[22] Klein, J. *et al.* Stark Effect Spectroscopy of Mono- and Few-Layer $MoS_2$. *Nano Lett.* **16**, 1554–1559 (2016).
[23] Roch, J. G. *et al.* Quantum-Confined Stark Effect in a $MoS_2$ Monolayer van der Waals Heterostructure. *Nano Lett.* **18**, 1070–1074 (2018).
[24] Leisgang, N. *et al.* Giant Stark splitting of an exciton in bilayer $MoS_2$. *Nat. Nanotechnol.* **15**, 901–907 (2020).
[25] Peimyoo, N. *et al.* Electrical tuning of optically active interlayer excitons in bilayer $MoS_2$. *Nat. Nanotechnol.* **16**, 888–893 (2021).
[26] Wang, Z., Chiu, Y.-H., Honz, K., Mak, K. F. & Shan, J. Electrical Tuning of Interlayer Exciton Gases in $WSe_2$ Bilayers. *Nano Lett.* **18**, 137–143 (2018).
[27] Wilson, N. P., Yao, W., Shan, J. & Xu, X. Excitons and emergent quantum phenomena in stacked 2D semiconductors. *Nature* **599**, 383–392 (2021).
[28] Mak, K. F. *et al.* Tightly bound trions in monolayer $MoS_2$. *Nat. Mater.* **12**, 207–211 (2013).
[29] Wang, Y. *et al.* Direct electrical modulation of second-order optical susceptibility via phase transitions. *Nature Electronics* 1–6 (2021).
[30] Saeed, M. *et al.* Graphene-based microwave circuits: A review. *Adv. Mater.* **34**, e2108473 (2022).
[31] Schwierz, F., Pezoldt, J. & Granzner, R. Two-dimensional materials and their prospects in transistor electronics. *Nanoscale* **7**, 8261–8283 (2015).
[32] Wu, Y. *et al.* High-frequency, scaled graphene transistors on diamond-like carbon. *Nature* **472**, 74–78 (2011).
[33] Fischer, H. & Martin, O. J. F. Engineering the optical response of plasmonic nanoantennas. *Opt. Express* **16**, 9144–9154 (2008).
[34] Singh, P., Zhang, J., Engel, D., Fingerhut, B. P. & Elsaesser, T. Transient terahertz stark effect: A dynamic probe of electric interactions in polar liquids. *J. Phys. Chem. Lett.* **14**, 5505–5510 (2023).
[35] Robertson, J. High dielectric constant oxides. *Eur. Phys. J. - Appl. phys.* **28**, 265–291 (2004).
[36] Palumbo, F. *et al.* A review on dielectric breakdown in thin dielectrics: Silicon dioxide, high- k , and layered dielectrics. *Adv. Funct. Mater.* **30**, 1900657 (2020).
[37] Cohen, R. W., Cody, G. D., Coutts, M. D. & Abeles, B. Optical Properties of Granular Silver and Gold Films. *Phys. Rev. B Condens. Matter* **8**, 3689–3701 (1973).
[38] Walther, M. *et al.* Terahertz conductivity of thin gold films at the metal-insulator percolation transition. *Phys. Rev. B Condens. Matter* 76, 125408 (2007).

[39] Kampfrath, T., Tanaka, K. & Nelson, K. A. Resonant and nonresonant control over matter and light by intense terahertz transients. *Nat. Photonics* **7**, 680–690 (2013).
[40] Hebling, J., Almasi, G., Kozma, I. & Kuhl, J. Velocity matching by pulse front tilting for large area THz-pulse generation. *Opt. Express* **10**, 1161–1166 (2002).
[41] Hirori, H., Doi, A., Blanchard, F. & Tanaka, K. Single-cycle terahertz pulses with amplitudes exceeding 1 MV/cm generated by optical rectification in $LiNbO_3$. *Appl. Phys. Lett.* **98**, 091106 (2011).
[42] Nahata, A., Weling, A. S. & Heinz, T. F. A wideband coherent terahertz spectroscopy system using optical rectification and electro-optic sampling. Appl. Phys. Lett. 69, 2321–2323 (1996).
[43] Dubietis, A., Tamošauskas, G., Šuminas, R., Jukna, V. & Couairon, A. Ultrafast supercontinuum generation in bulk condensed media. *Lith. J. Phys.* **57**, (2017).
[44] Miller, D. A. B. *et al.* Band-edge electroabsorption in quantum well structures: The quantum-confined stark effect. *Phys. Rev. Lett.* **53**, 2173–2176 (1984).
[45] Trallero Giner, C. & López Gondar, J. Exact wave functions and energy levels for a quantum well with an applied electric field. *Physica* **138**, 287–294 (1986).
[46] Zhao, P. *et al.* Evaluation of border traps and interface traps in $HfO_2$ /$MoS_2$ gate stacks by capacitance–voltage analysis. *2d Mater.* **5**, 031002 (2018).
[47] Song, X., Xu, J., Liu, L., Lai, P.-T. & Tang, W.-M. Improved interfacial and electrical properties of few-layered $MoS_2$ FETs with plasma-treated $Al_2O_3$ as gate dielectric. *Appl. Surf. Sci.* **481**, 1028–1034 (2019).
[48] Zhao, P. *et al.* Understanding the Impact of Annealing on Interface and Border Traps in the Cr/$HfO_2$/$Al_2O_3$/$MoS_2$ System. *ACS Appl. Electron. Mater.* **1**, 1372–1377 (2019).
[49] Aguirre, F. *et al.* Hardware implementation of memristor-based artificial neural networks. *Nat. Commun.* **15**, 1974 (2024).
[50] Lui, C. H. *et al.* Observation of interlayer phonon modes in van der Waals heterostructures. *Phys. Rev. B Condens. Matter* **91**, 165403 (2015).
[51] Miao, X., Zhang, G., Wang, F., Yan, H. & Ji, M. Layer-Dependent Ultrafast Carrier and Coherent Phonon Dynamics in Black Phosphorus. *Nano Lett.* **18**, 3053–3059 (2018).
[52] Castellanos-Gomez, A. *et al.* Deterministic transfer of twom-dimensional materials by all-dry viscoelastic stamping. *2D Mater.* **1**, 011002 (2014).
[53] Nie, Z. *et al.* Ultrafast carrier thermalization and cooling dynamics in few-layer $MoS_2$. *ACS Nano* **8**, 10931–10940 (2014).

## Methods

### 1. Device Fabrication

The hybrid 2D-3D THz nanoantenna was fabricated as follows: (1) A 1.1-mm thick alkaline earth boro-aluminosilicate glass (Corning Eagle XG) substrate was prepared. (2) An 8-nm-thick Au layer was deposited via electron-beam lithography and sputtering to form the bottom electrode. (3) A 42-nm-thick Au layer was added for the connection line and contact pad of the bottom electrode via electron-beam lithography and sputtering. (4) A 75-nm-thick $Al_2O_3$ layer was grown via atomic-layer deposition (ALD) as the bottom half of the spacer layer. (5) A few-layer $MoS_2$ flake was fabricated via mechanical exfoliation technique. (6) The $MoS_2$ flake was transferred above the square of the bottom electrode using the viscoelastic stamp technique [52]. (7) A 75-nm-thick $Al_2O_3$ layer was grown via ALD as the top half of the spacer layer. (8) The bottom contact pad was exposed by etching the top $Al_2O_3$ layer. (9) A 3-nm-thick Cr and a thick Au layer were deposited via electron-beam lithography and sputtering to form the top electrode. The thickness of the thick Au layer was 25 nm for $\alpha$-batch and 50 nm for $\beta$-batch. Samples of $\beta$-batch were annealed at 250°C for 1 hour in ambient conditions after step (9). The layer number of the $MoS_2$ flake was determined via Raman microscopy before step (6). The shape of the flake was monitored with an optical microscope to confirm that the desired layer number had been transferred to the target location.

### 2. Device Characterization

The leak current and dielectric-breakdown threshold for devices without $MoS_2$ flake were measured via a two-terminal method, as shown in Supplementary Section E. Although the samples from $\alpha$-batch had large sample-to-sample variation of leak current and breakdown threshold, these devices typically had a leak current of up to 8 to 40 μA under a DC voltage of 110 to 150 V and underwent dielectric breakdown at larger voltages. The samples from $\beta$-batch had much less variation, and had a leak current of 200 pA under a DC voltage of 20 V. The similar TPOP response observed for samples from $\alpha$- and $\beta$-batch regardless of the large difference of the insulation of the leak current indicates that the TPOP response is not caused by the leak current.

Photoluminescence (PL) spectra of the $MoS_2$ at the antenna gap were measured via PL microscopy. For each device, a PL peak corresponding to A-exciton luminescence was observed. This result supports our assignment of the A-dip to the A-exciton transition (for detail, see Supplementary Section G.)

**3. Optical Measurements**

We used femtosecond laser from a Ti;Sapphire regenerative amplifier (Solstice, Spectraphysics) with center wavelength of 800 nm, spectral bandwidth of 50 nm (FWHM), repetition rate of 1 kHz, pulse energy of 6 mJ for the reflection microspectroscopy, TPOP reflection microspectroscopy, and electro-optic (EO) sampling measurement. The laser was divided into THz-generation and optical-probe pulse via a beam splitter. The probe pulse attenuated to approximately 1 mW was loosely focused on a 4-mm-thick sapphire crystal to generate a white-light pulse as a probe of the microspectroscopy covering the range of 1.75-2.2 eV. The white light was split into sample and reference beam using a cube beam splitter to obtain the reflectance spectra. An imaging spectrometer (Shamrock 303i, Andor) with a grating of 150 l/mm and a CCD camera (iKon-M 934, Andor) was used to simultaneously measure the spectra of the sample and reference beam focused in the entrance slit of spectrometer. The spectral resolution was characterized as 0.8 nm (FWHM) by measuring a spectrum of a HeNe laser.

For the reflection microspectroscopy, the sample beam attenuated to 100 fJ/pulse was focused on the field-enhancement region of the antenna from the backside of the substrate, using a 20x objective lens (LCD Plan Apo NIR 20x (t1.1), Mitsutoyo.). The size of the focused beam was approximately 2 μm as radius of the dark ring of Airy disk, corresponding to peak fluence of 3 $\mu J/cm^2$. The spectra of the reflected sample beam and reference beam were measured simultaneously. A reflection spectrum of protected silver mirror (P01, Thorlabs) was measured to calibrate the spectral transfer function of the setup and to derive the absolute static reflectance spectra $R_{\mathrm{st}}$ of the devices.

For the TPOP measurement, the THz-generation pulse of approximately 2 W was used for THz-pulse generation in a $LiNbO_3$ with a tilted-pulsefront scheme [41,42]. The THz pulse was focused on the antenna device from the front side of the substrate. The sample beam attenuated to 900 fJ/pulse was focused on the field-enhancement region using a 10x objective lens. The radius of dark ring of Airy disk was approximately 5 μm, corresponding to the fluence of 4 $\mu J/cm^2$. By measuring the spectra of the reflected sample beam and reference beam while chopping the THz pump, transient differential reflectance spectra $\Delta R/R$ under the THz pump was obtained. The time delay between the pump and probe was swept to obtain the transient differential reflectance. The amplitude of THz pulse was attenuated with high resistivity silicon wafers to approximately 100 kV/cm to avoid the dielectric breakdown of the devices which was observed in the higher field. The transient absolute reflectance spectra were obtained from $R_{\mathrm{st}}$ and $\Delta R/R$ as $R = R_{\mathrm{st}}(1 + \Delta R/R)$. We note that the out-of-plane THz field in the field-enhancement region has slight spatial inhomogeneity, and the observed spectral change is the average of the effect of the inhomogeneous field. However, the inhomogeneity is small enough and does not affect our observation in any significant way (see Supplementary Section A6).

An EO sampling measurement [42] at the sample position in a transmission geometry with a 1-mm thick ZnTe was performed to measure the THz-field waveform. For more detail on the measurements with the setup geometry, see Supplementary Section H.

**4. De-chirping of the Optical Probe Pulse**

The chirp of the optical probe pulse was separately characterized, and the numerical chirp correction was applied in the data processing in order to ensure the maximum time resolution in our measurements of ca 100 fs. We note that due to a very low pulse fluence of the order of 1 $\mu J/cm^2$, the probe field interaction with the antenna structure and the 2D material remains linear for both chirped and compressed probe pulses [53] with and without the THz field present. Therefore, numerical chirp correction of the probe pulse, performed at the stage of the data processing, cannot influence the dynamics of the presented results. See Supplementary Section C for more detail.

**5. Fitting of Transient Reflectance Spectra**

In the fitting of transient absolute reflectance spectra, we fitted the reflectance spectra at each TPOP time delay with the following equation: $R(E,t) = f(E) - \mathrm{Lor_A}(E,t) - \mathrm{Lor_B}(E,t)$, where $f(E)$ is the third-order polynomial which does not vary in time, and $\mathrm{Lor_A}(E,t)$ and $\mathrm{Lor_B}(E,t)$ are Lorentzian shape corresponding to A- and B-dips. Amplitudes, widths, and center energies of $\mathrm{Lor_A}(E,t)$ and $\mathrm{Lor_B}(E,t)$ are fitting parameters at each time delay. Global fitting was used to fit all the spectra with the common time-constant background $f(E)$ and the time-dependent Lorentzian parameters.

**6. Fitting of Exciton Transition Energy**

In the fitting of exciton transition energy against incident THz field in Fig. 5, to fit the data point without significant saturation in the lower-energy side while maximizing the available data points, we fitted in the following way: For Fig 5a, we used the trace (1) and limited the fitting range to the range without significant saturation in the lower-energy side. In the fitting of Fig. 5c and e, we used multiple traces without significant saturation on the lower-energy side without limiting the range and performed a global fitting by linking all the fitting parameters. The obtained fitting parameters are as follows: From the fitting of trace (1) in Fig. 5b, $E_{A,\mathrm{init}} = 1.8977 \pm 0.0008$ eV, $F_{z,\mathrm{bi}} = 2.4 \pm 0.1$ MV/cm, $k = 15 \pm 2$, from the global-fitting of traces (2) and (3) in Fig. 5d, $E_{A,\mathrm{init}} = 1.8701 \pm 0.0004$ eV, $F_{z,\mathrm{bi}} = 1.5 \pm 0.1$ MV/cm, $k = 25 \pm 4$, and from the global-fitting of traces (1)-(3) in Fig. 5f, $E_{A,\mathrm{init}} = 1.8898 \pm 0.0001$ eV, $F_{z,\mathrm{bi}} = 0.94 \pm 0.07$ MV/cm, $k = 15 \pm 1$.

**Supplementary Information**

# Terahertz field effect in a two-dimensional semiconductor $MoS_2$

Tomoki Hiraoka[1], Sandra Nestler[2], Wentao Zhang[1], Simon Rossel[1], Hassan A. Hafez[1], Savio Fabretti[1], Heike Schlörb[2], Andy Thomas[2,3], and Dmitry Turchinovich[1]

[1]Fakultät für Physik, Universität Bielefeld, 33615 Bielefeld, Germany
[2]Leibniz-Institut für Festkörper- und Werkstoffforschung, Helmholtzstraße 20, 01069 Dresden, Germany
[3]Institut für Festkörper- und Materialphysik, Technische Universität Dresden, Haeckelstraße 3, 01069 Dresden, Germany

**Contents**

**Section A. Simulation of THz-field response of 2D-3D hybrid THz antenna**

We have simulated the THz-field response of the hybrid 2D-3D antenna structure with a finite difference time domain (FDTD) method. Section A1 shows the actual dimensions of the antenna structure, which was also used for the simulation. Section A2 shows the characterization of THz-conductivity of 8-nm-thick gold film, which was necessary for the simulation. Sections A3, A4, and A5 show the settings, convergence test, and result of the simulation, respectively.

**A1. Dimensions of 2D-3D hybrid THz antenna**

Figure A1-1 shows the whole device structure with information on the thickness of the layers. Figure A1-2 and 1-3 show the in-plane dimensions of the bottom and top electrodes, respectively.

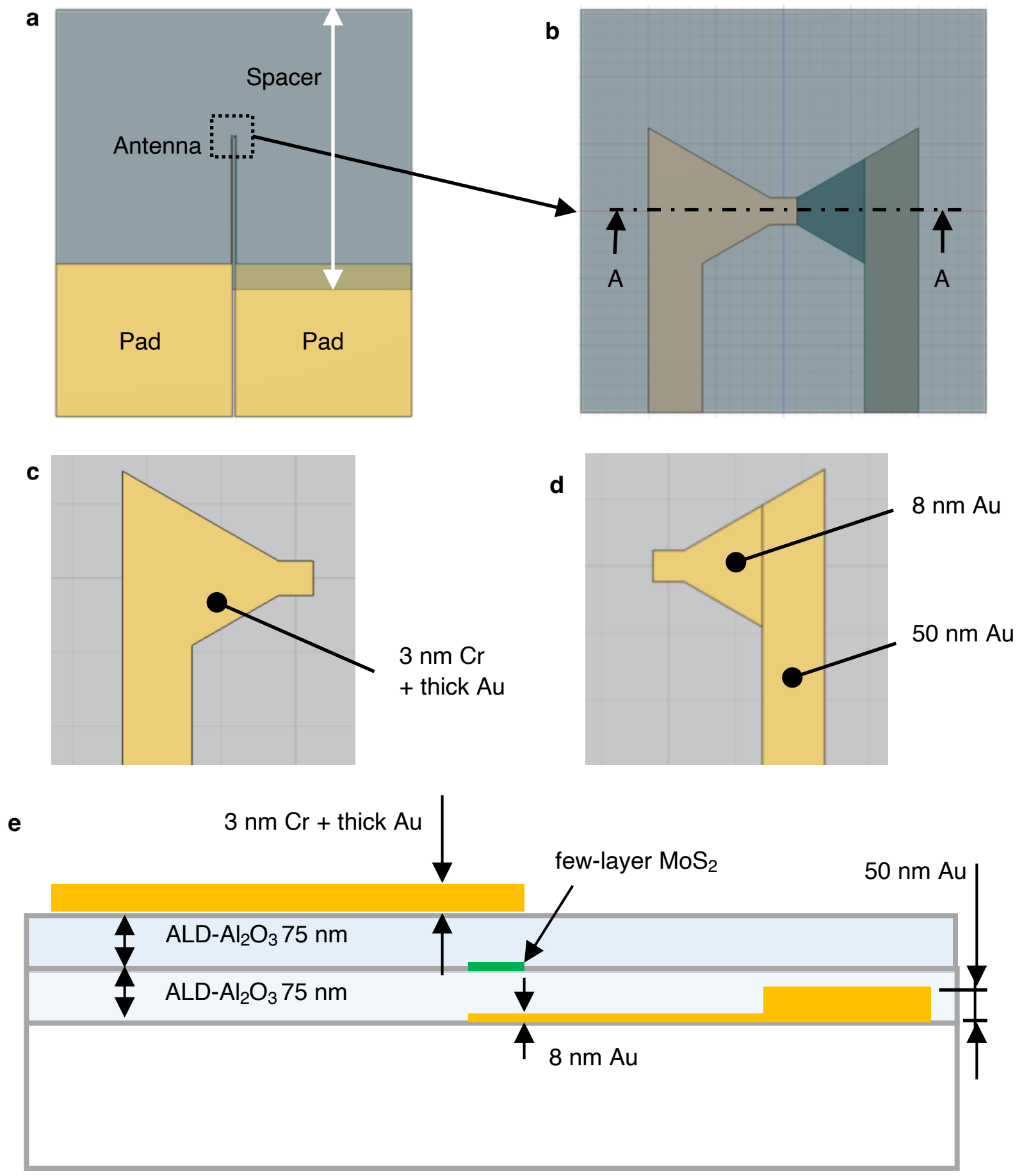

**Figure A1-1.** The whole device structure with information on the thickness of layers. **a,** Top view of the entire structure on the substrate. The two parts noted as Pad are the contact pads for DC-biasing. **b,** Enlarged view of the antenna. The top and bottom electrodes are connected to the contact pads via connection lines. **c,** Enlarged view of the top electrode only. The thickness of the top electrode is 25 nm for α-batch and 50 nm for β-batch devices, respectively. **d,** Enlarged view of the bottom electrode only. **e,** A-A section of the device with information on thickness of the layers.

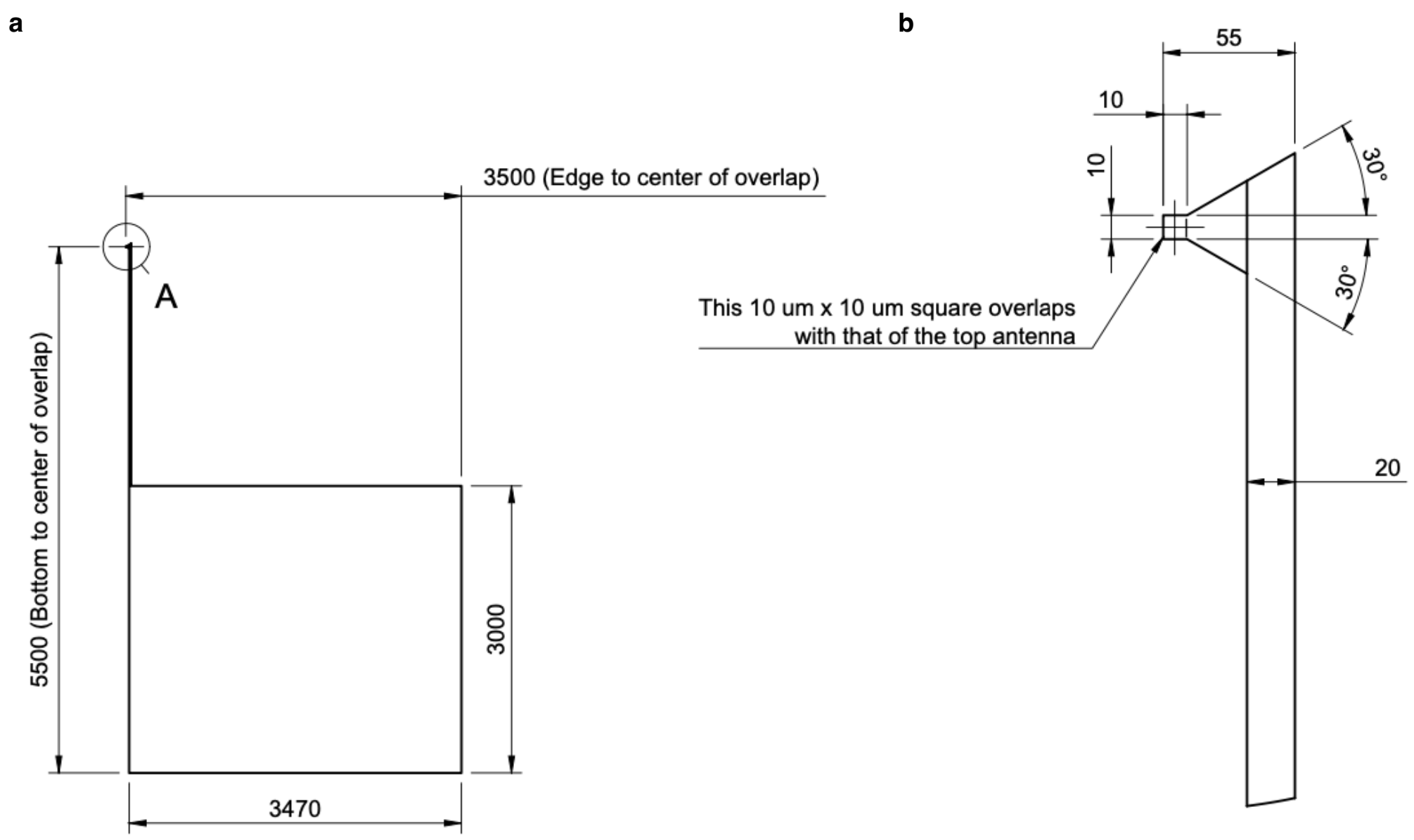

**Figure A1-2.** In-plane dimension of the bottom electrode. **a**, Whole structure. **b**, Enlarged view of A in **a**. The unit of dimensions is μm.

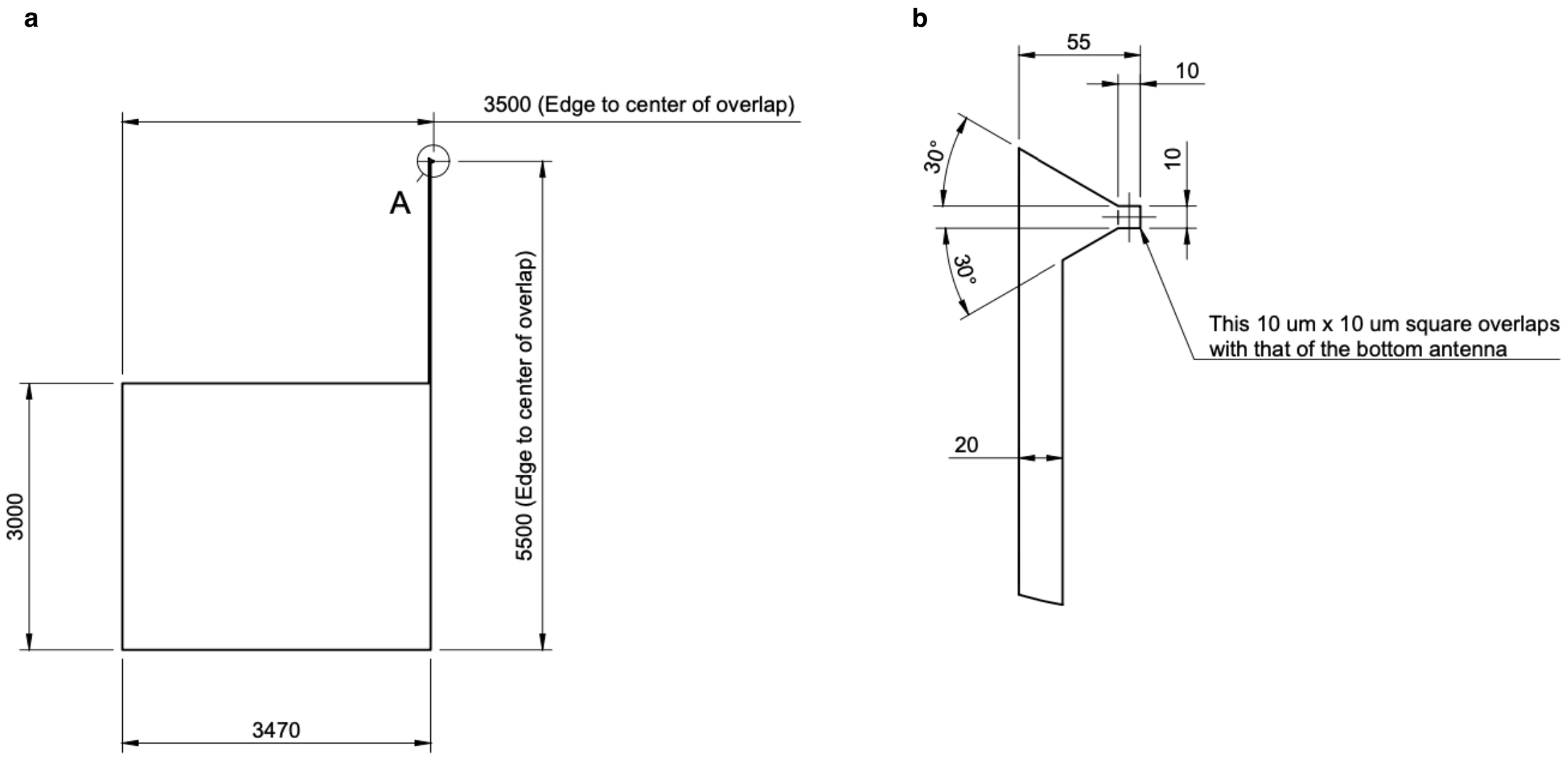

**Figure A1-3.** In-plane dimension of the top electrode. **a**, Whole structure. **b**, Enlarged view of A in **a**. The unit of dimensions is μm.

## A2. THz conductivity of 8-nm-thick Au film

To measure the conductivity of 8-nm-thick Au film, we performed the THz time-domain spectroscopy (THz-TDS [1] on a uniform 8-nm-thick Au film sputtered on the corning Eagle XG, the glass substrate which is used for actual devices. In this measurement, we measured THz-field transmission through a glass substrate with 8-nm-thick Au film and a reference glass substrate, as shown in Fig. A2-1.

The conductivity spectrum was derived with an assumption that the 8-nm-thick Au film is metallic in the following way: We used Tinkham's formula [2], which gives the complex conductivity spectra of conductive thin film $\tilde{\sigma}(\omega) = \sigma_1(\omega) + i\sigma_2(\omega)$ from the complex field transmission spectra $\tilde{t}(\omega) = E_{\mathrm{Substrate+film}}(\omega)/E_{\mathrm{substrate}}(\omega)$ as

$$\tilde{\sigma}(\omega) = \frac{1+n}{Z_0 d}\left(\frac{1}{\tilde{t}(\omega)} - 1\right). \qquad (\mathrm{A1})$$

To relate the complex transmission spectra to the complex conductivity, we need both the amplitude and phase of the transmission spectra. However, in this measurement, we could not obtain the phase of transmission spectra due to the large uncertainty of the thickness of substrates for sample and reference (tolerance of the substrate thickness was ±20 μm). To overcome this point, we used the approximation of the Hagen-Rubens regime [3], which is valid when the film thickness is much smaller than the skin depth. Since the skin depth of the gold in the THz frequency range is approximately 100 nm, the 8-nm-thick Au film is in this regime, where the real conductivity is nearly constant, and the imaginary conductivity is negligible. Then, we can relate the amplitude transmission to the real conductivity as

$$\sigma_1(\omega) = \frac{1+n}{Z_0 d}\left(\frac{1}{|\tilde{t}(\omega)|} - 1\right). \qquad (A2)$$

By using eq. (A2), we derived the real conductivity of the 8-nm-thick Au film as $1.597 \times 10^7 \pm 5 \times 10^4/\Omega\mathrm{m}$, as shown in Fig. A2-2. This value is in the same order as that of the conductive Au nanofilms in a previous study [4], which also does not contradict the assumption of the conductive film and Hagen-Rubens regime.

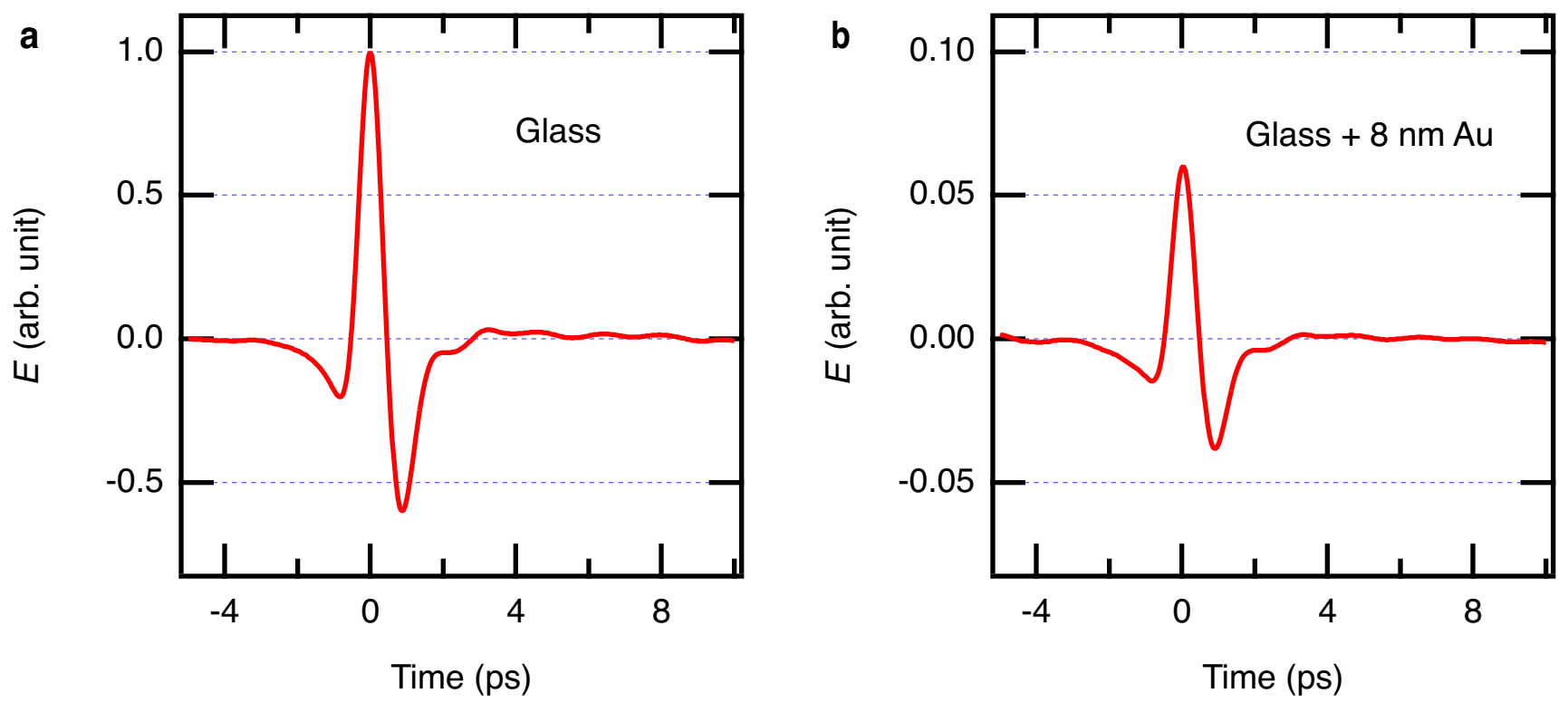

**Figure A2-1.** Measured THz field waveforms. **a**, THz field through the reference glass substrate. **b**, THz field through the glass substrate with 8-nm-thick Au film.

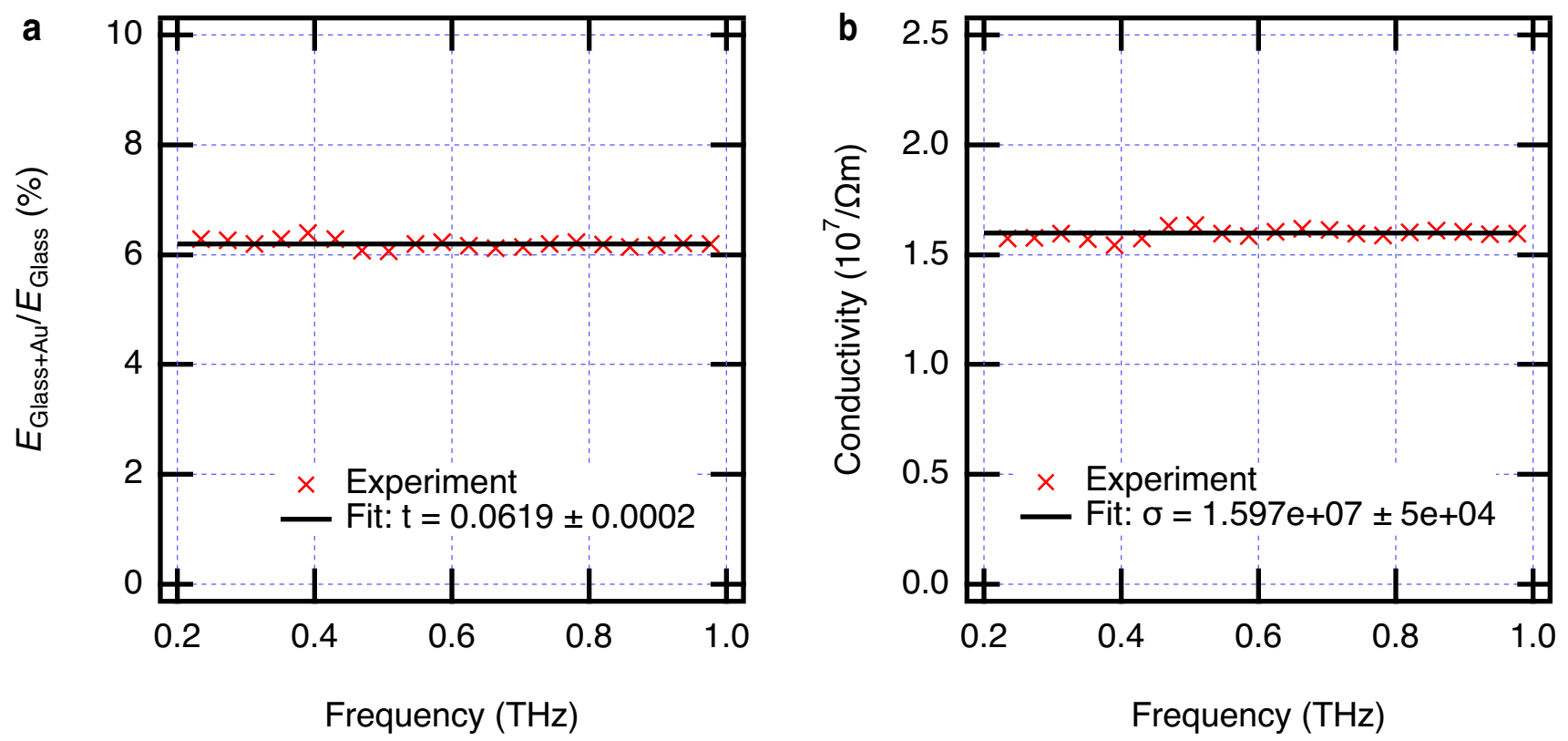

**Figure A2-2.** Obtained spectra from THz-TDS measurement of 8-nm-thick Au film. **a**, Field-transmission spectrum. **b**, Conductivity spectrum derived from **a** via Tinkham equation and Drude model.

## A3. FDTD simulation setting

For the FDTD simulation, we used the commercial software Ansys Lumerical 2024 R2.2 Finite Difference IDE. We simulated the entire device structure without $MoS_2$ with the dimension shown in Supplementary Section A1. The properties of materials were given as shown in Table A3-1. The refractive index of the glass substrate in the THz frequency range was measured via THz-TDS. The conductivity of 8-nm-thick Au film was obtained via THz-TDS, as shown in Supplementary Section A2, and was used as the conductivity of the entire top and bottom electrodes. The refractive index of ALD-grown $Al_2O_3$ was obtained from literature as a square root of its dielectric constant [5].

Figures A3-1 a and b show the entire simulation domain, location of the device and the field monitor, and location of the incident beam source. The center of the field-enhancement region of the device is set as the origin of the coordinate in this simulation, and the THz-field waveforms were monitored at this point. All the boundaries of the simulation domain are perfectly matched layers (PML). The boundaries are far enough from the origin to eliminate the simulation artifact around the PML. Also, it is far enough to delay the reflected wave at the simulation boundaries, which occurs even with the PML boundaries, returning to the monitor. Fig. A3-1 c and d show the close-in antenna structure and the domains with additional finer mesh settings. The details of the mesh settings are discussed in the next section.

The incident THz wave source is a Gaussian beam source with a 1/e full width (in intensity) of 1.4 mm located 600 μm above the origin and focused to the origin, as shown in Fig. A3-1 b and Fig. A3-2. Its waveform is made from the actual pump THz waveform in our experiment, which was measured via EO sampling [6]. To eliminate a simulation artifact caused by the incident waveform, we made the incident waveform for the simulation by subtracting a DC offset from the measured waveform and moderating the signal cutoff via a Hanning window, as shown in Fig. A3-3.

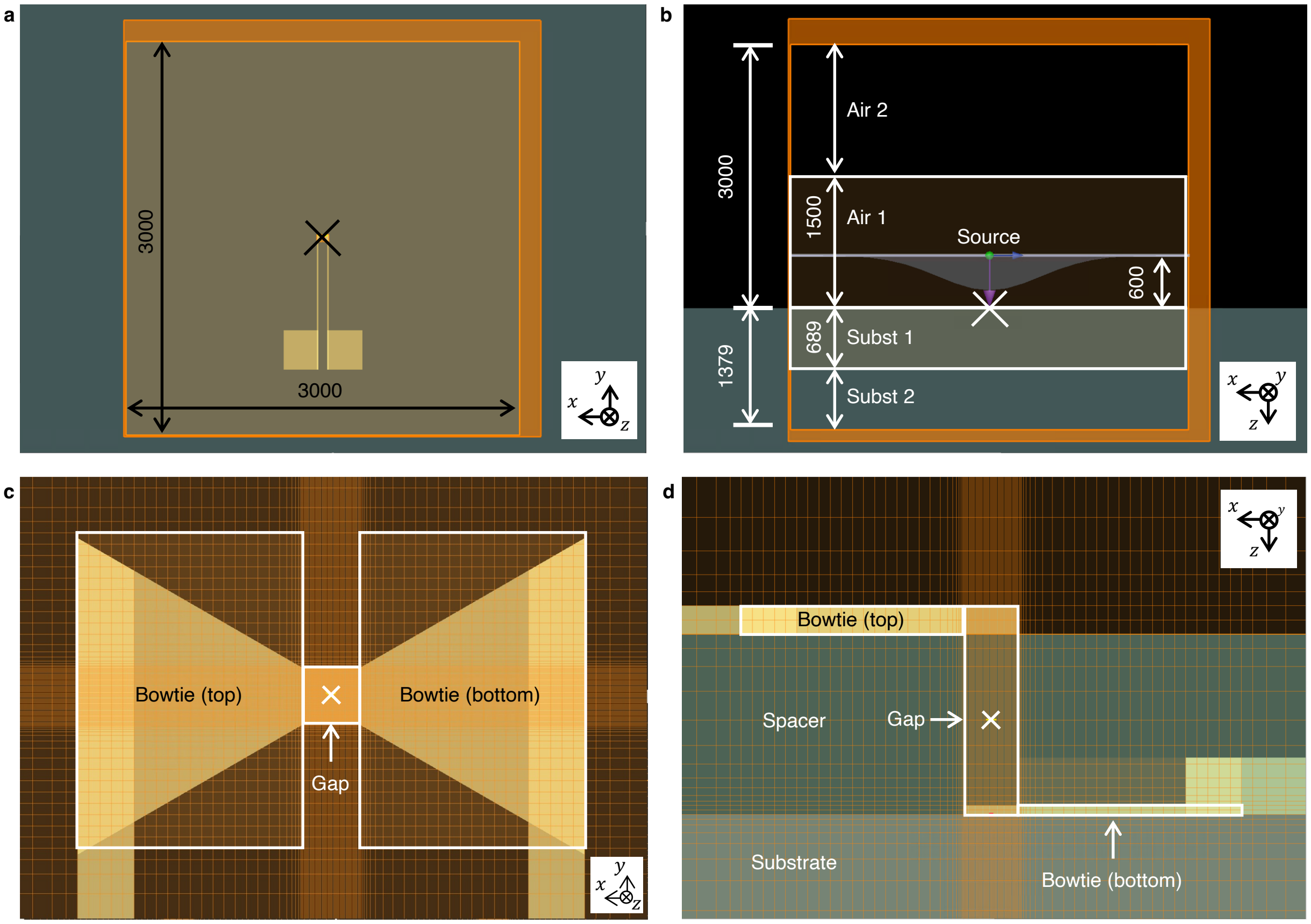

**Figure A3-1.** Entire simulation domain and domains with additional finer mesh settings. The cross marks are the origin of the coordinate where the THz waveforms are recorded using an electric field monitor. The white lines illustrate the domains with additional finer mesh settings. The unit of length is μm. **a**, An xy-view of the entire simulation domain. The orange-colored squared domain of 3000 μm × 3000 μm is the entire simulation domain. PML boundary conditions are applied to all boundaries. **b**, An xz-view of the entire simulation domain. The extent of simulation domain in z-direction is 3000 and 1379 μm in the air and substrate, respectively. A Gaussian beam source centered at (0 μm, 0 μm, 600 μm) with the focus at the center of the antenna gap (0 μm, 0 μm, 0 μm) is used as the incident beam. Additional finer mesh settings were applied in Air 1 and Subst 1. Air 2 and Subst 2 are additionally added space to delay back reflected waves at the simulation boundary coming back to the field monitor. The mesh sizes of these domains were not specified and automatically set by the simulation software. These domains were omitted in the simulations where we do not need to monitor the waveform after 15 ps. **c**, An xy-view of the close-in antenna structure and the domains with additional finer mesh settings on the antenna: Bowtie (top), Bowtie (bottom), and Gap. **d**, An xz-view of the close-in antenna structure and the domains with additional finer mesh settings on the antenna.

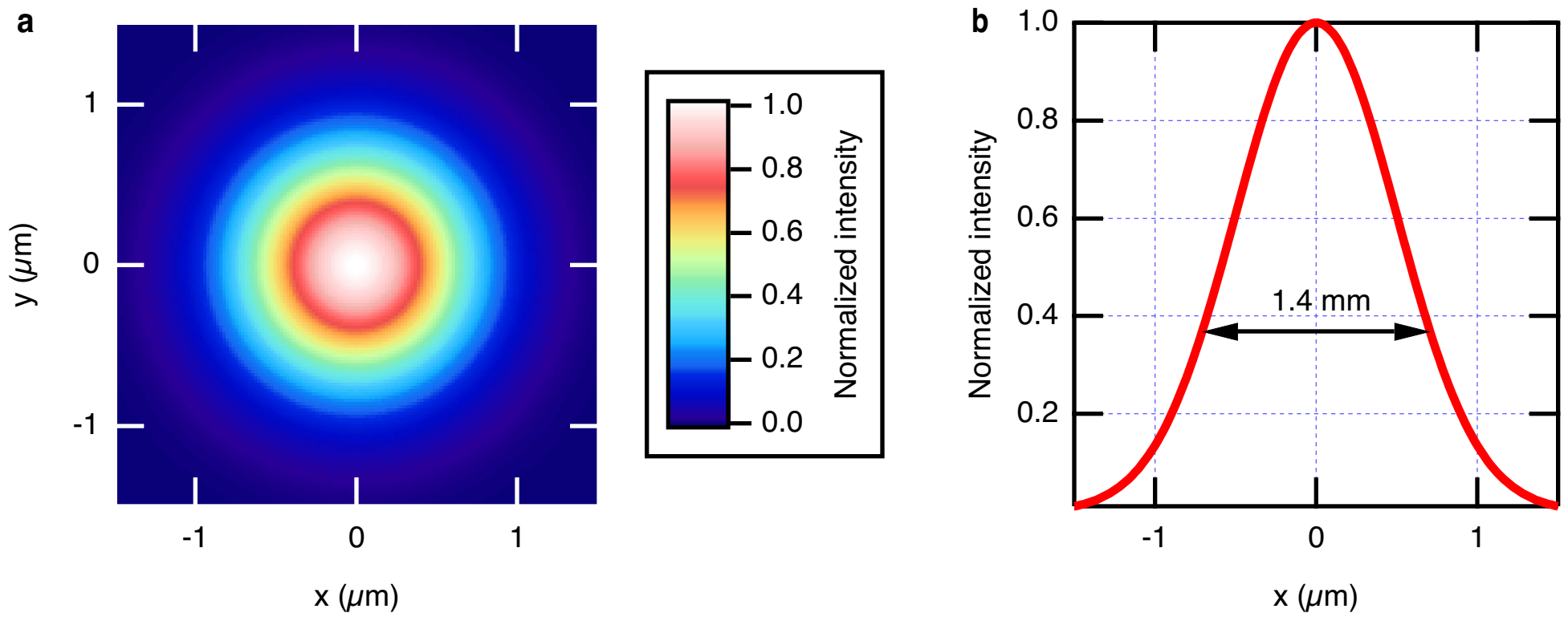

**Figure A3-2.** Spatial distribution of the Gaussian beam source at the source plane $z = 600$ μm. **a**, Color plot of the beam intensity. **b**, Section of the beam distribution at y = 0 μm. The 1/e full width is 1.4 mm in the intensity.

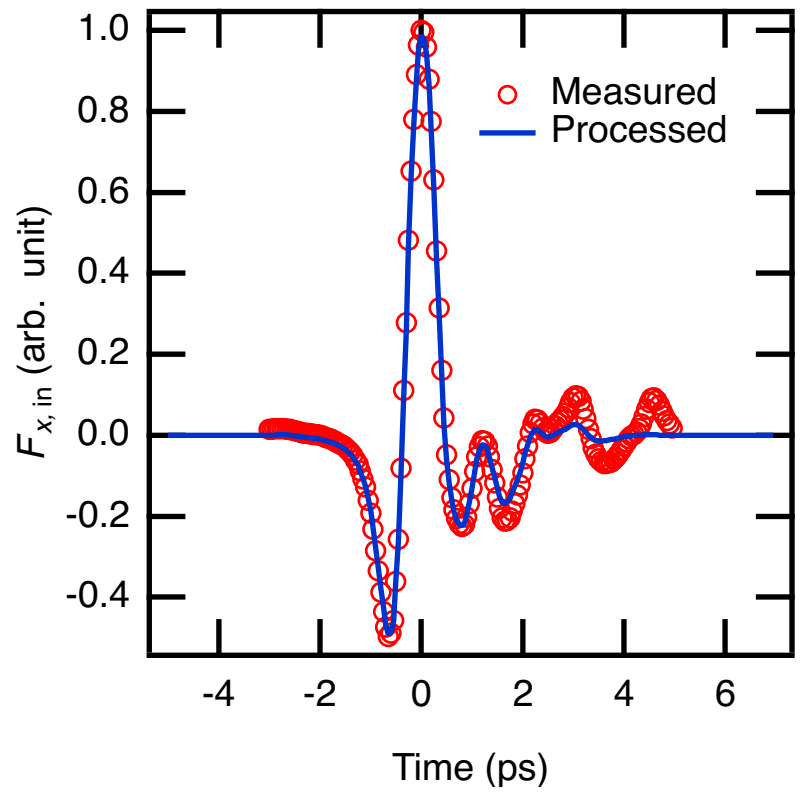

**Figure A3-3.** Experimentally measured THz waveform and preprocessed THz waveform used as the incident wave in the simulation.

**Table A3-1**. Properties of materials in the THz frequency domain used in the FDTD simulation.

| | Material | Model | Reflactive index | Conductivity (/Ωm) |
|---|---|---|---|---|
| Background | Air | Dielectric | 1 | - |
| Substrate | Glass | Dielectric | 2.176 | - |
| Electrodes | Au thin film | Metal | - | $1.597 \times 10^7$ |
| Spacer layer | ALD-grown $Al_2O_3$ | Dielectric | 3 | - |

**A4. Convergence test**

In FDTD simulation, the simulated waveform gets altered in shape and amplitude if we do not use appropriate mesh settings. Therefore, we confirmed the convergence of the result by changing the mesh settings over the air, glass substrate, and the fine structure of the antenna. For the finer mesh settings, we set the upper limit of the mesh size in the specified domain. In some cases, the software may automatically make the meshes smaller than the mesh size limit so that there is no spatially abrupt change in their size. To avoid redundancy, we call the mesh-size limit as mesh size later.

First, we simulate the propagation of the Gaussian beam in the absence of the antenna to judge the meshing of the background and glass substrate. Figure A4-1a shows the incident electric field recorded by the field monitor at (0 µm, 0 µm, 0 µm) simulated with only air in the entire simulation domain. Figure A4-1b shows the field simulated with air covering the entire domain of z > 75 nm and a glass substrate covering the entire domain of z < 75 nm. These figures show that the maximum amplitude of the incident field depends on the mesh settings.

Figure A4-2 shows the dependence of the maximum value of incident in-plane field amplitude, corresponding to the peak field amplitude around 7 ps in Fig. A4-1, to the mesh sizes in the air and glass. In each graph, the amplitude is normalized to the amplitude obtained with the mesh size adopted for the later simulations. We fit the data points in each graph with a linear function. The section of this line corresponds to the error due to meshing, i.e., the ratio of the amplitude obtained with infinitely small meshes to that obtained with the adopted mesh settings. We have adopted the mesh settings that make the error reasonably small and do not make the simulation time too long. Only for the result in Fig. A4-2b, the section is still as large as 0.9. However, this error is resolved by setting finer mesh over the antenna structure, as discussed in the next paragraph. Based on this convergence test, the mesh sizes in the air and substrate were decided, as shown in Table A4-1.

Figure A4-3 shows the result of the convergence test with the antenna structure with additional mesh settings. The graphs show the normalized maximum amplitude of the enhanced field in the z-direction. The simulated enhanced field amplitude is approximately linearly dependent on the mesh size. Based on the simulation time and estimated error, i.e., section of the linear fit, we adopted the mesh size shown in Table A4-1. The estimated error in the adopted mesh settings is only a few percent. In Fig. A4-2b, there is a large error since the mesh size was too large to separate the surface of the glass and the monitor point. However, this was solved by setting a finer mesh around the monitor point defined by the convergence test, which will be discussed in Fig. A4-3.

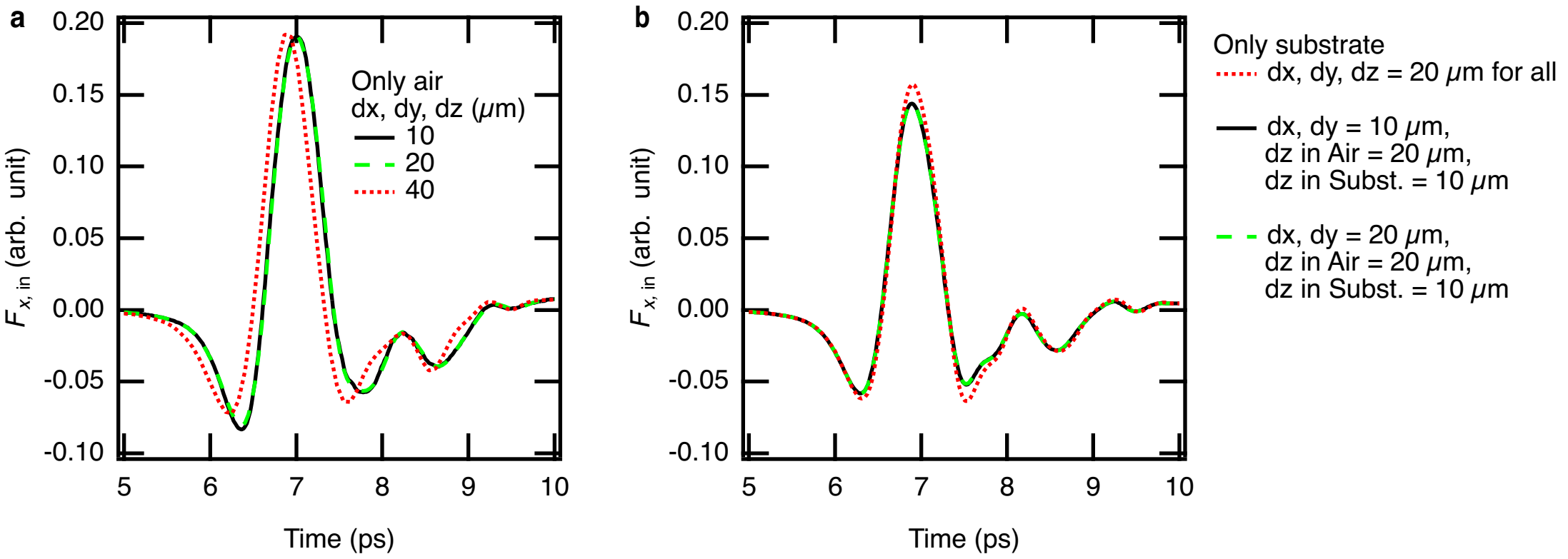

**Figure A4-1**. Effect of varying mesh size on the incident THz waveform sampled at the origin (0 μm, 0 μm, 0 μm) without the antenna structure. **a**, Simulation with only air with varying the mesh size for the entire simulation domain as dx = dy = dz. **b**, Simulation with the air and a glass substrate with varying the mesh settings (dx and dy for all the simulation domain and dz for the glass.)

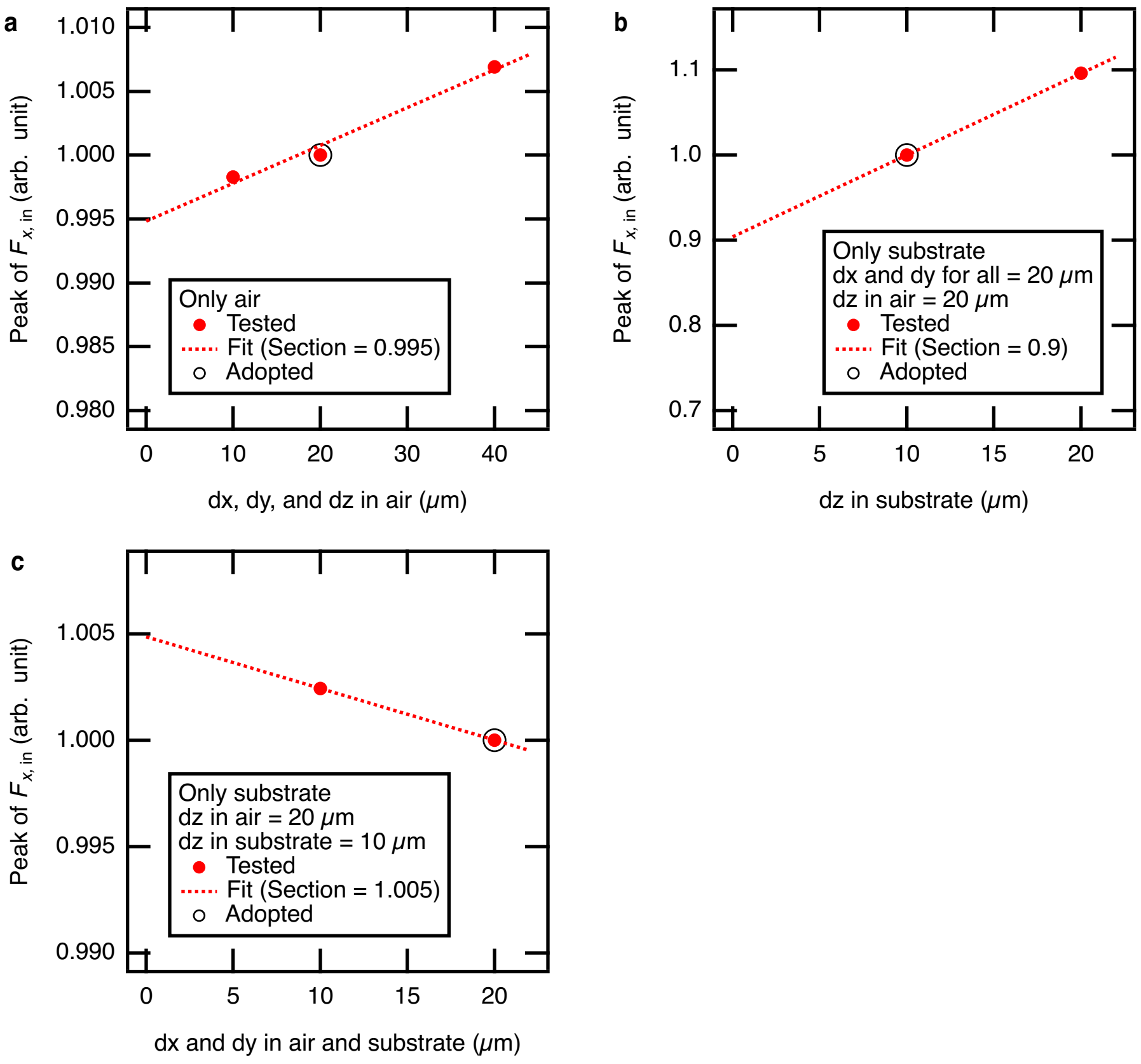

**Figure A4-2**. Maximum incident-field amplitude to the varying mesh size in air and substrate. **a**, Test with only air with varying mesh size in the air (dx = dy = dz.) **b**, Test with air and substrate with varying vertical mesh size (dz) in the substrate. **c**, Test with air and substrate with varying the horizontal mesh size (dx = dy) in air and substrate. In each graph, the red dots show the test results, and the open black circle shows the mesh settings

adopted for the later simulations. The field amplitude in each graph is normalized to the field amplitude in the adopted settings. The red dotted line shows the linear fit of the test results. The section of the fit curve is the expected field amplitude in case infinitely small meshes are used for the simulation. The detailed mesh settings for the convergence tests are also shown in the text boxes in the graphs.

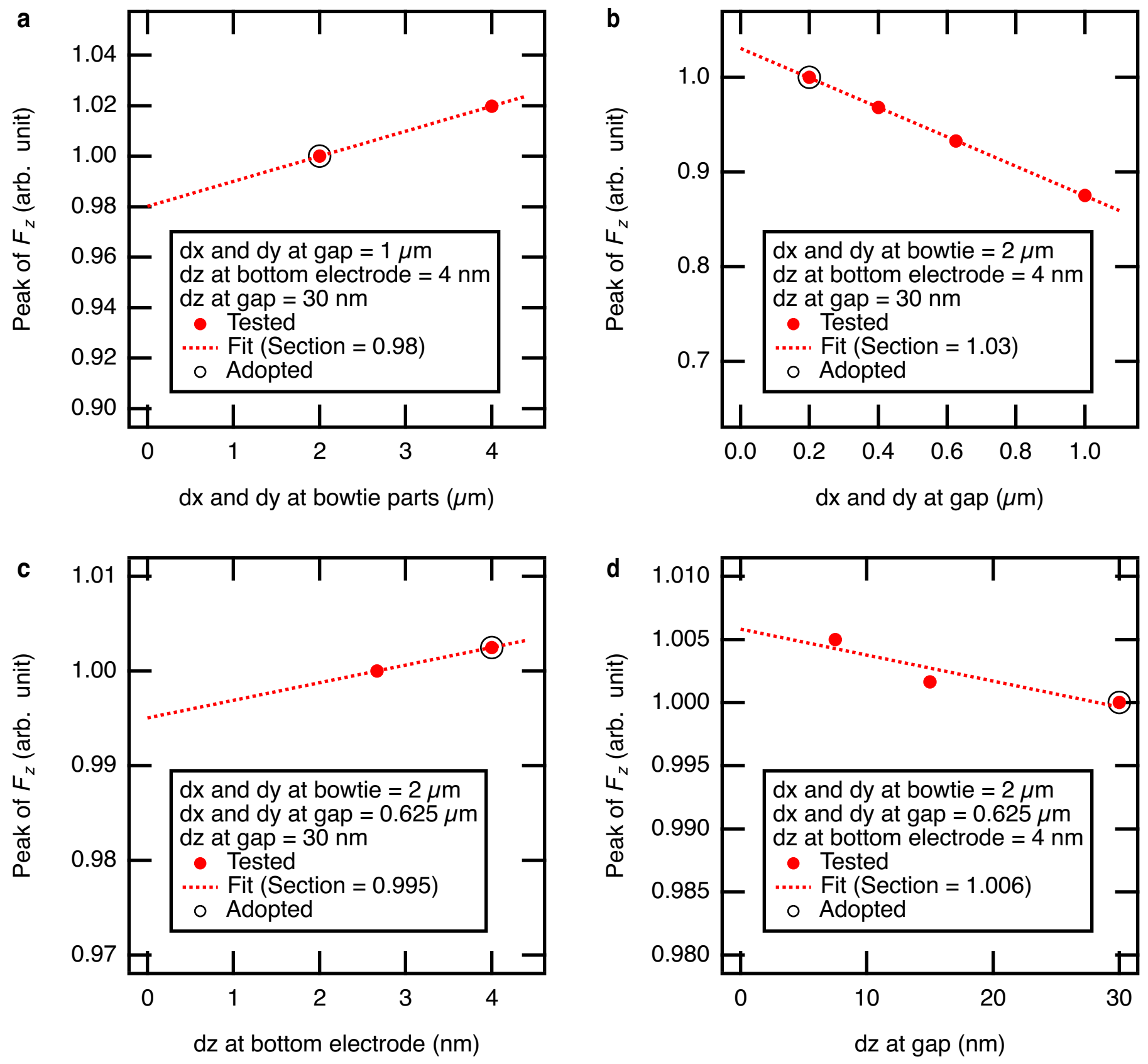

**Figure A4-3**. Maximum incident-field amplitude to the mesh settings at the antenna structure, plotted in the same way as Fig. A4-1. **a**, With varying the horizontal mesh size (dx = dy) at bowtie parts of the top and bottom electrodes. **b**, With varying the horizontal mesh size (dx = dy) at the gap of the antenna structure. **c**, With varying the vertical mesh size (dz) at the bottom and top electrodes. When the vertical mesh size for the bottom electrode was 4 and 2.67 nm, the vertical mesh size for the top electrode was 12.5 and 6.25 nm, respectively. **d**, With varying the vertical mesh size (dz) at the gap.

**Table A4-1.** Mesh size adopted based on the convergence test.

| Direction | Domain | Mesh size (Upper limit in the domain) |
|---|---|---|
| x, y | Entire simulation domain | 20 μm |
| z | Air 1 | 20 μm |
| z | Subst 1 | 10 μm |
| z | Air 2 | Not specified |
| z | Subst 2 | Not specified |
| x, y | Bowtie (bottom and top) | 2 μm |
| x, y | Gap | 0.2 μm |
| z | Bowtie (bottom) | 4 nm |
| z | Bowtie (top) | 12.5 nm for device with 25-nm top electrode and 25 nm for device with 50-nm top electrode |
| z | Gap | 30 nm |

### A5. FDTD simulation results: Field enhancement

Figure A5-1 shows the simulated waveforms of an incident in-plane THz field $F_{x,\mathrm{in}}$ without the antenna structure and the enhanced out-of-plane plane THz field $F_z$ at the gap of 2D-3D hybrid antenna. Figure A5-1a shows the incident THz waveform in a long temporal window of 0 to 28 ps. The signal after 25 ps is the THz wave reflected at the simulation-domain boundary. Figure A5-1b shows the out-of-plane THz waveform in the long temporal window. This waveform has two components: One is a fast component dominant in the time range of 5-10 ps and resembles the incident waveform. This is a resonance to the bowtie shape of the antenna. The other is a slow component dominant in the time range of 10-25 ps and does not resemble the incident waveform. This is a resonance to the large-scale structure, including the contact pad. Figure A5-1c shows the incident in-plane THz field and out-of-plane THz field plotted with a time offset to adjust the peak at 0 ps. The amplitude of the out-of-plane field is 13.3 times larger than the incident in-plane field. Although the out-of-plane field has a slightly lower frequency compared to the incident field due to the antenna's resonance frequency, the overall waveform of the fast component of the out-of-plane field resembles the incident waveform.

Figure A5-2 shows the spectral information of the simulated waveforms. Figure A5-2a shows the intensity spectra of the in-plane and out-of-plane THz fields. Figures A5-2b and c show the response function of the antenna structure as the ratio of the amplitude spectra and the difference of the phase spectra between the in-plane and out-of-plane THz fields.

Figure A5-3 shows the difference of the out-of-plane THz field caused by the antenna structures with a thickness of the top electrode of 25 nm and 50 nm. The peak amplitude of the fast component at 7.5 ps has almost no difference in these devices, and it is only 0.8% higher for the device with a 50-nm-thick top electrode compared to a 25-nm-thick top electrode.

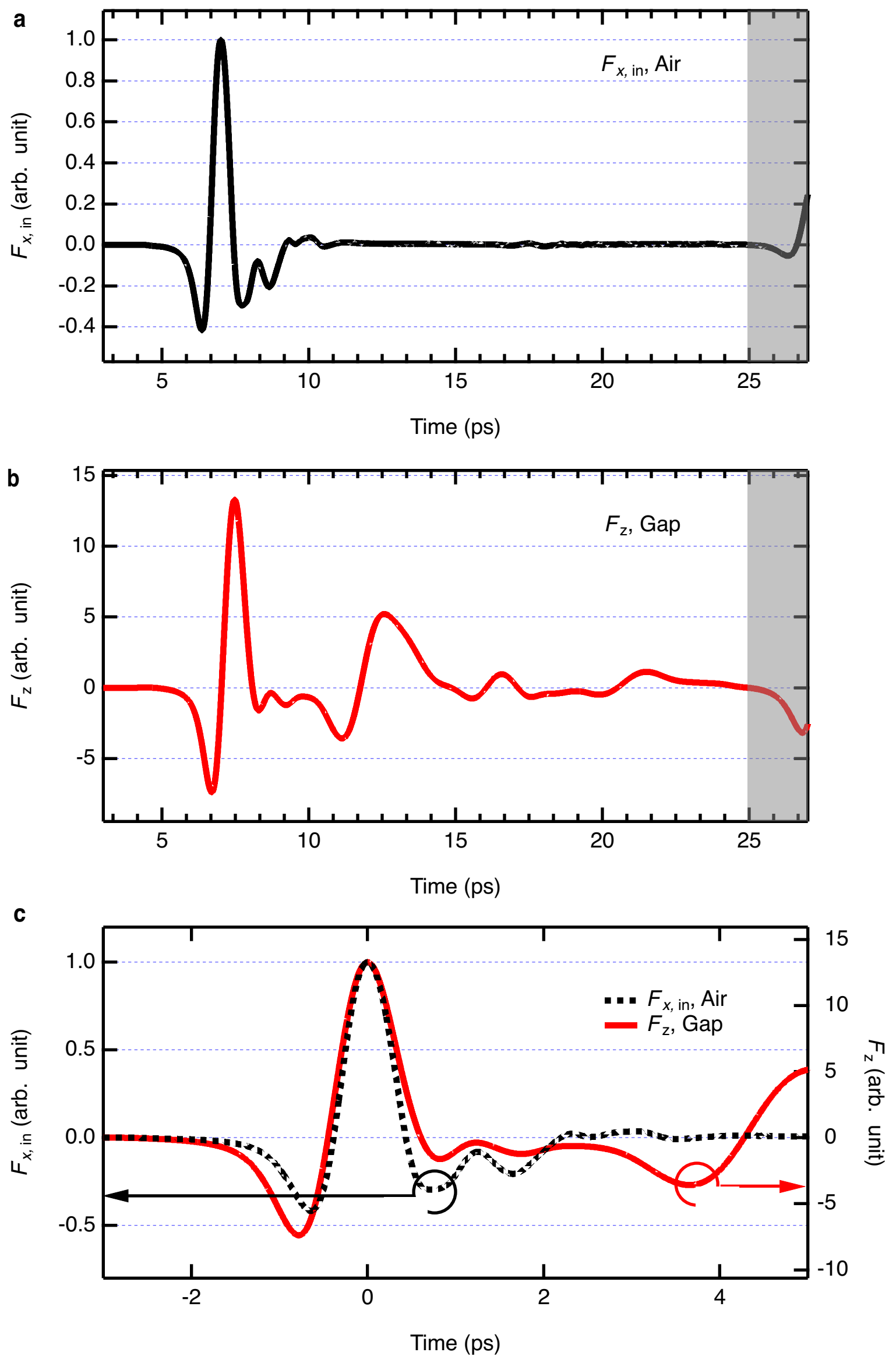

**Figure A5-1.** Simulated waveforms of incident in-plane THz field and out-of-plane THz field caused by the antenna structure. **a**, Incident in-plane THz field without the antenna structure. **b**, Out-of-plane THz field caused by the antenna structure. The gray-colored area after 25 ps is affected by the back-reflected waves at the simulation boundaries. **b**, Incident in-plane field and out-of-plane field plotted together to show the time offset of the waveforms. **c**, Incident in-plane field and out-of-plane field plotted with a time offset adjusted to the field peak. In all the plots, the field amplitude is normalized to the peak of the incident in-plane field.

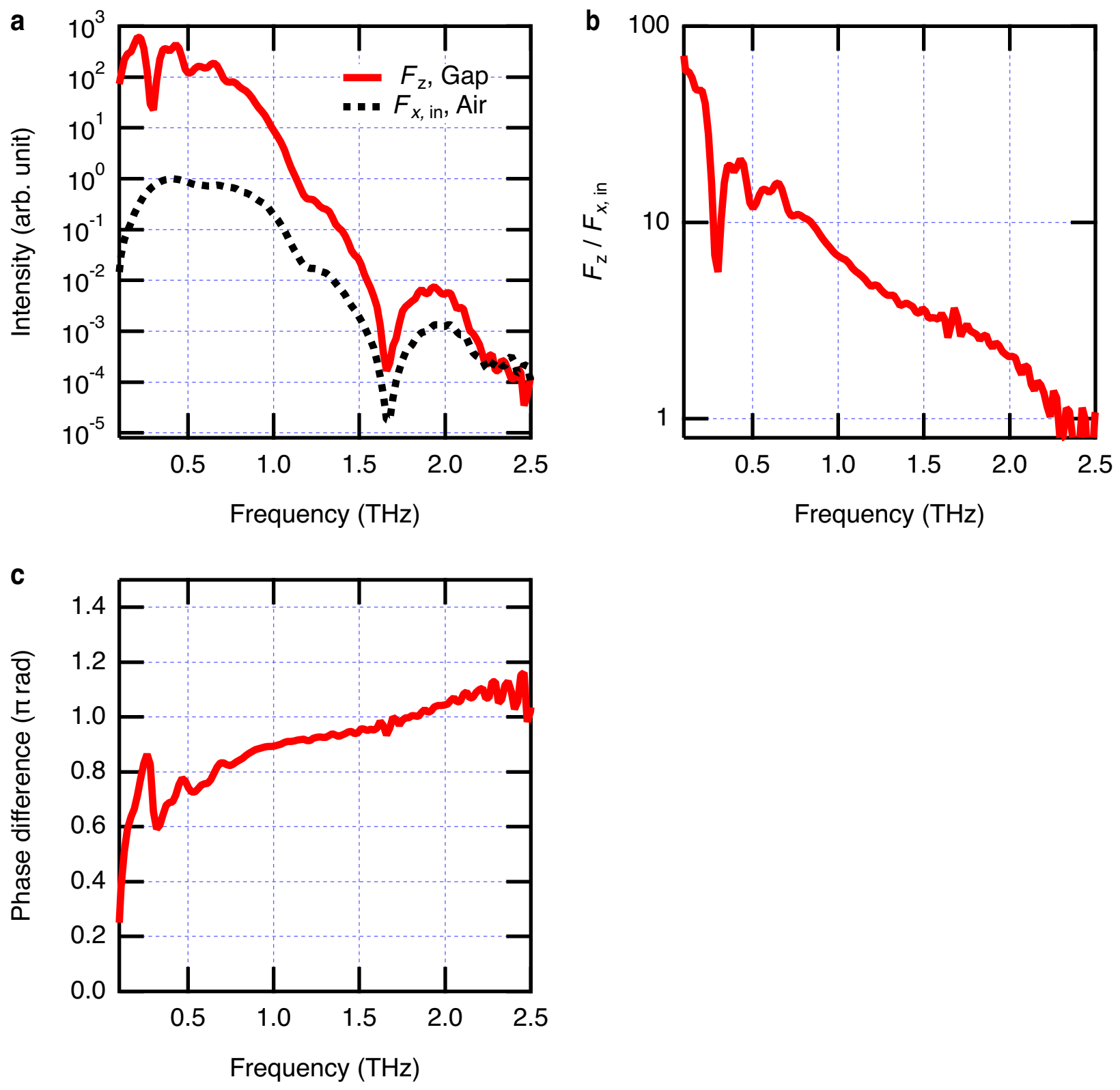

**Figure A5-2.** Spectral information of the simulated waveforms. **a**, Intensity spectra of the incident in-plane THz field and out-of-plane THz field caused by the antenna. **b**, Raito of amplitude spectra of the incident in-plane THz field and out-of-plane THz field. **c**, Difference of phase spectra of the incident in-plane THz field and out-of-plane THz field.

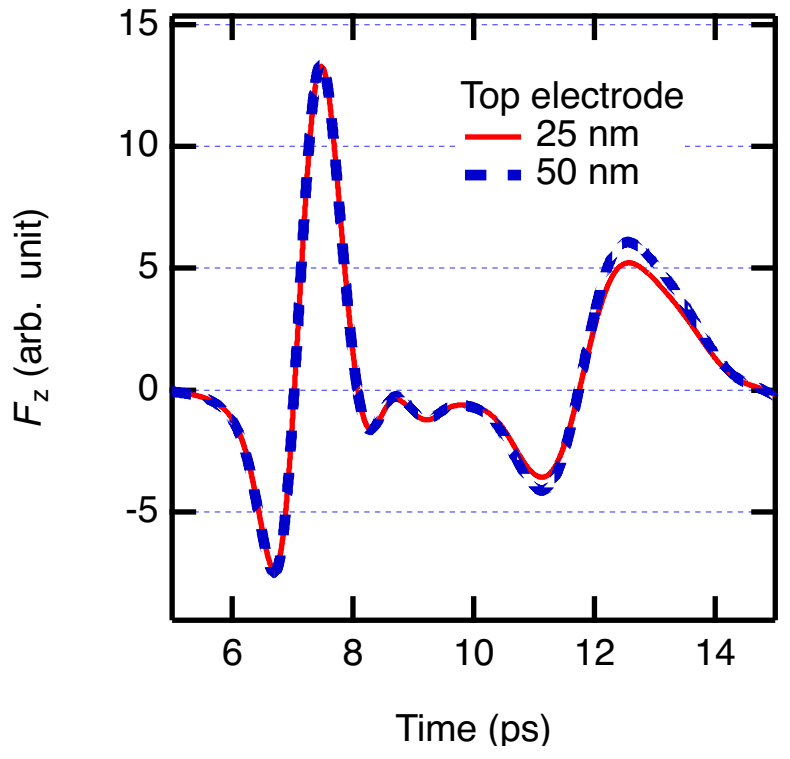

**Figure A5-3.** Simulated out-of-plane field waveform for antenna devices with the top-electrode thickness of 25 nm and 50 nm. The scaling of the vertical axis is the same as that of Fig. A5-1.

## A6. FDTD simulation results: Spatial inhomogeneity

Our antenna generates the out-of-plane field in the gap with spatial inhomogeneity that is unique to this system: Due to the spatial charge distribution defined by the antenna's resonance, the out-of-plane field in the field-enhancement region has a slight inhomogeneity in the *x*-direction as shown in Fig. A6-1.

Figure A6-2 shows the waveforms of the THz field probed at three different points in the antenna gap. The waveforms have a delay of 160 fs and a peak amplitude difference of 12 % for the two points located on the opposite side of the gap in the x-direction, representing the points of maximum field inhomogeneity.

In our measurement, the effect of such a spatially distributed THz field is averaged by the optical probe spread over the gap region. However, as we see in Fig. 2a and b, the difference in the waveforms is not so large, and it does not significantly affect the observed result.

In addition, in relation to the spatial inhomogeneity of the charge distribution, we have checked the amplitude of the in-plane THz field in the gap. As shown in Fig. A6-3, the in-plane component is smaller than the out-of-plane component by two orders of magnitude. Therefore, the effect of the inhomogeneity on the in-plane field component is also negligible. We do not see the in-plane component from the incident field in this plot. It would be because the incident field is blocked at the top electrode and it does not diffract into the gap.

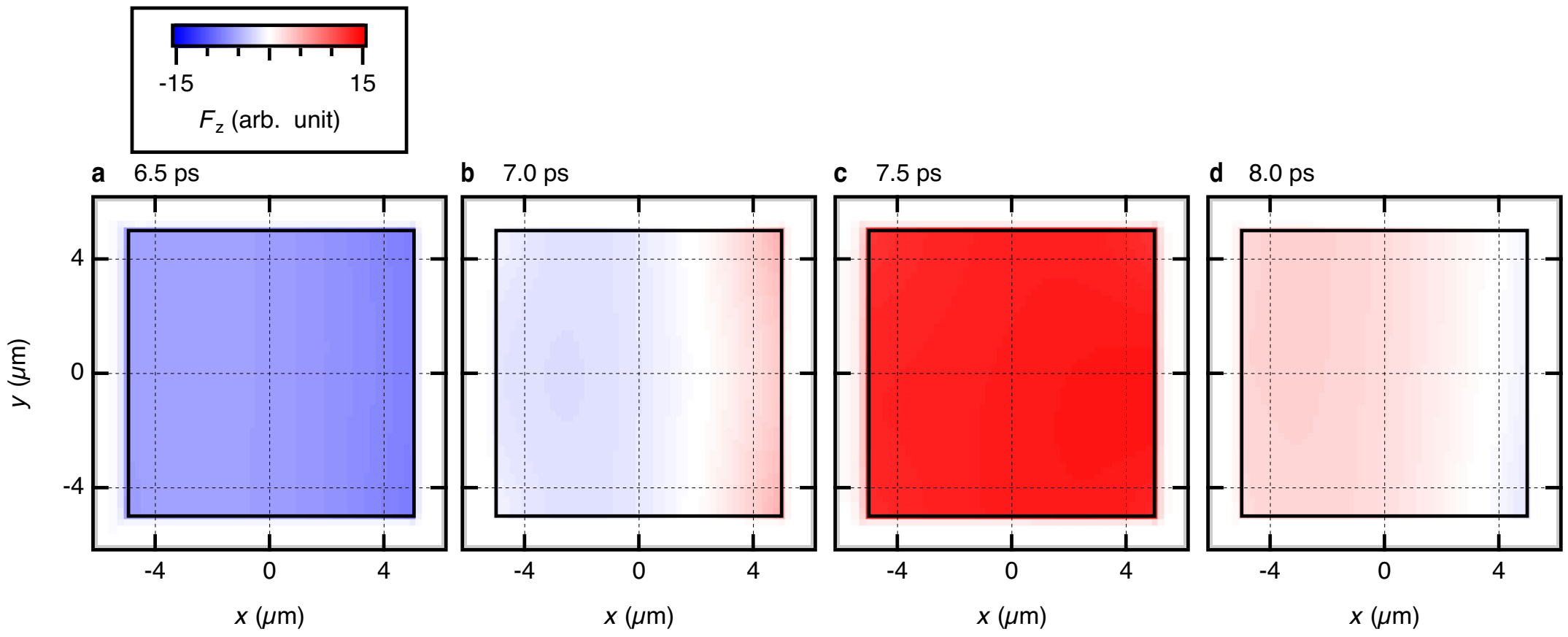

**Figure A6-1.** Spatial distribution of simulated out-of-plane THz field around the field-enhancement region at $z = 0$ plane plotted for various timing around the main peak of the THz pulse. The time for each frame is **a**: 6.5 ps, **b:** 7.0 ps, **c:** 7.5 ps, and **d:** 8.0 ps. The field-enhancement region of 10 μm × 10 μm is shown by the solid square in each frame.

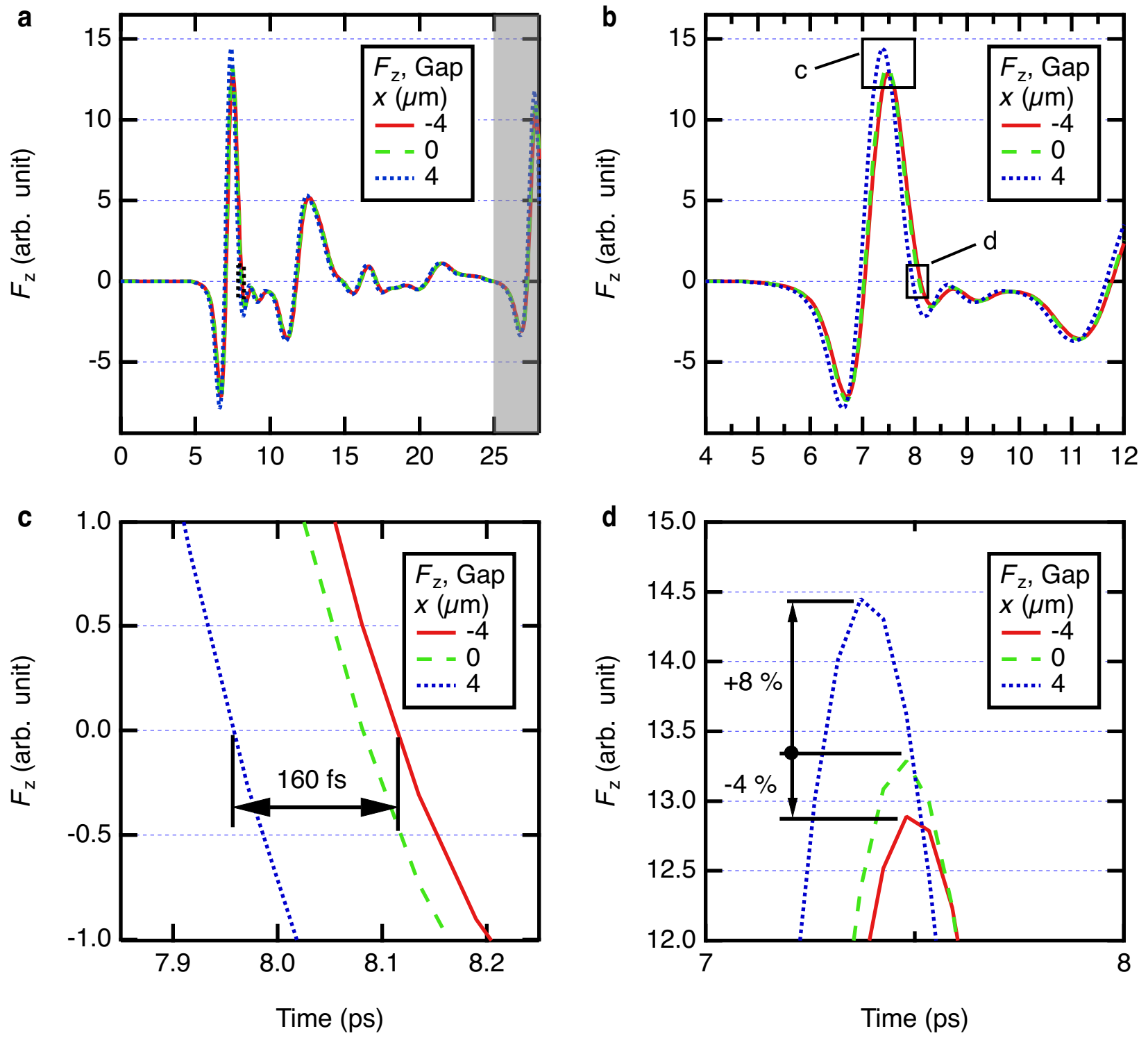

**Figure A6-2.** Simulated out-of-plane THz field waveform in the field-enhancement region probed at different points. Coordinate of the probed points are ($x$,$y$,$z$) = (-4 μm, 0 μm, 0 μm), (0 μm, 0 μm, 0 μm), (4 μm, 0 μm, 0 μm), respectively. **a**, Temporal waveforms plotted over 0-28 ps. **b**, Temporal waveforms plotted over 4-12 ps. The boxes labelled as c and d indicate the region plotted in **b** and **c**. **c**, Magnified view of the waveforms around their zero-crossing points around 8 ps. **d**, Magnified view of the waveforms around their peaks around 7.5 ps.

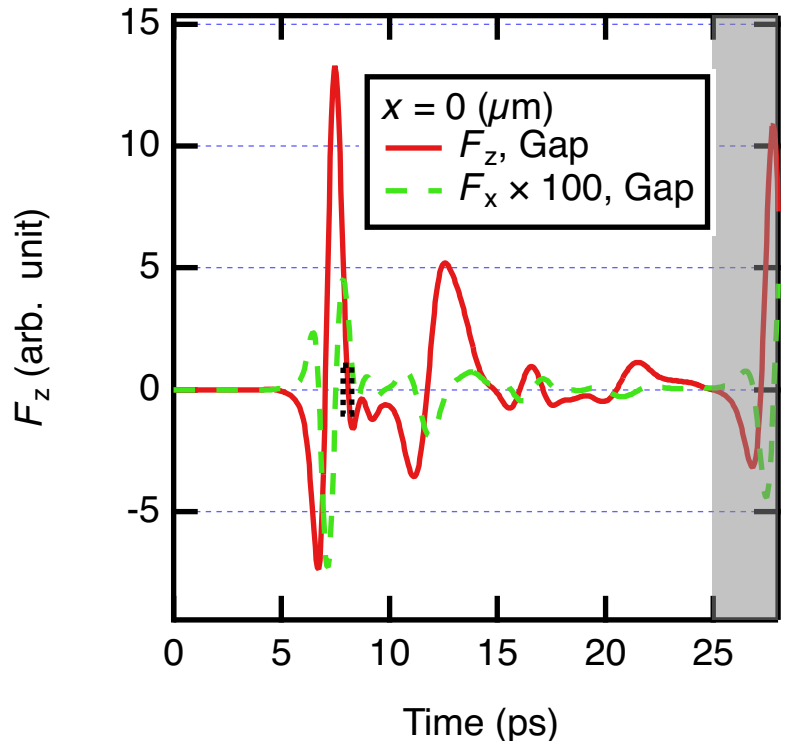

**Figure A6-3.** Comparison of the ouf-of-plane and in-plane THz field in the antenna gap probed at ($x$, $y$, $z$) = (0 μm, 0 μm, 0 μm). The in-plane component is smaller than the out-of-plane component by two orders of magnitude.

**Section B. Simulation of optical reflectance spectra**

**B1. Reflectance spectra of antenna devices with and without $MoS_2$**

We use the electromagnetic simulation software CST Studio Suite (Frequency domain solver) to simulate the reflectance spectra of the antenna devices. Here we model the central part of the antenna shown in Fig. B1-1 and we apply unit cell boundary conditions in transverse directions (XZ &YZ planes) to simplify the simulation. In the simulation, we use a plane wave at normal incidence (propagating in +z-direction from Port 1) as excitation source. This may result in slightly different behaviour, because in the experiment a focused Gaussian beam is used. However, this simplified simulation is sufficient for the qualitative discussion in this section. Further, Port 1 is placed in direct contact with the glass substrate. This approach eliminates reflections at the backside of the substrate and allows the substrate thickness to be modelled as much smaller, while avoiding introducing unrelated Fabry-Perot (FP) effects. This way the propagation through the complete 1.1 mm glass substrate does not have to be simulated. The glass-air interface is highly transmissive, so neglecting the reflection at this interface will not make much difference. For the material properties in the optical frequency range for the simulation, the following literature values were used:

- Top electrode (50 nm) – Dielectric function of gold film reported by Olmon et al. [7]
- Bottom electrode (8 nm) – Dielectric function of 7.5 nm gold reported by Ze et al. [8]
- $Al_2O_3$ – Refractive index $n = 1.76$ (corresponding the dielectric function $\varepsilon = 3$) [9]
- Glass – Refractie index $n = 1.5$
- $MoS_2$ – Dielectric function of 3L $MoS_2$ reported by Yu et al. [10]

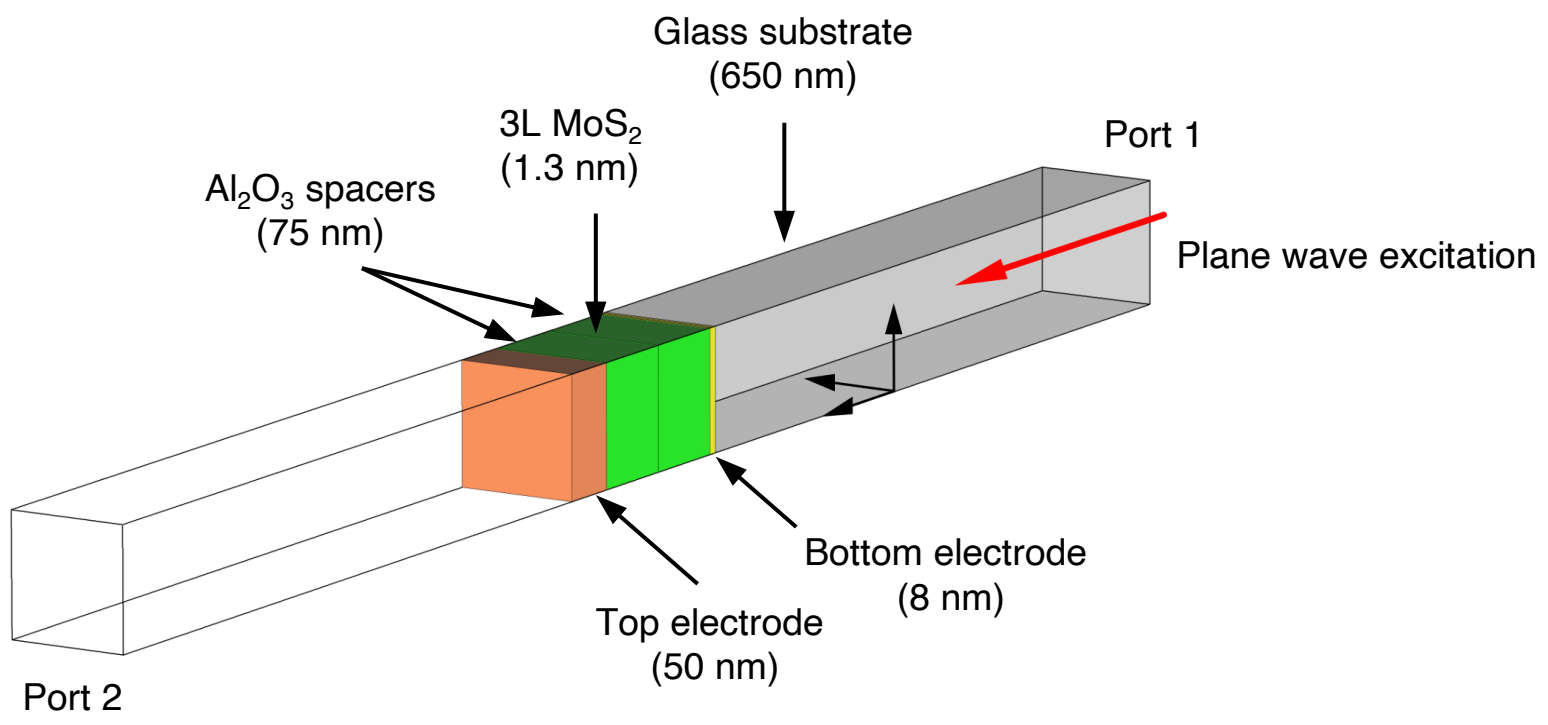

**Figure B1-1.** Modelled center part of the antenna. The numbers indicate the layer thicknesses.

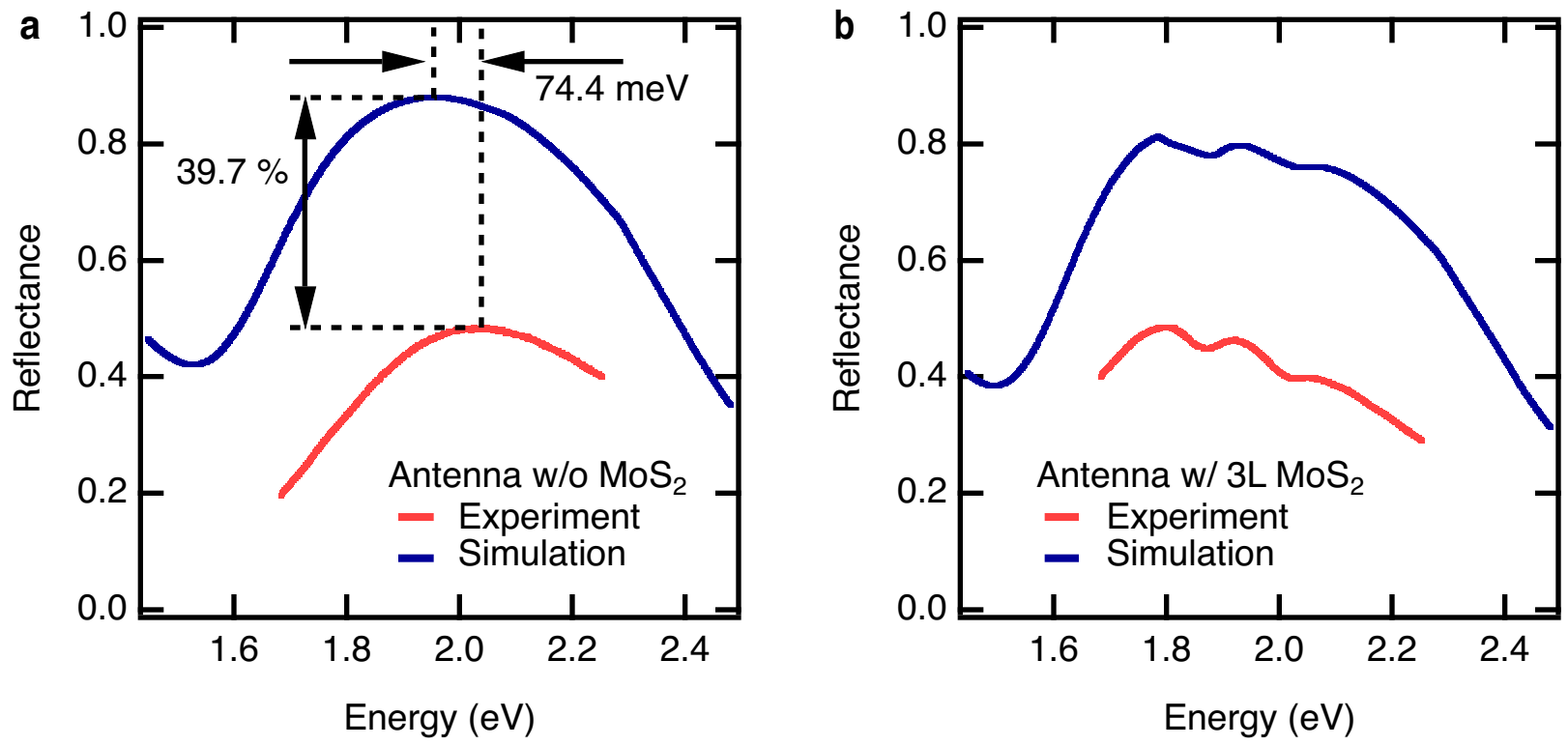

**Figure B1-2. (a)** Simulated and measured optical reflectance spectra of antenna structure without $MoS_2$, and **(b)** that of the antenna structure with 3 layers (3L) of $MoS_2$. There is only 74.4 meV energy difference of simulated and measured FP peak in (a).

Figure B1-2 (a) shows the simulated and measured reflectance spectra of the antenna without $MoS_2$. Both spectra show a curve originated from a FP resonance of low-Q factor caused by multi-reflection between the top and bottom electrodes. Absorption within the layers also affects this curve. There is a vertical offset of 39.7% of the simulated result compared to the experiment. This is most likely due to the focused Gaussian beam used in the experiment. In a previous study, it has been shown that the FP effect is sensitive to the beam shape [11]. In particular, the divergence of the Gaussian beam leads to decreased FP peak height. Fig. B1-2 (b) shows the simulated and measured reflectance spectra with 3 layers of $MoS_2$. The reflectance dips caused by the $MoS_2$ absorption are clearly visible in both spectra. There is a vertical offset of the simulated data for the same reasons in Fig. B1-2 (a).

Figure B1-3 (a) summarises the simulated reflectance spectra of antenna without and with $MoS_2$ from Fig. B1-2(a) and (b). The redshift of the FP resonance peak in the antenna with 3L $MoS_2$ is due to the increased effective cavity length. Fig. B1-3(b) shows the energy loss relative to the incident energy for the antenna simulation with the 3L $MoS_2$. The two $MoS_2$ absorption peaks coincide with the dips in the reflectance spectrum. Therefore, the reflectance dips are caused by the $MoS_2$ absorption peaks. The two absorption peaks correspond to A- and B- exciton absorption ($\varepsilon_i$ peak of literature [10]).

We note that there is a small deviation of the reflectance A-dip minima (1.8765 eV) and the absorption A-peak maxima (1.8791 eV). This should be due to the FP background in the reflectance spectra. In addition, it is affected by the fact that the absorption curves (Fig. B1-3(b)) arise not only from the intrinsic material parameters but also from the spatial overlap of FP modes with the material. Further, the additional loss contributions of the gold layers can also affect the reflectance dip energy.

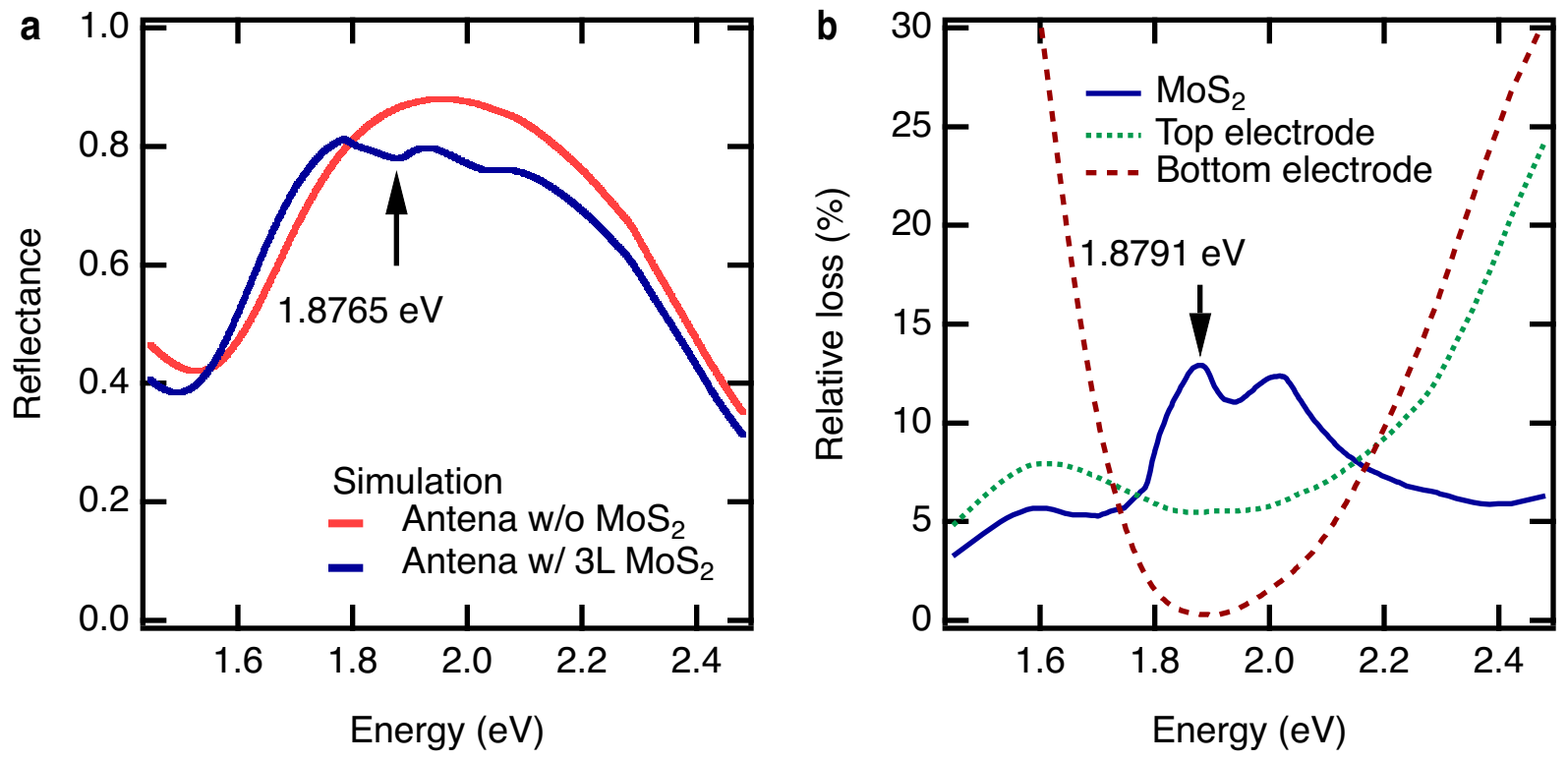

**Figure B1-3. (a)** Simulated optical reflectance spectra of antenna with and without a 3L $MoS_2$ (from Fig.2(a)-(b)), and **(b)** the energy loss relative to the incident energy in each of the materials of the simulation with a 3L $MoS_2$.

## B2. Effect of material-property change in $Al_2O_3$ spacer layer and gold electrodes

In this section, we show that (I) the broadband TPOP response of the antenna without $MoS_2$ most likely originates from the change of refractive index of $Al_2O_3$ spacer and (II) the shift of the reflectance dips of the antenna with $MoS_2$ is originated from the shift of absorption peaks in $MoS_2$. To address these points, we simulated two possible effects in addition to the absorption-peak shift of $MoS_2$, which could potentially lead to a shift of the reflectance dips when the strong THz is present in the antenna gap:

A. The change of dielectric function of the $Al_2O_3$ spacers due to electro-optic effect or carrier trap
B. The change of conductivity of the gold electrodes due to a modulation of charge density when the strong THz field is present in the antenna gap (Further explained in Supplementary Section B3)

We numerically investigate the effects of the change of material properties of $Al_2O_3$ as well as the gold on the antenna reflectance spectrum in the following. In the subsequent simulation sweeps, the material parameters are swept in a range that leads to a noticeable change in the reflectance spectrum. It should be noted that the change of material parameters required for this is much larger compared to the realistic strength of the underlying effects discussed above.

### B2-A. Change of dielectric function of $Al_2O_3$ spacers

The dielectric function of $Al_2O_3$ may be expressed by $\varepsilon(\omega) = \varepsilon_0\varepsilon_r + i\,\sigma/\omega$, where $\varepsilon_0, \varepsilon_r$ and $\sigma$ are the vacuum permittivity, real part of the relative permittivity, and conductivity, respectively. The simulated results for different $\varepsilon_r$ of the $Al_2O_3$ is shown in Fig. B2-1. The increase $\varepsilon_r$ leads to a redshift of the FP resonance due to an increase of the effective cavity length. It results in the broadband change of reflectance, which is an increase in a certain energy range (1.5-2.15 eV) and a decrease in another range (> 2.15 eV) as $\varepsilon_r$ increases, as shown in Fig. B2-1(a). This behaviour qualitatively matches with the experimentally observed TPOP response of an antenna without $MoS_2$ (see Fig. 2a of the main text), which is an increase of reflectance for the energy lower than about 1.95 THz and decrease for the energy higher than that, while the amount of the reflectance change is only on the order of $\pm$ 1%. Therefore, slight change of $\varepsilon_r$ and the FP resonance can be the origin of the TPOP response of antenna without $MoS_2$.

A shift of A-dip minima of $\Delta_A$ = 10.3 meV can be observed when $\varepsilon_r$ is increased from 2.84 to 3.16 ($\Delta\varepsilon_r$ = 0.32), as shown in Fig. B2-1(b). However, at the same time, the FP-mode shifts as much as 84.8 meV in the lower energy side of the FP resonance. Such a large alteration of the FP curve is not observed in the experiment, in which only a local change in the reflectance dips is observed. Therefore, a change of $\varepsilon_r$ of $Al_2O_3$ spacer layer is not the origin of reflectance dip shift observed in the TPOP measurements.

The simulated reflectance spectra of the antenna with increased conductivity $\sigma$ of the $Al_2O_3$ is shown in Fig. B2-2(a)-(b). The increasing absorption in the $Al_2O_3$ strongly alters the FP background spectrum: the reflectance decreases as the conductivity increases in the entire spectral range. This behaviour of the FP background does not match with the experimentally observed TPOP response of the antenna without $MoS_2$, which shows an increase and decrease in a different energy range. Therefore, a change of conductivity in $Al_2O_3$ spacer layer is not the origin of reflectance dip shift.

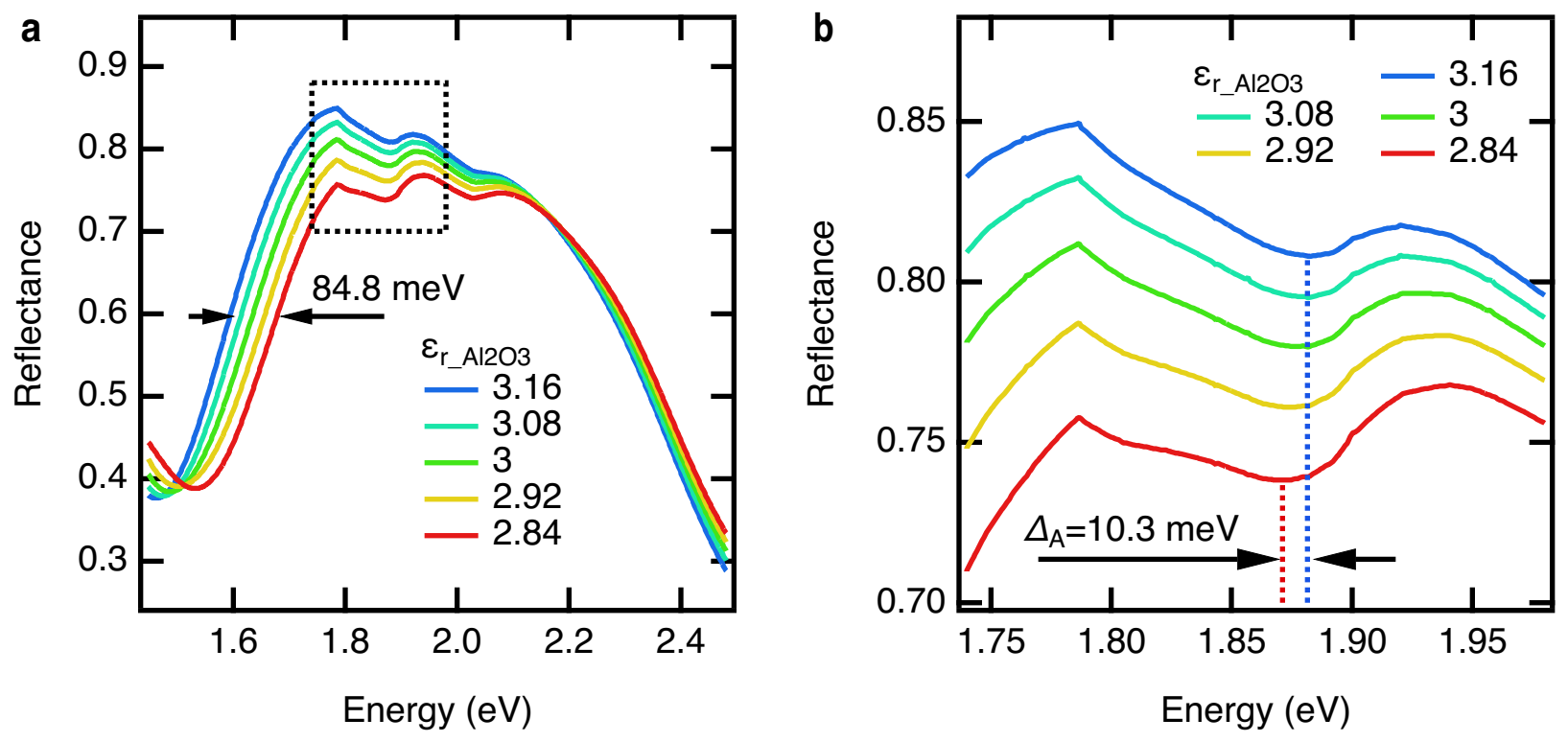

**Figure B2-1. (a)** Simulated optical reflectance of antenna with 3L $MoS_2$ with different real-valued permittivity of the $Al_2O_3$ spacers. **(b)** Magnified plot of (a) corresponding to the region shown by the dotted box in (a).

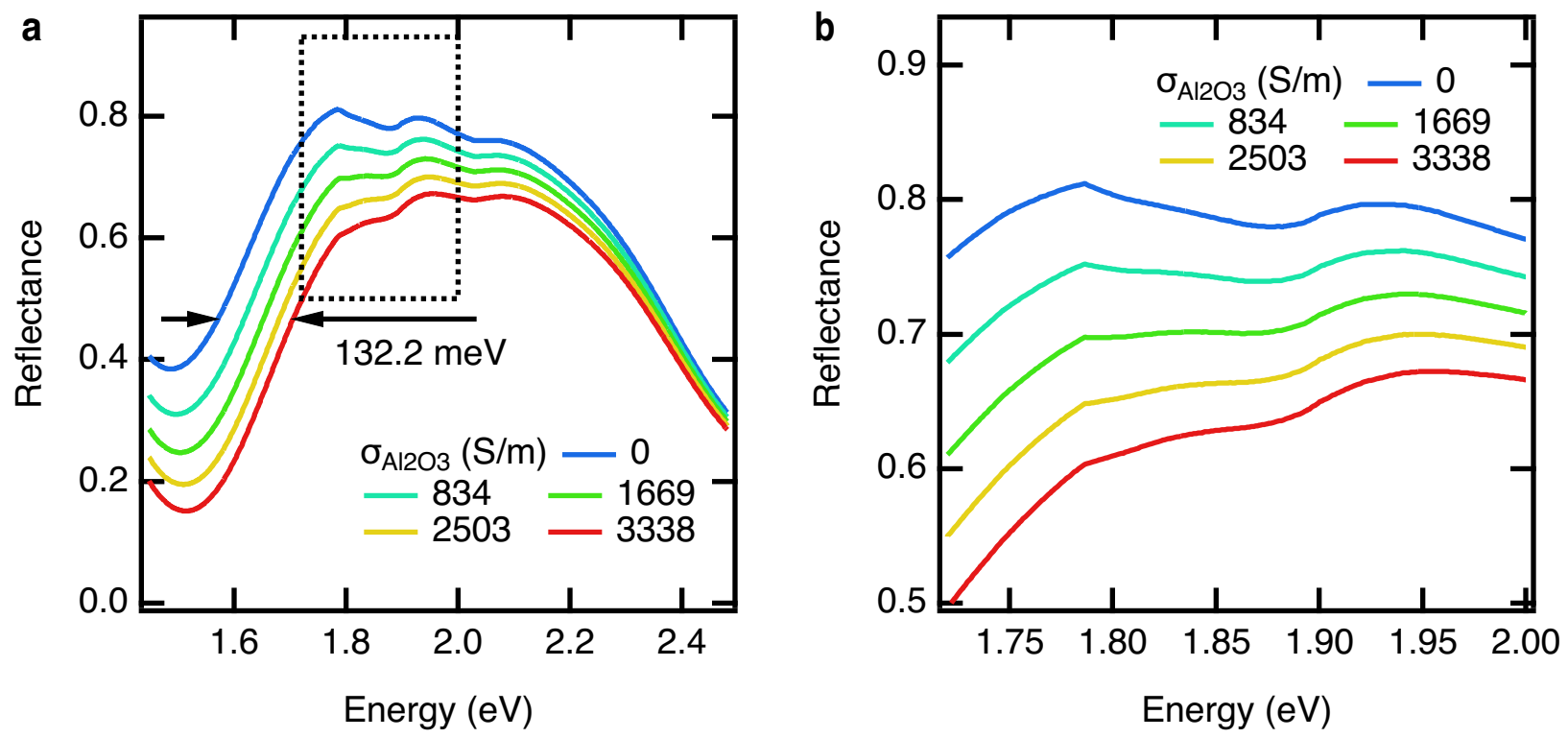

**Figure B2-2. (a)** Simulated optical reflectance of antenna with 3L $MoS_2$ with different conductivities of the $Al_2O_3$ spacers. **(b)** Magnified plot of (a) corresponding to the region shown by the dotted box in (a).

**B2-B. Change of conductivity of gold electrodes**

Figure B2-3(a)-(b) show the simulated reflectance spectra for different dielectric functions of the top electrode layer. Here the literature dielectric function $\varepsilon_{\mathrm{Au,top}}$ of the gold is scaled by multiplying a real-valued coefficient $c$ on both real and imaginary part. The scaled dielectric functions are shown in Fig. B2-3(c)-(d).

Figure B2-4(a)-(b) show the simulated reflectance spectra for the scaled dielectric functions of the bottom electrode layer. Here the literature dielectric function $\varepsilon_{\mathrm{Au,bottom}}$ of 7.5-nm thick gold is scaled by a real-valued coefficient $c$. The scaled dielectric functions are shown in Fig. B2-4(c)-(d).

A shift of A-dip minima of $\Delta_{\mathrm{A}}$ = 11.6 meV and 14.9 meV can be observed when the scaling of the dielectric function $c$ increases from 0.6 to 1.4. However, a large modulation of the FP background, which was not observed experimentally, should occur at the same time. Also, the dielectric function of gold needs to be more than doubled ($c$ = 0.6 to 1.4) to realize the simulated dip shift, which is too large and unrealistic. We estimate the change of conductivity of gold electrodes when the strong THz field is present in the antenna gap and conclude that it should be negligibly small in Supplementary Section B3. Therefore, a change of the dielectric function of top and bottom electrodes is not the origin of the shift of reflectance dips in the TPOP measurements.

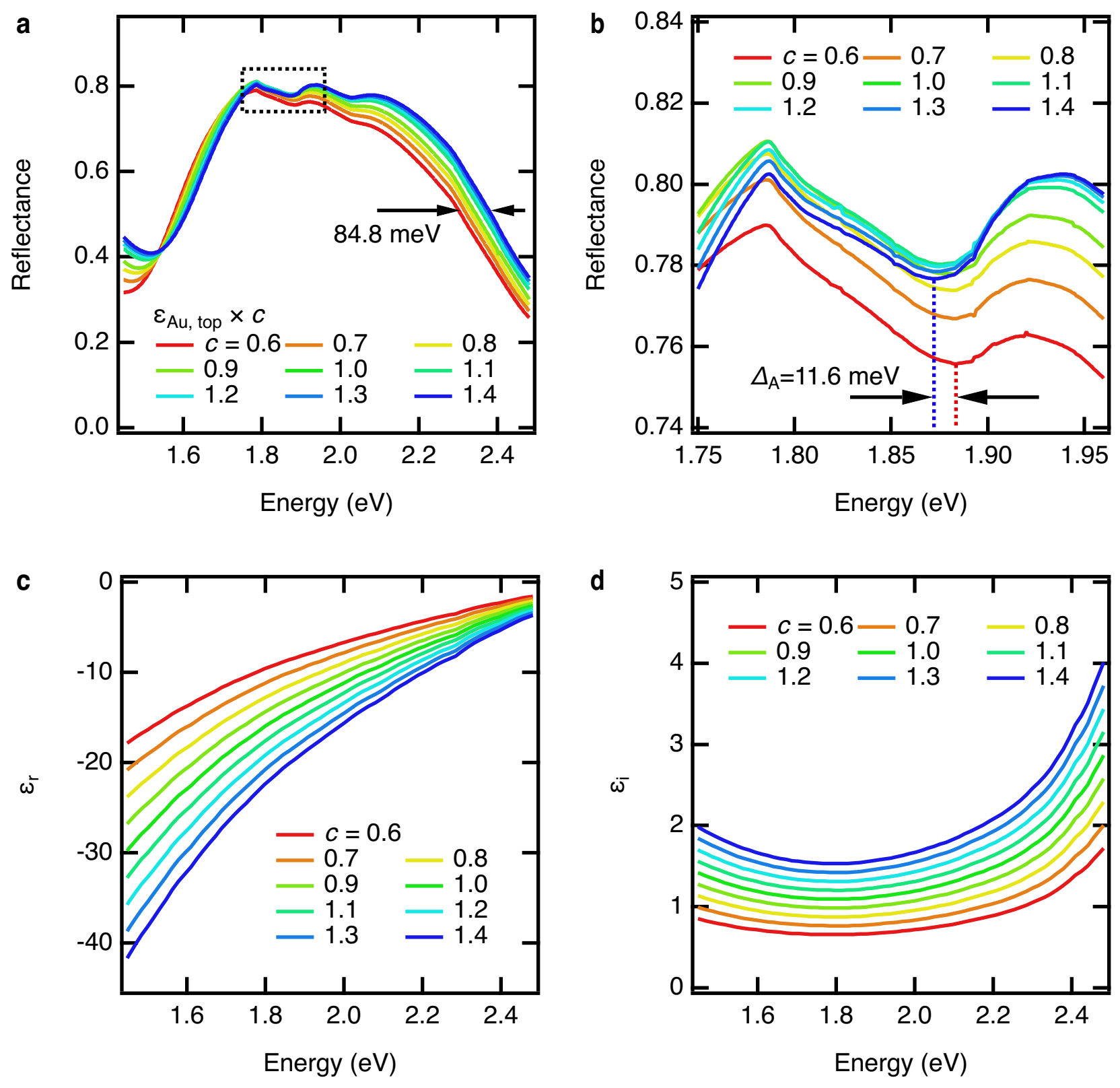

**Figure B2-3. (a)** Simulated optical reflectance spectra of antenna with 3L $MoS_2$ with scaled dielectric function of the top electrode layer. **(b)** A magnified plot of (a) around the dips. **(c)** Real part of the corresponding scaled dielectric functions of the top electrode. **(d)** Imaginary part of the corresponding scaled dielectric functions of the top electrode.

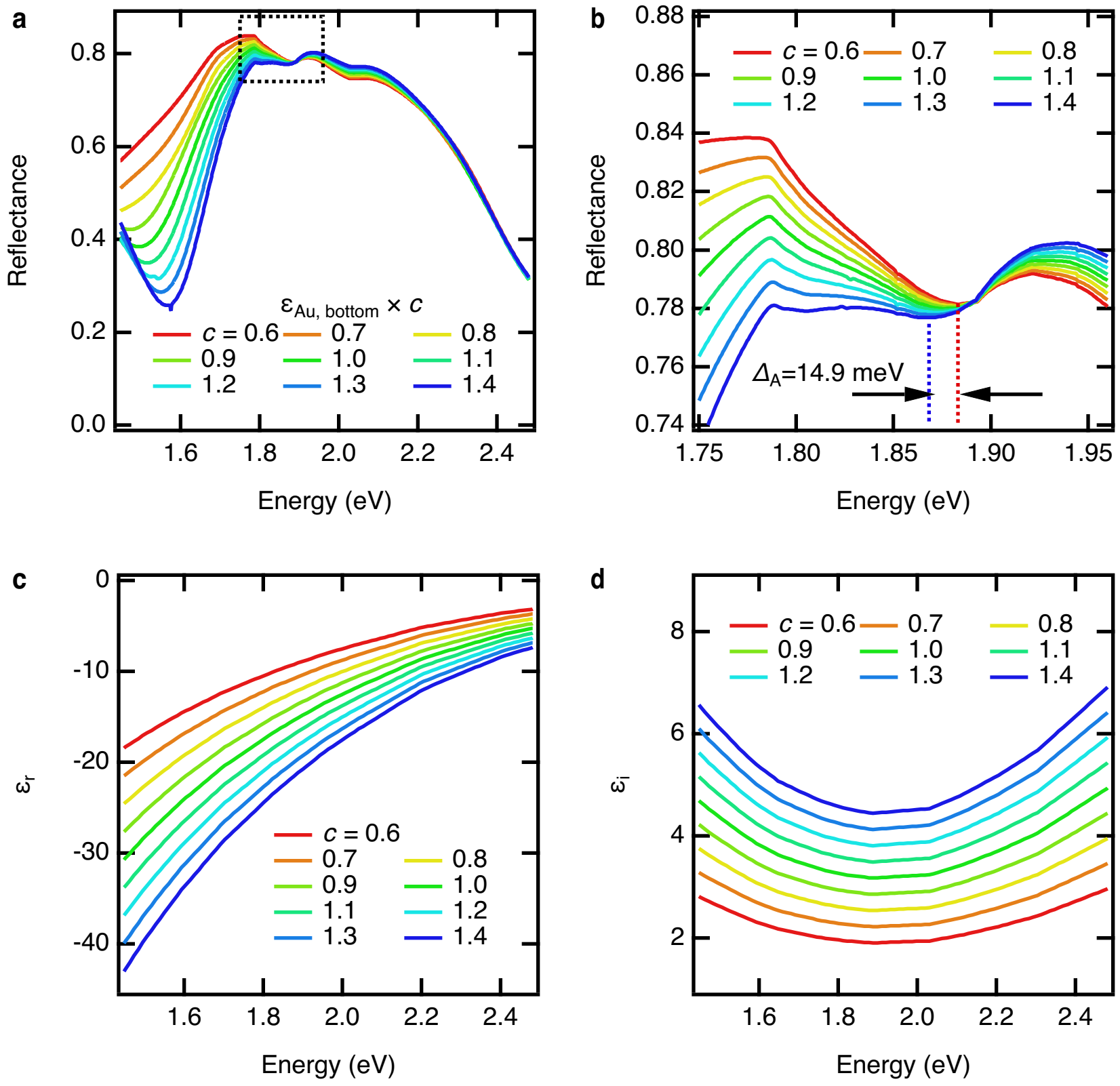

**Figure B2-4. (a)** Simulated optical reflectance spectra of antenna with 3L $MoS_2$ with scaled dielectric function of the bottom electrode layer. **(b)** A magnified plot of (a) around the dips. **(c)** Real part of the corresponding scaled dielectric functions of the bottom electrode. **(d)** Imaginary part of the corresponding scaled dielectric functions of the bottom electrode.

**Summary of Section B2**

From the results shown in Section B2, we conclude as following:

(I) Slight change of permittivity of $Al_2O_3$ spacer layer causes the FP resonance shift, which is most likely the origin of the experimentally observed TPOP response of the antenna without $MoS_2$ shown in Fig. 2a.

(II) Shift of the reflectance dips cannot be caused by change of dielectric function of $Al_2O_3$ spacer nor gold electrodes without inducing unrealistically large alternation of the FP background spectrum. Therefore, the origin of the observed reflectance dip shift in TPOP experiment is the absorption peak shift of $MoS_2$.

**B3. Estimation of change of conductivity due to accumulation of charge on gold electrodes**

The-field enhancement at our 2D-3D hybrid THz antenna occurs due to a capacitor effect: When an in-plane THz field is incident to the antenna, it causes a modulation of the charge density in the top and bottom electrodes at the centre part of antenna. These electrodes act as two capacitor plates as illustrated in Fig. B3-1 and causes the strong field in the field-enhancement section. In this section, we estimate the modulation of charge density.

First, we combine the two known capacitor equations

$$C = \frac{Q}{V} = \frac{Q}{E_z d} \qquad (B1)$$

and

$$C = \frac{\varepsilon_0 \varepsilon A}{d} \qquad (B2)$$

where $C$ is the capacitance, $Q$ the accumulated charge, $E_z$ the enhanced THz electric field, $d$ the distance between the plates, $\varepsilon$ the permittivity of $Al_2O_3$, and $A$ the area of the capacitor (simplified to include only the centre patch of the antenna). In the first equation, the voltage $V$ was substituted by $V = E_z d$ under the assumption that the field in the gap is homogeneous. By combining above equations, one obtains

$$Q = \varepsilon_0 \varepsilon A E_z. \qquad (B3)$$

Under the assumption of spatially homogenous electron density on both plates one can write as well

$$Q = tAe\Delta n, \qquad (B4)$$

where $t$ is the electrode thickness, $\Delta n$ the electron-density modulation for the capacitor effect, and $e$ the electron charge. By combining the two previous equations, we obtain

$$\Delta n = \varepsilon_0 \varepsilon A E_z \frac{1}{tAe} = \frac{\varepsilon_0 \varepsilon E_z}{te}. \qquad (B5)$$

The permittivity of $Al_2O_3$ is $\varepsilon = 3$ and the THz field in the gap is $|E_z| = k|E_{x,in}|$, where $|E_{x,in}| \approx 10^7\ \mathrm{V/m}$ is the incident THz field strength and $k = 13.3$ is the field enhancement factor, which was obtained through FDTD simulation in Supplementary Section A. The thickness of the bottom and top electrodes is 8 and 50 nm, respectively. Using these values, the electron-density modulation of bottom electrode $\Delta n_{\mathrm{bottom}}$ and that of top electrode $\Delta n_{\mathrm{top}}$ can be calculated as

$$\Delta n_{\mathrm{bottom}} \approx 2.8 \times 10^{24}/\mathrm{m}^3 \qquad (B6)$$

and

$$\Delta n_{\mathrm{top}} \approx 4.4 \times 10^{23}/\mathrm{m}^3\,. \qquad (B7)$$

These electron-density modulation potentially lead to an increase of conductivity. The increase can be estimated using the Drude conductivity model. The electron density affects the plasma frequency $\omega_p$ as

$$\omega_p = \sqrt{\frac{(n + \Delta n)e^2}{\varepsilon_0 m^*}} = \sqrt{\frac{ne^2}{\varepsilon_0 m^*} + \frac{\Delta n e^2}{\varepsilon_0 m^*}}, \qquad (B8)$$

where $n$ is the electron density of the gold in absence of the THz field (equilibrium) and $m^*$ is the effective electron mass. For gold the effective mass is approximately $m^* = 1.1 \times m_e$ [7], where $m_e$ is the electron mass, and the plasma frequency of gold is in the order of $\omega_p \approx 10^{16}\ \mathrm{s}^{-1}$ [7]. This means that the first term in the square root, which gives the plasma frequency of the gold in absence of the THz (equilibrium), is of the order $10^{32}$. The second term constitutes the change of the plasma frequency due to the electron-density change on the capacitor plates when the enhanced THz field is present. Using the previous values for $\Delta n$, we obtain the following values for the second term $\left(\frac{\Delta n e^2}{\varepsilon_0 m^*}\right)$ for each electrode:

$$\frac{\Delta n_{\mathrm{bottom}}\, e^2}{\varepsilon_0 m^*} \approx 8.0 \times 10^{27}\ \mathrm{s}^{-2} \tag{B9}$$

and

$$\frac{\Delta n_{\mathrm{top}}\, e^2}{\varepsilon_0 m^*} \approx 1.3 \times 10^{27}\ \mathrm{s}^{-2}. \tag{B10}$$

These terms are at least four orders smaller than the first term ($10^{32}$) in the plasma frequency formula.

Thus, the change of the plasma frequency $\omega_p$ due to the capacitor effect is negligible. The plasma frequency lies also far above the observed frequency range in the experiment. Thus, our measurement is not sensitive to small changes of $\omega_p$. We conclude that the change of conductivity in the gold electrodes will be negligible in the observed TPOP response.

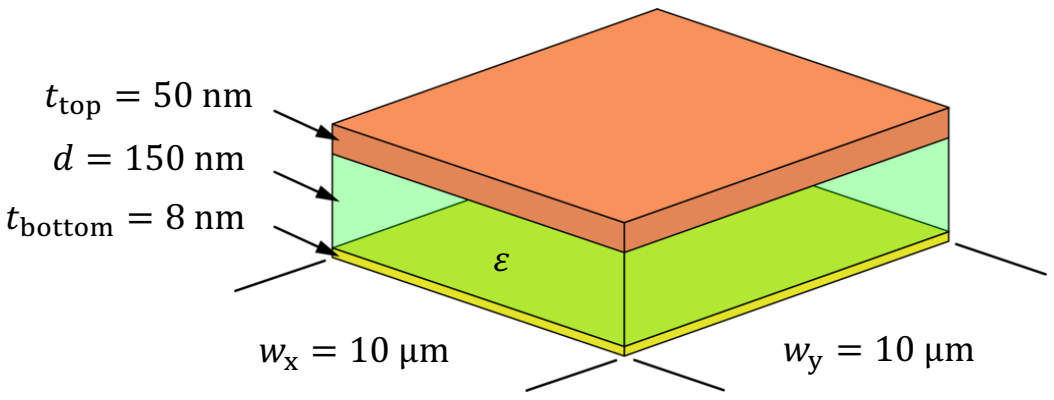

**Figure B3-1.** Illustration of the centre part of the antenna structure without $MoS_2$.

**B4. Injecting carrier to 2D material via direct contact to an electrode**

We could make a similar device structure without one of the top or bottom spacer layers to electrically contact the 2D material layer and an electrode for the ultrafast carrier density control of 2D material. The change of surface charge density corresponding to that of eq. (B6) and eq. (B7) is $2.2 \times 10^{12}\ \mathrm{cm}^{-2}$. With some optimization, such as reducing capacitance, it will be possible to realize a modulation of $10^{13}\ \mathrm{cm}^{-2}$, thus causing a significant modification of the properties of 2D materials [12][13]. Therefore, it will be a promising platform for ultrafast 2D material control through charge density.

## Section C. Chirp correction of TPOP measurement data

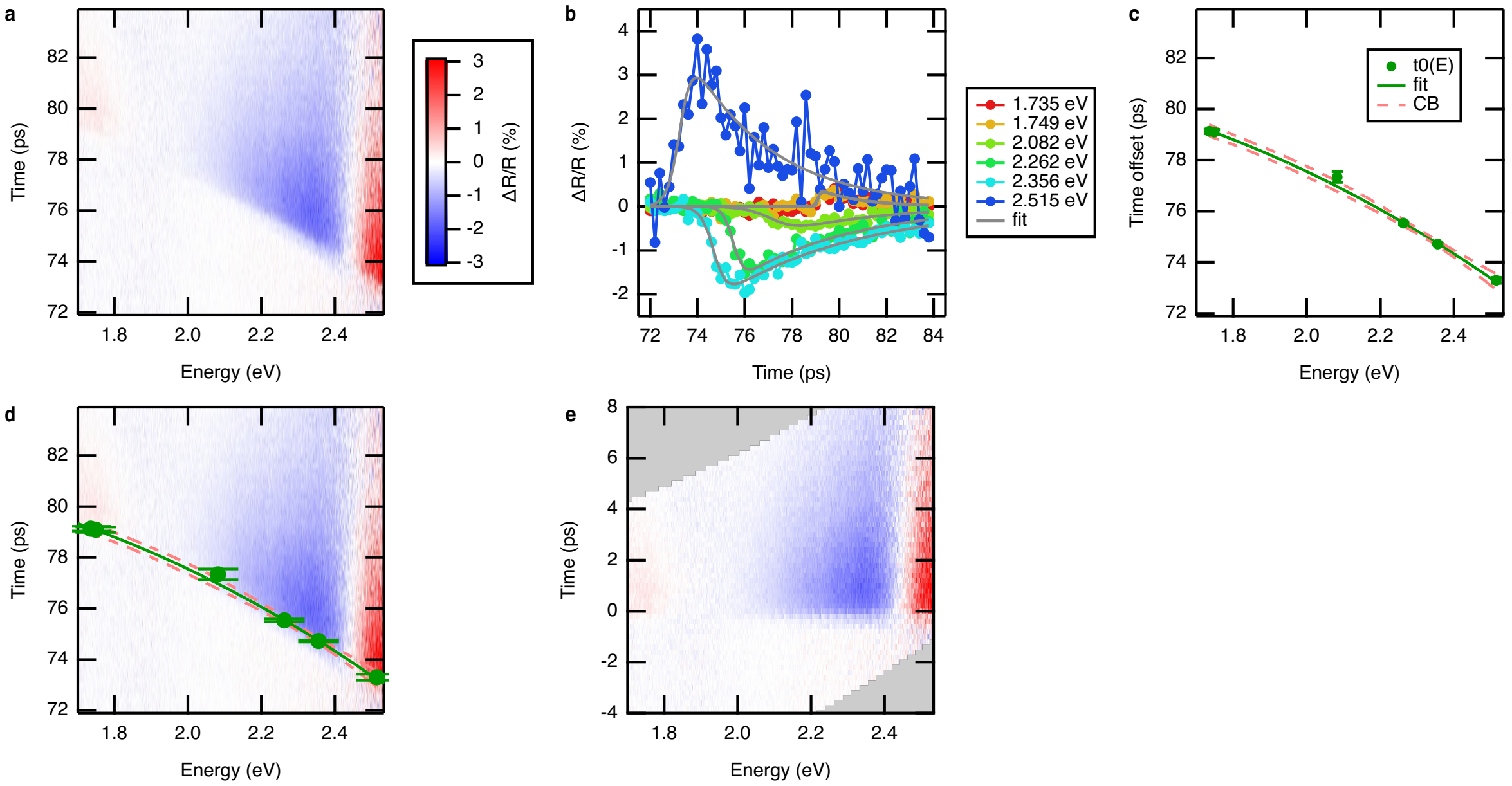

**Figure C1.** Method of chirp correction. **a,** A transient reflectance-change spectrum of an antenna without $MoS_2$ before chirp correction. The vertical axis is the time delay defined by the optical delay stage. **b**, Transient reflectance change at several probe-photon energies. Gray lines are fitting curves. **c,** Time offset plotted against photon energy obtained via fitting of the transient reflectance shown in Fig. C1b. The error bar of the points are the standard deviation as a fitting coefficient. The solid line is the fitting curve to the time offset with a parabolic curve (time-offset curve.) Confidence band (CB) with a confidence level of 95% is also plotted to show the range where the true time-offset curve may exist, which has only about 0.2-0.3 ps width. **d**, The transient reflectance-change spectrum of an antenna without $MoS_2$ with the time-offset curve and its confidence band. **e**, The transient reflectance-change spectrum after chirp correction. The vertical axis is the time delay after THz pump. The data points with gray color did not exist in the original data. They are treated as the points which does not have a value.

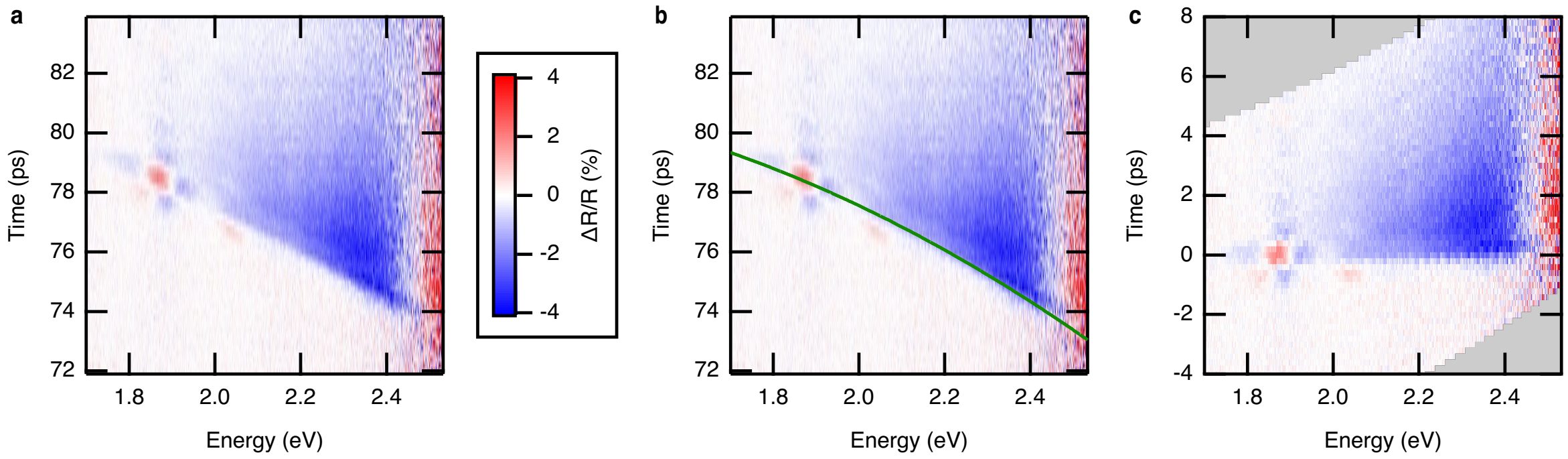

**Figure C2.** Chirp correction for the TPOP data of antenna devices with $MoS_2$. **a,** A transient reflectance-change spectrum of an antenna without $MoS_2$ before chirp correction. **b**, The transient reflectance-change spectrum of an antenna without $MoS_2$ with time-offset curve obtained in the analysis shown in Fig. C1. **c**, The transient reflectance-change spectrum after chirp correction.

In the original data of TPOP measurement, a probe-energy-dependent delay was observed in the transient reflectance change of antenna devices without $MoS_2$ as shown in Fig. C1a. The THz-pump response of the antenna without $MoS_2$ is originated from the change of dielectric properties of $Al_2O_3$ spacer layer as discussed in Supplementary Section B. Such a response is broadband and not have a probe-energy-dependent response time. Therefore, the origin of the probe-energy-dependent delay of the signal is the chirp of the probe pulse, i.e., each energy component of the probe beam has different timing offset to the THz pump.

To obtain the reflectance change as a function of the actual time delay after THz pump independent of the probe energy, we performed a chirp correction as follows: (i) Plot the transient reflectance change ($\Delta R/R$) at several energies as a function of the time delay in the experiment $t$ as shown in Fig. C1b. Then, fit the transient reflectance change with the following phenomenological formula:

$$\Delta R/R = A\ \exp(-t/T_2)/[1+\exp(-(t-t_0)/T_1)] \qquad (C1)$$

Here, $A, T_1, T_2$, and $t_0$ are amplitude, rising time, decay time, and time offset, respectively. (ii) Plot the time offset $t_0$ against the probe photon energy as shown in Fig. C1c. Then, fit the plot with a parabolic curve, assuming the chirp up to the second order. We call the fitting curve as time-offset curve since it gives the time offset at every probe photon energy. Figure C1d is the plot of the transient reflectance change plotted with the time-offset curve. (iii) Shifted the time axis of transient-reflectance spectrum at every energy point by the value of the time-offset curve. The shifted transient-reflectance spectrum is shown in Fig. C1e. This results in the broadband TPOP signal without the probe-photon-energy dependent delay.

The TPOP signal of an antenna with a $MoS_2$ flake is also affected by the probe chirp. However, it is more complicated due to the probe-energy dependent response of $MoS_2$, as shown in Fig. C2a. Therefore, for the chirp correction of the devices with $MoS_2$ flakes, we used the time-offset curve obtained from a measurement of an antenna without $MoS_2$ flake in the same experimental condition, as shown in Fig. C2b. By shifting the time axis at every photon energy using the time-offset curve, we obtained the chirp-corrected data as shown in Fig. C2c.

**Section D. Adjustment of time offset of EO-sampling and TPOP-measurement data**

In this study, we performed the TPOP measurements with a reflection probe and the EO sampling measurements with a transmission probe, as shown in Supplementary Section J. Therefore, the position of the optical translation stage in these two measurements are different, and the measurement results have an unknown time offset. In this section, we describe how we adjusted the time offset when we make the plot of dip position against input THz field.

Figure D1 shows all the Lorentzian parameter of the transient reflectance spectra obtained from a TPOP measurement (Fig. D1a-f) and the input THz field obtained from EO sampling (g and h). Here, the time offset of the THz waveform of Fig. D1g and h are adjusted so that the timing of peak THz field is roughly the same as that of the maximum dip shift. Furthermore, the time offset for Fig. D1g is adjusted by our fine-tuning policy described below. The offset for Fig. D1h is shifted by 200 fs compared to that of Fig. D1g for comparison.

Figure D2 and D3 are the plot of Lorentzian parameters vs input THz field based on the time offset of Fig. D1g and Fig. D1h, respectively. As shown in Fig. D2e and f, the trace of dip position vs incident THz field is nearly single valued for the fine-tuned time delay. On the other hand, as shown in Fig. D3e and f, the trace is no more single valued, and it gets an apparent hysteresis loop when the time delay is off from the optimal one. If the dip shift is an caused by QCSE as we model, it should be instantaneous to the incident field and does not have a hysteresis. Therefore, to obtain the optimal time offset, we plotted the dip against incident field for various offset value and chose the one which gives the minimal apparent hysteresis loop.

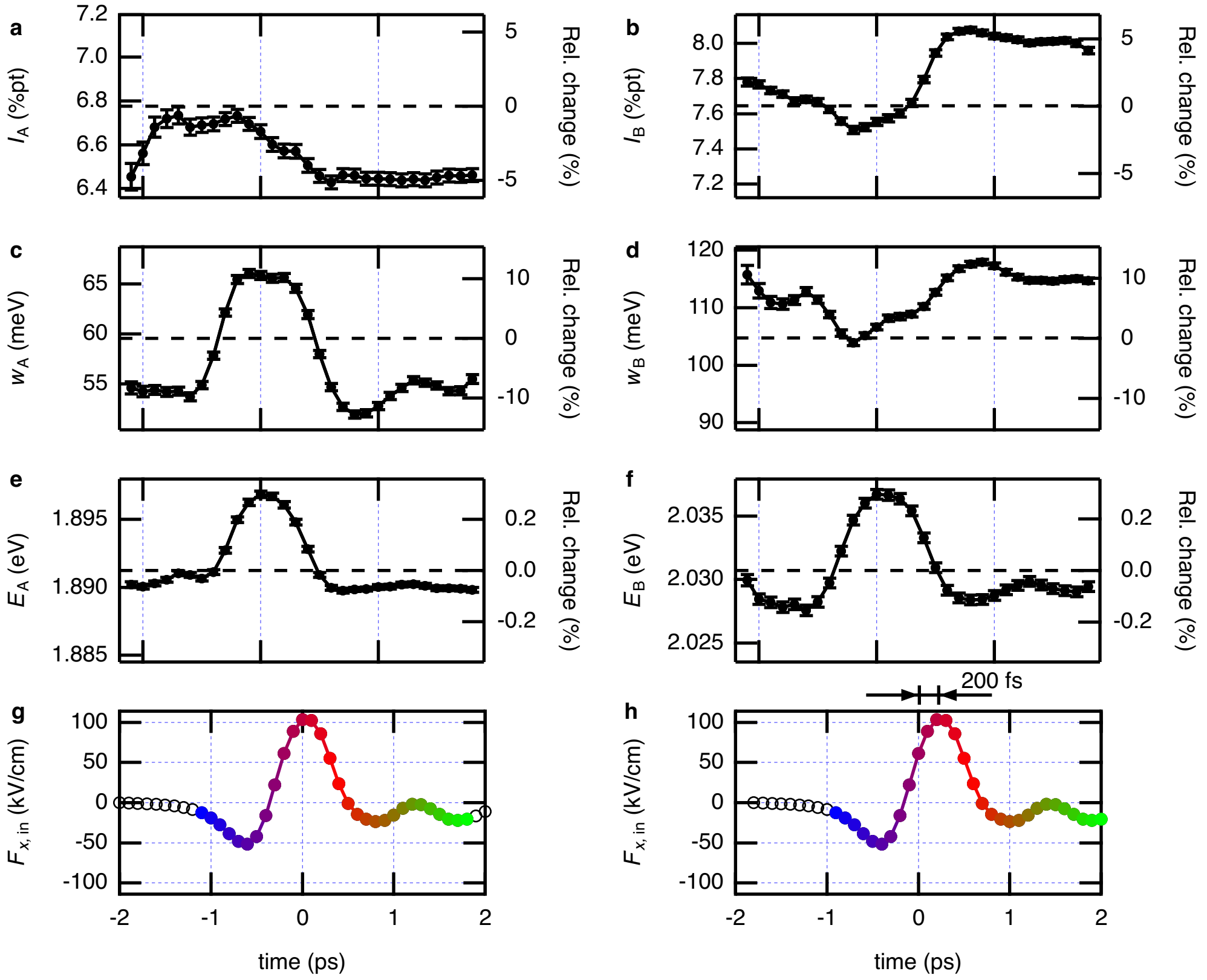

**Figure D1.** All the Lorentzian-fit parameter of the transient reflectance spectra obtained from a TPOP measurement (a-f) and the input THz waveform obtained from EO sampling (g and h) as a function of delay time. **a**, Amplitude, **c**, width, and **e**, center of the A-dip. **b**, Amplitude, **d**, width, and **f**, center of the B-dip. **g**, Input THz waveform with an optimal time offset. **h**, Input THz waveform which is shifted by 200 fs from that of **g**.

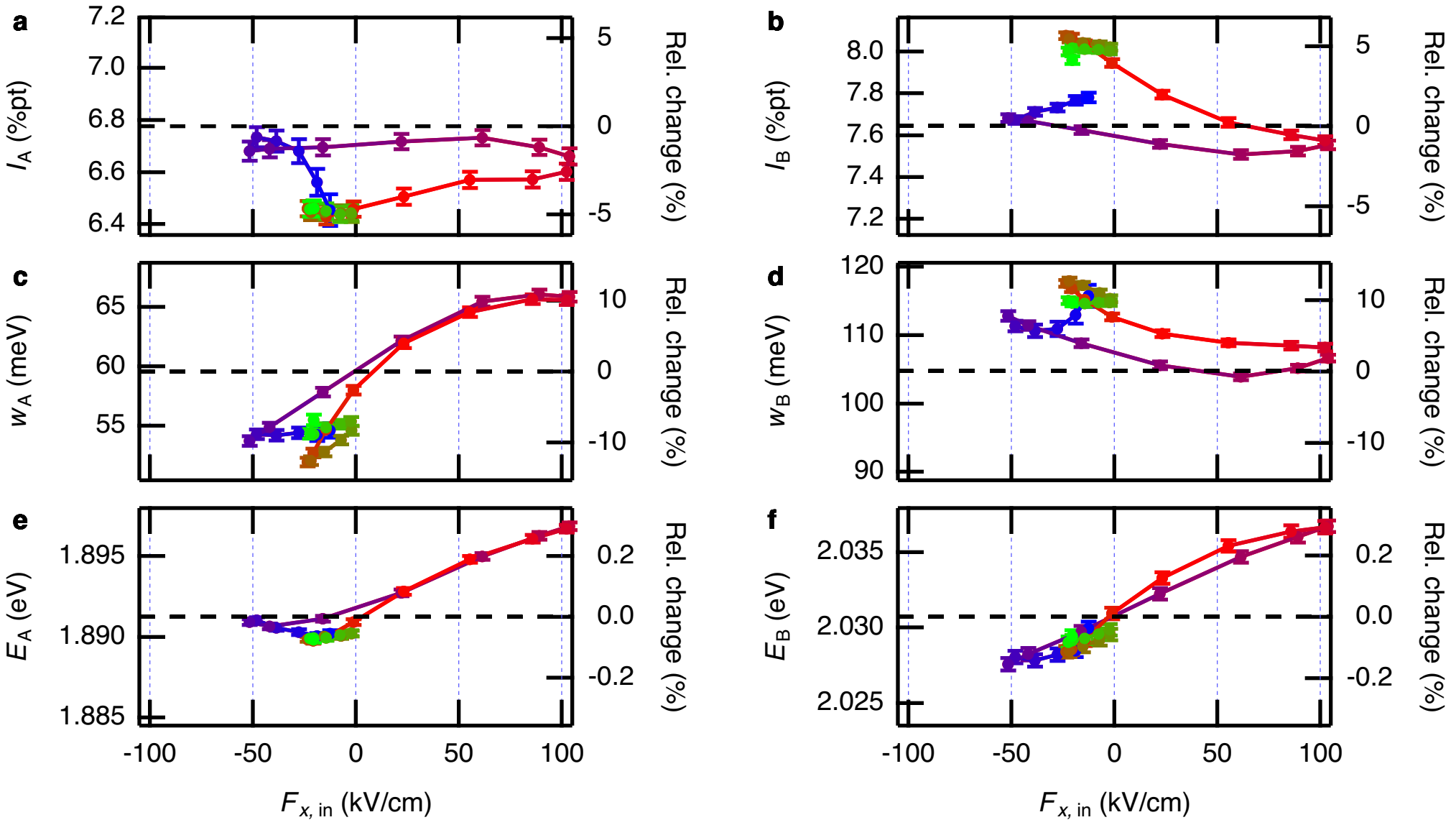

**Figure D2.** Lorentzian parameter plotted against the incident THz field with the optimal time offset of Fig. D1g. **a**, Amplitude, **c**, width, and **e**, center of the A-dip. **b**, Amplitude, **d**, width, and **f**, center of the B-dip. Right axes are the relative change to the values without THz pump.

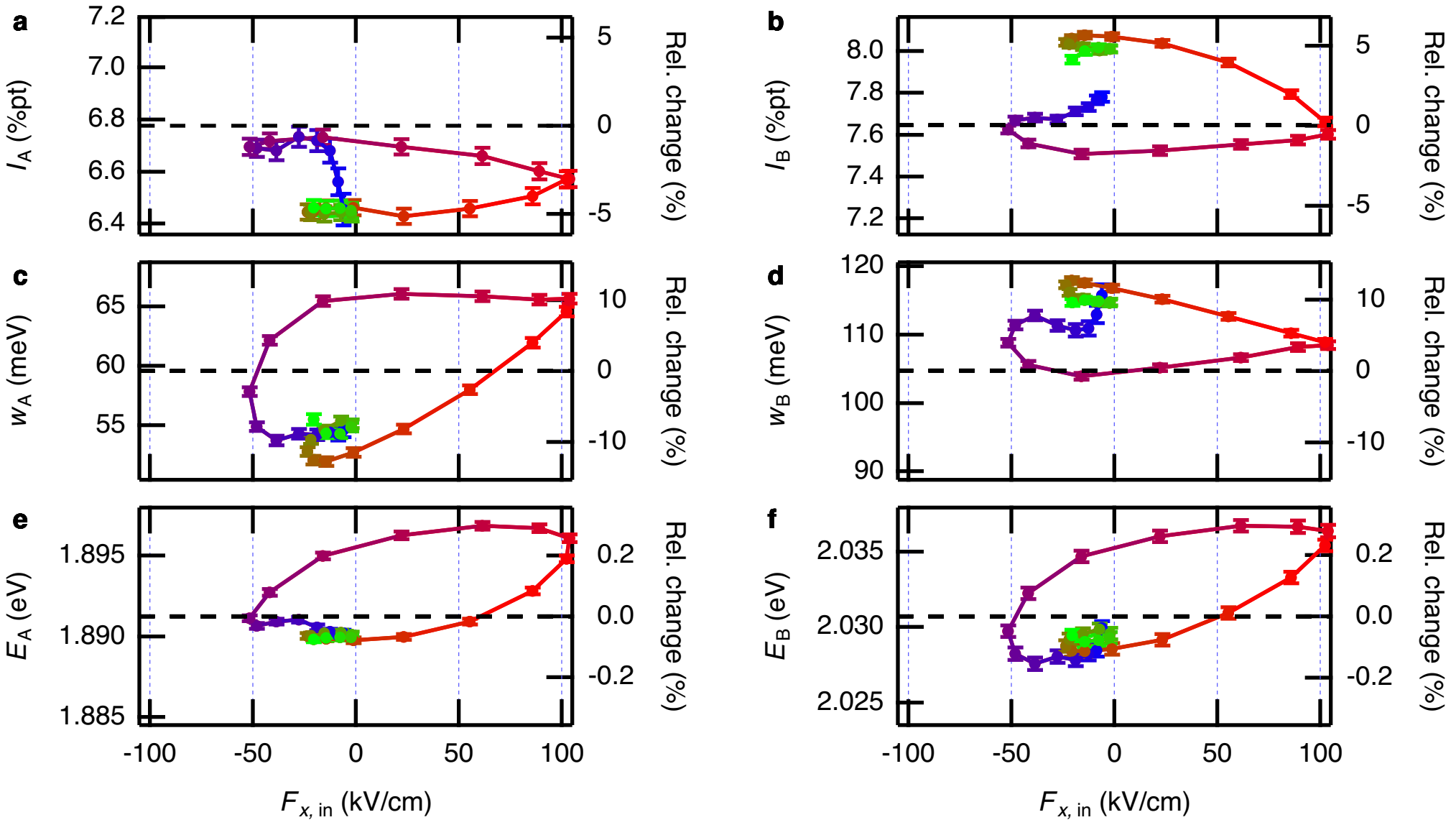

**Figure D3** Lorentzian parameter plotted against the incident THz field with the time offset of Fig. D1h, which is 200 fs off from the optimal value. **a**, Amplitude, **c**, width, and **e**, center of the A-dip. **b**, Amplitude, **d**, width, and **f**, center of the B-dip. Right axes are the relative change to the values without THz pump.

**Section E. DC biasing**

**E1. Leak current and dielectric-breakdown threshold measurement of devices without $MoS_2$ flakes**

We measured the leak current and the dielectric-breakdown threshold under a DC bias voltage for four 2D-3D hybrid THz antenna devices without $MoS_2$ flakes. We have connected a source meter (SMU 2450, Keithley) to the two contact pads of the device with a bonding wire and measured the current-voltage curve with two-terminal method. The tested samples were fabricated in the condition of α-batch.

The antennae without $MoS_2$ from α-batch had large sample-to-sample variation of the leak current and the breakdown threshold. The leak current was typically up to 8 to 40 μA at 110 to 150 V, and they went through the breakdown under the larger voltage. Figure E1-1 shows the results for the antenna without $MoS_2$, fabrication number #1 from α-batch. Fabrication number is an original index given to all the samples during fabrication, which is noted here for traceability of fabrication and experimental records. This sample had a leak current up to 8 μA at 145 V and went through the dielectric breakdown. After the breakdown, the antenna, connection lines between antenna to contact pads, a part of contact pads, and bond contact were burned out. Figure E1-2 shows the results for the antenna without $MoS_2$, fabrication number #2 from α-batch. It had a leak current up to 8 μA at 150 V and went through the dielectric breakdown. Before the breakdown, two discontinuity of the current-voltage curve was observed. These discontinuities most likely correspond to burn-down of a short-cut path of current, such as dusts connecting the two contact pads. Figure E1-3 shows the results for the antenna without $MoS_2$, fabrication number #3 from α-batch. It had a leak current of only less than 500 pA and did not go through the breakdown of the device. This might be due to very high quality of the device occasionally achieved. Figure E1-4 shows the results for the antenna without $MoS_2$ #4 from α-batch. It had a leak current up to 40 μA at 110 V and went through a breakdown of the bond contact.

Figure E1-5 shows the results for an antenna without $MoS_2$, fabrication number #17 from β-batch, repeatedly measured for three times. The voltage was continuously swept in the order of 0V, +20V, -20V, and 0V as indicated by the arrows in the figure. This sample from β-batch had a leak current of only about 200 pA at 20V. Also, the reproducible I-V property shows the high stability of the spacer dielectric. Each current-voltage curve shows hysteresis loop, which may be originated by charge transfer from the electrodes to the trap states in the oxide layer.

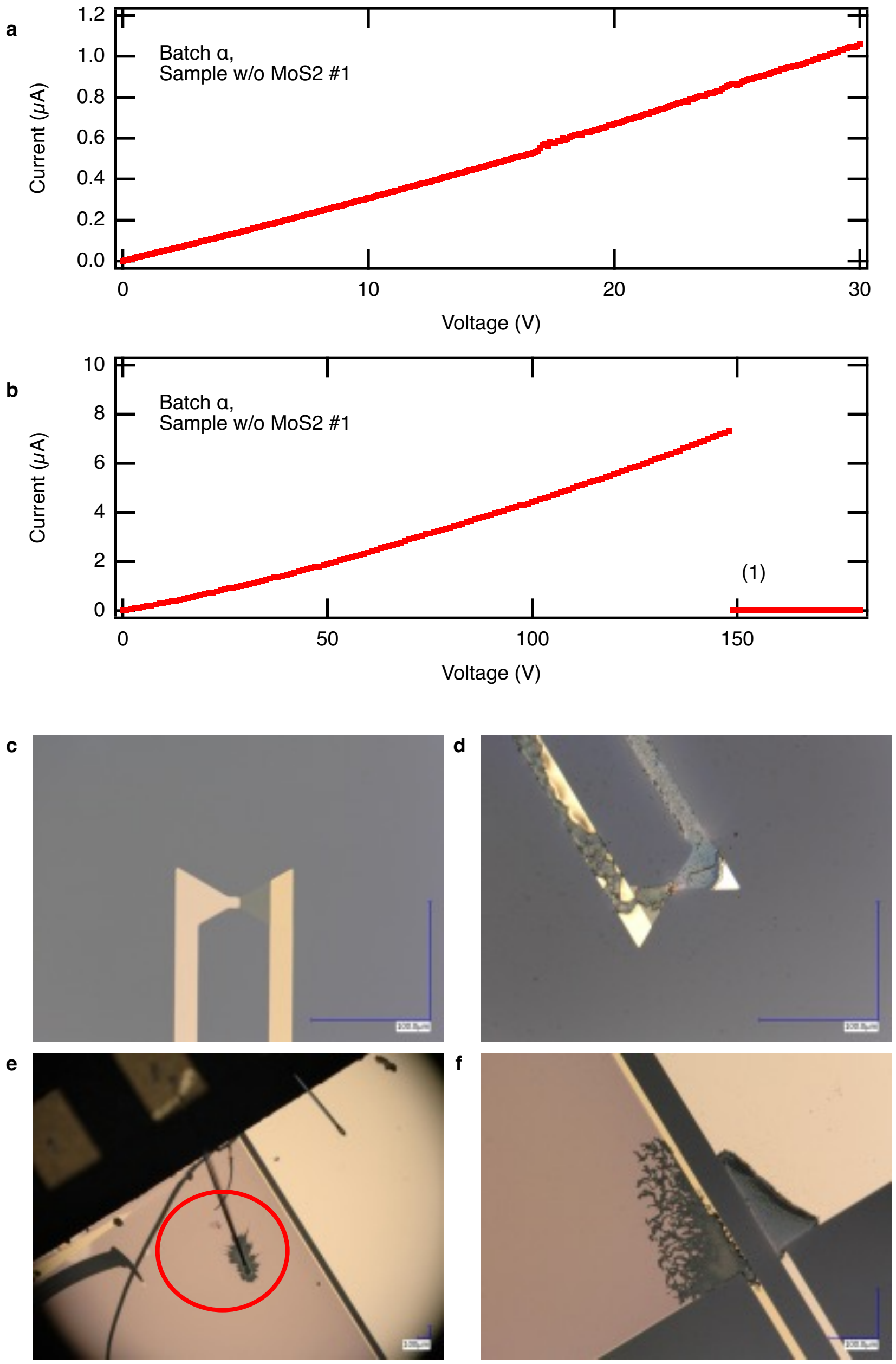

**Figure E1-1.** Result of the leak current measurement and breakdown test for antenna without $MoS_2$, fabrication number #1 from α-batch. **a**, Current-voltage curve up to 30 V. **b**, Current-voltage curve up to 180 V. The point (1) corresponds to the breakdown of the device. **c**, Antenna as prepared. **d**, Antenna after the test. **e**, Bond contact after the test. **f**, Contact pad after the test.

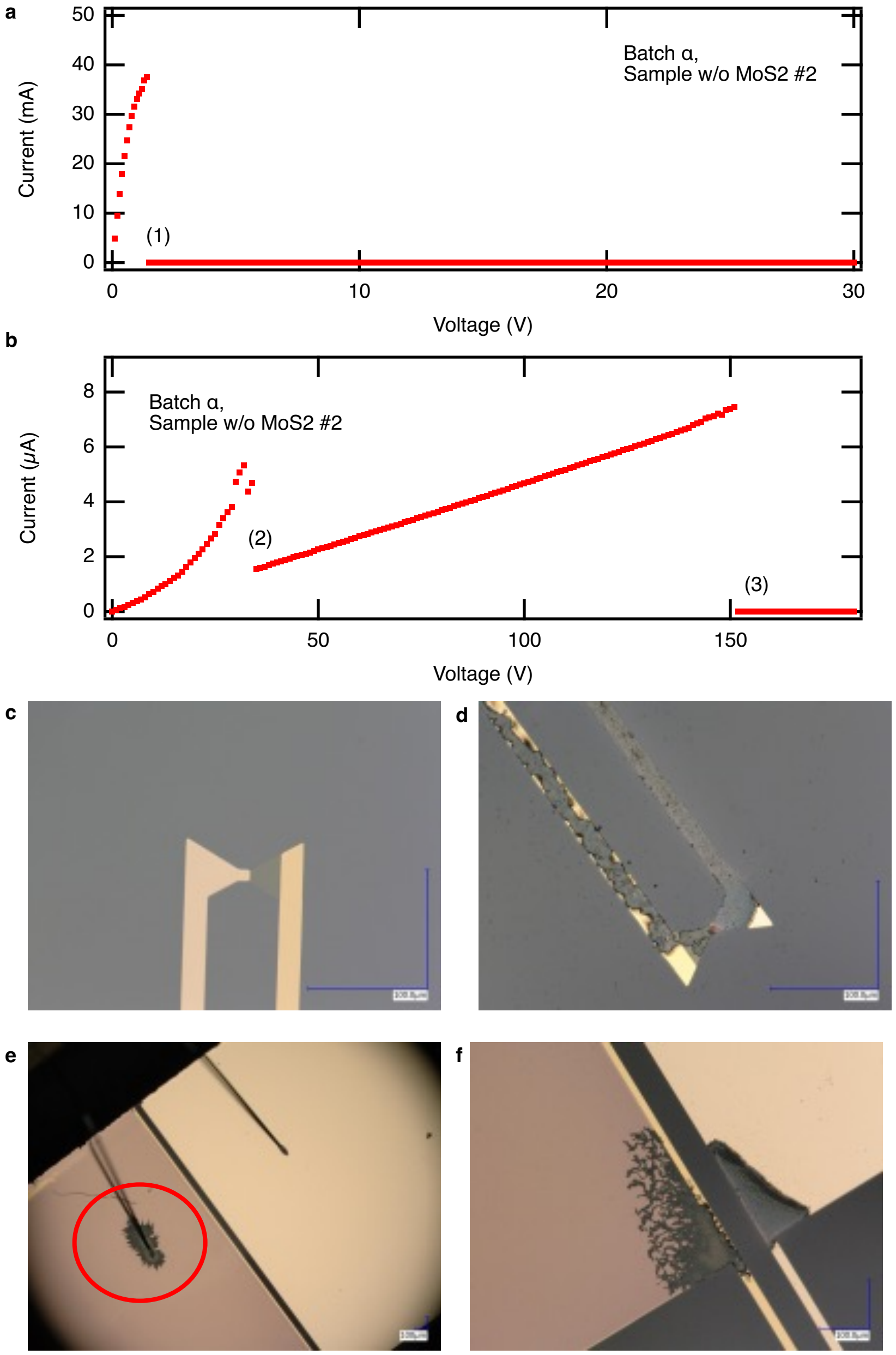

**Figure E1-2.** Result of the leak current measurement and breakdown test for antenna without $MoS_2$, fabrication number #2 from α-batch. **a**, Current-voltage curve up to 30 V. **b**, Current-voltage curve up to 180 V. The points (1) and (2) probably correspond to burn-out of short-cut path for the current, such as a dust bridging the two contact pads. The point (3) corresponds to the dielectric breakdown of the device. **c**, Antenna as prepared. **d**, Antenna after the test. **e**, Bond contact after the test. **f**, Contact pad after the test.

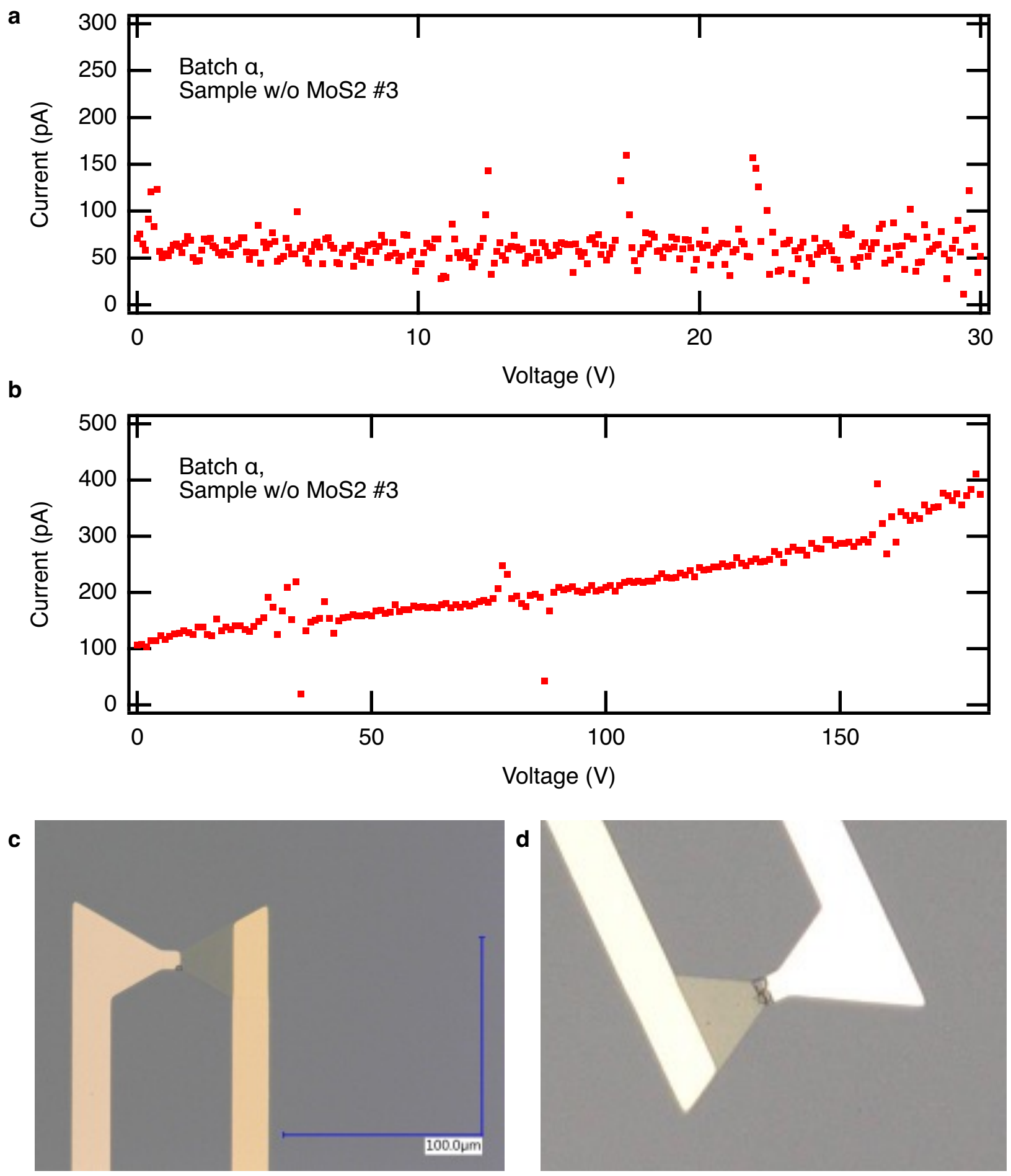

**Figure E1-3.** Result of the leak current measurement and breakdown test for antenna without $MoS_2$, fabrication number #3 from α-batch. **a**, Current-voltage curve from the first measurement 1. **b**, Current-voltage curve from the second measurement up to 180 V. **c**, Antenna as prepared. **d**, Antenna after the test.

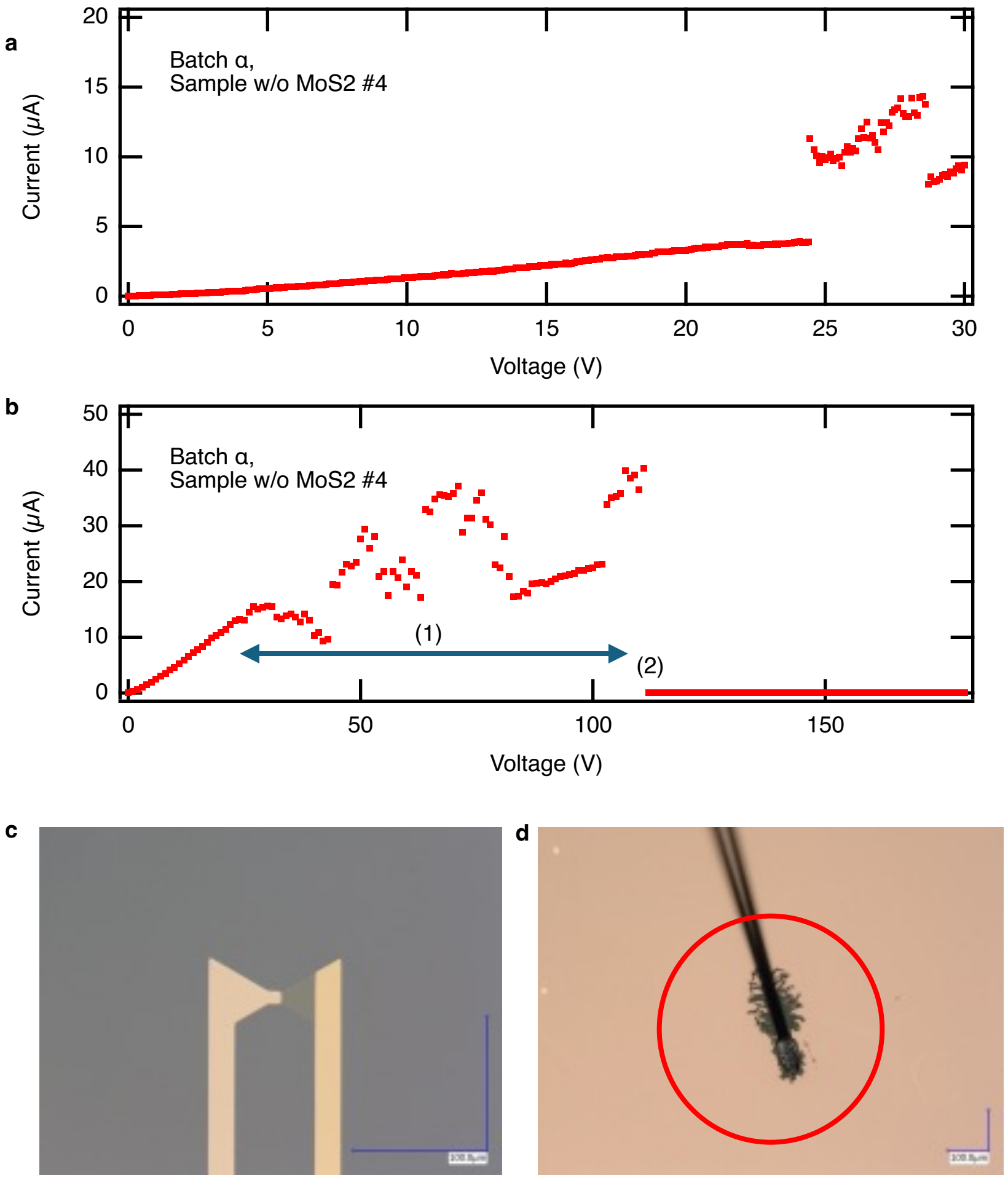

**Figure E1-4.** Result of the leak current measurement and breakdown test for antenna without $MoS_2$, fabrication number #4 from α-batch. **a**, Current-voltage curve up to 30 V. **b**, Current-voltage curve up to 180 V. The regime (1) and point (2) probably correspond to destabilization and breakdown of the bond contact, respectively. **c**, Antenna as prepared. No change of antenna was observed after the point (2). **d**, Bond contact after the test.

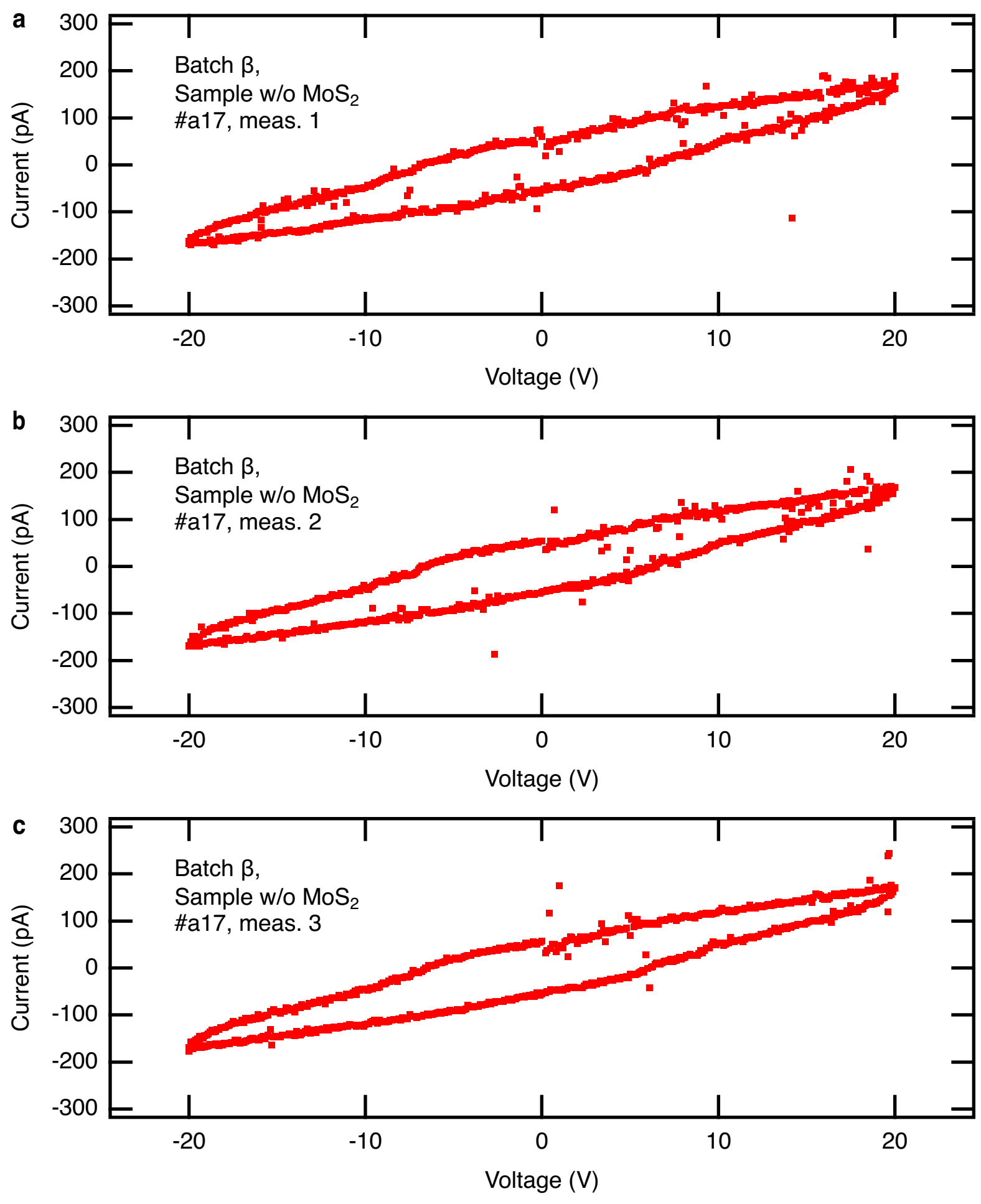

**Figure E1-5.** Result of the leak current measurement for antenna without $MoS_2$, fabrication numbers #a17 from β-batch repeatedly. **a**, First measurement, **b**, Second measurement, **c**, Third measurement.

### E2. Unintentional shortcuts and dielectric breakdown of the devices under DC bias

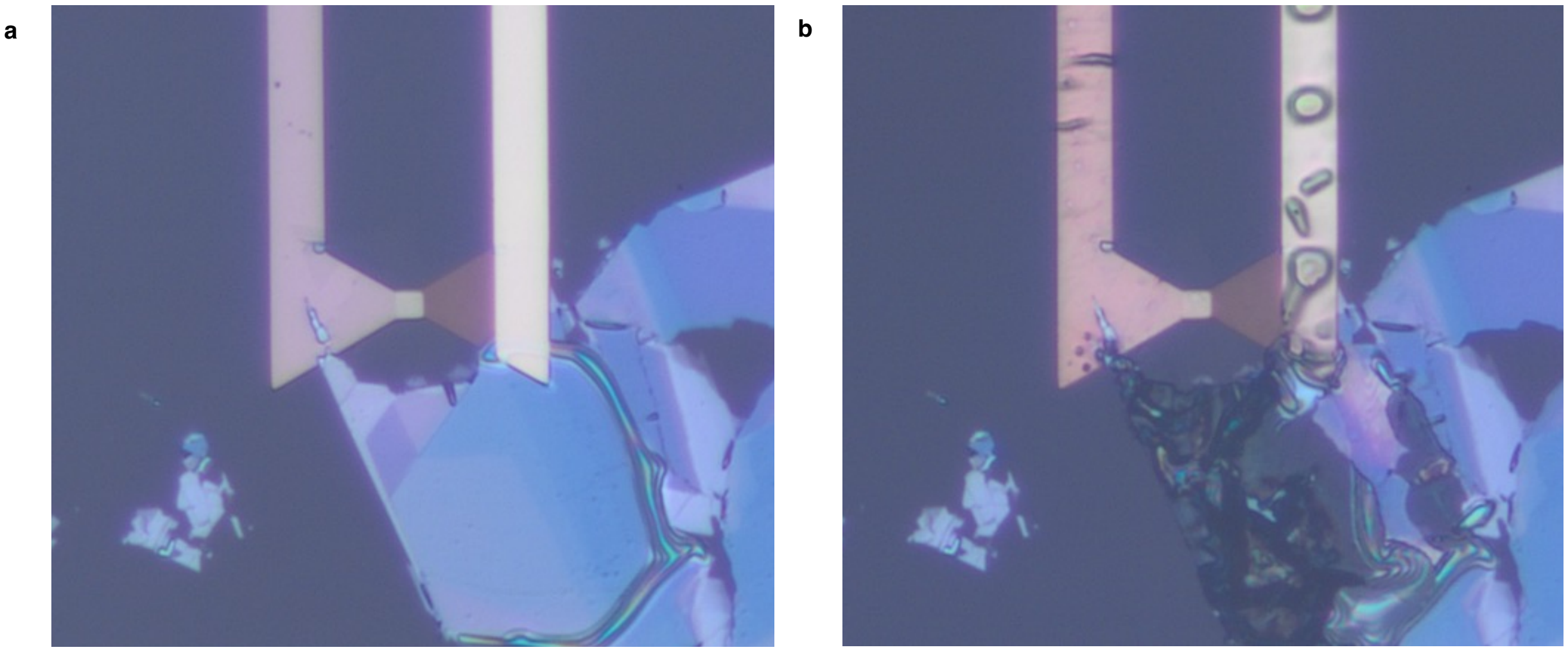

**Figure E2-1**. Microscope image of an antenna device with $MoS_2$ flake before and after DC-biasing test. **a**, Before the test. A few-layer $MoS_2$ flake is located at the antenna gap even it is not clearly seen. In addition, thicker $MoS_2$ flake is located around the antenna structure. **b**, After the test. A leak current pathway from top electrode to bottom electrode through the thick $MoS_2$ flake is burned, and the line is deformed due to Joule heating.

Our devices had contact pads and lines connected to the top and bottom electrodes. We have applied DC-bias voltage on the devices to measure the DC-bias induced dip shift in reflectance spectra for calibration of the THz field amplitude. However, it did not work due to unexpected leak-current paths.

Figure E2-1 shows an example of a device before and after such a measurement with DC biasing. In this device, not only the few-layer $MoS_2$ flake at the gap but also thicker $MoS_2$ flake was placed around the antenna structure due to the mechanical exfoliation and transfer process. After the DC biasing experiment, a change of appearance in the thick $MoS_2$ flake and the connection line was observed. This indicates that a leak current from the top to bottom electrodes through thick $MoS_2$ flew, and the dielectric breakdown took place along this leak-current path. Since most of our devices had such a leak pathway, we could not measure an effect of the DC-bias voltage. Removal of such unwanted leak and breakdown pathways will enable the MV/cm - scale DC biasing in the future studies.

**Section F. All the TPOP measurement data and analysis result**

In Supplementary Figures, the results of all the TPOP measurements and their analysis are presented in several types of figures. In this section, we give the caption of these figures using an example of measurement (1) for device 1 shown in Figs. F1-F6. The device name, measurement number, and antenna angle are given in the folder name.

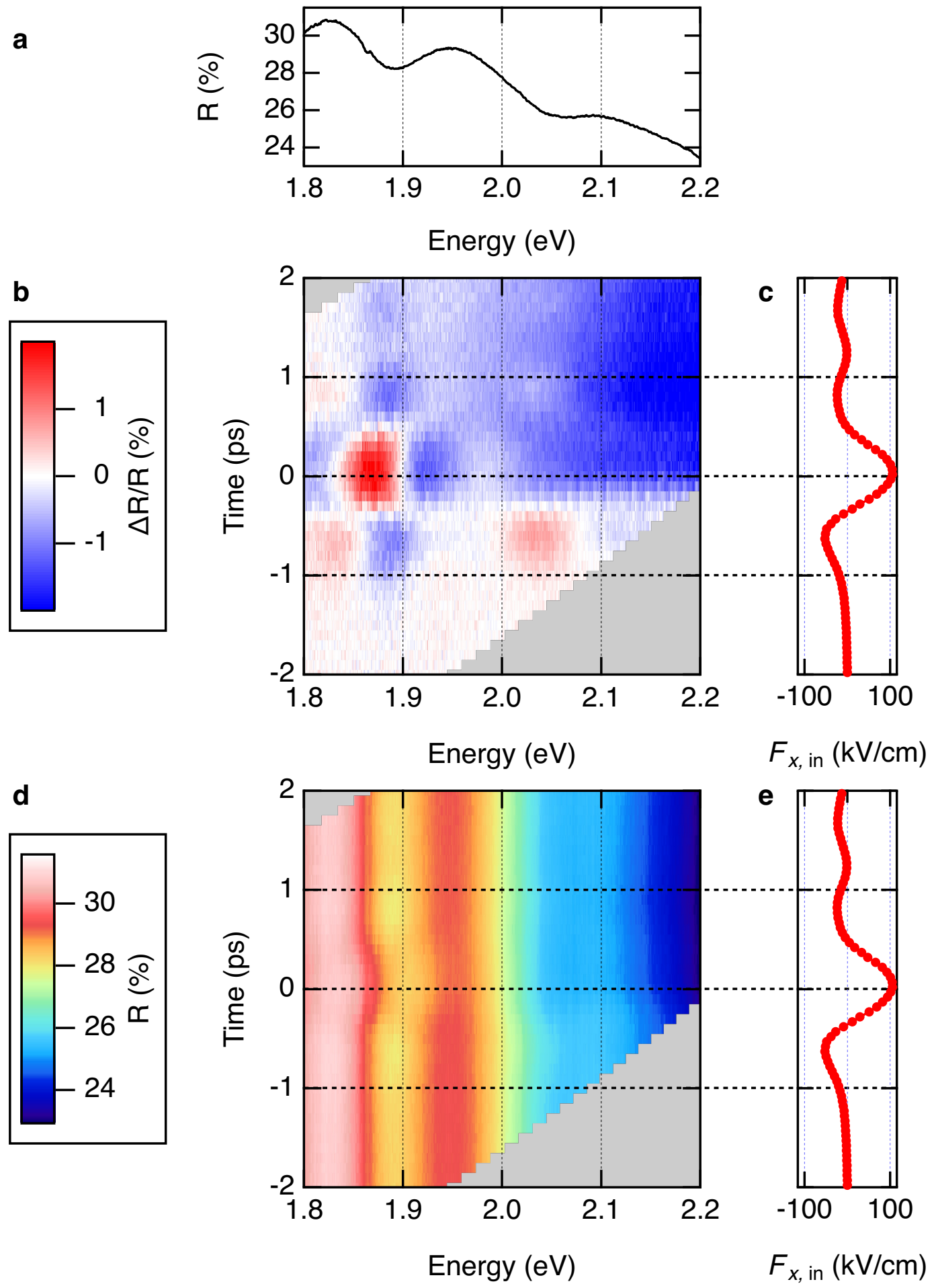

**Figure F1.** An example of a figure named "XXX_summary_measTPOPdata," that shows the measured TPOP response for each antenna. XXX is an identifier for each experiment. **a**, Measured absoltue reflectance spectrum. **b**, Transient reflectance change observed in the TPOP measurement. **c**, Incident THz waveform for the TPOP measurement measured with EO sampling. **d**, Transient reflectance obtained from the reflectance spectrum and transient reflectance change shown in **a** and **b**. **e**, Incident THz waveform same as **c**.

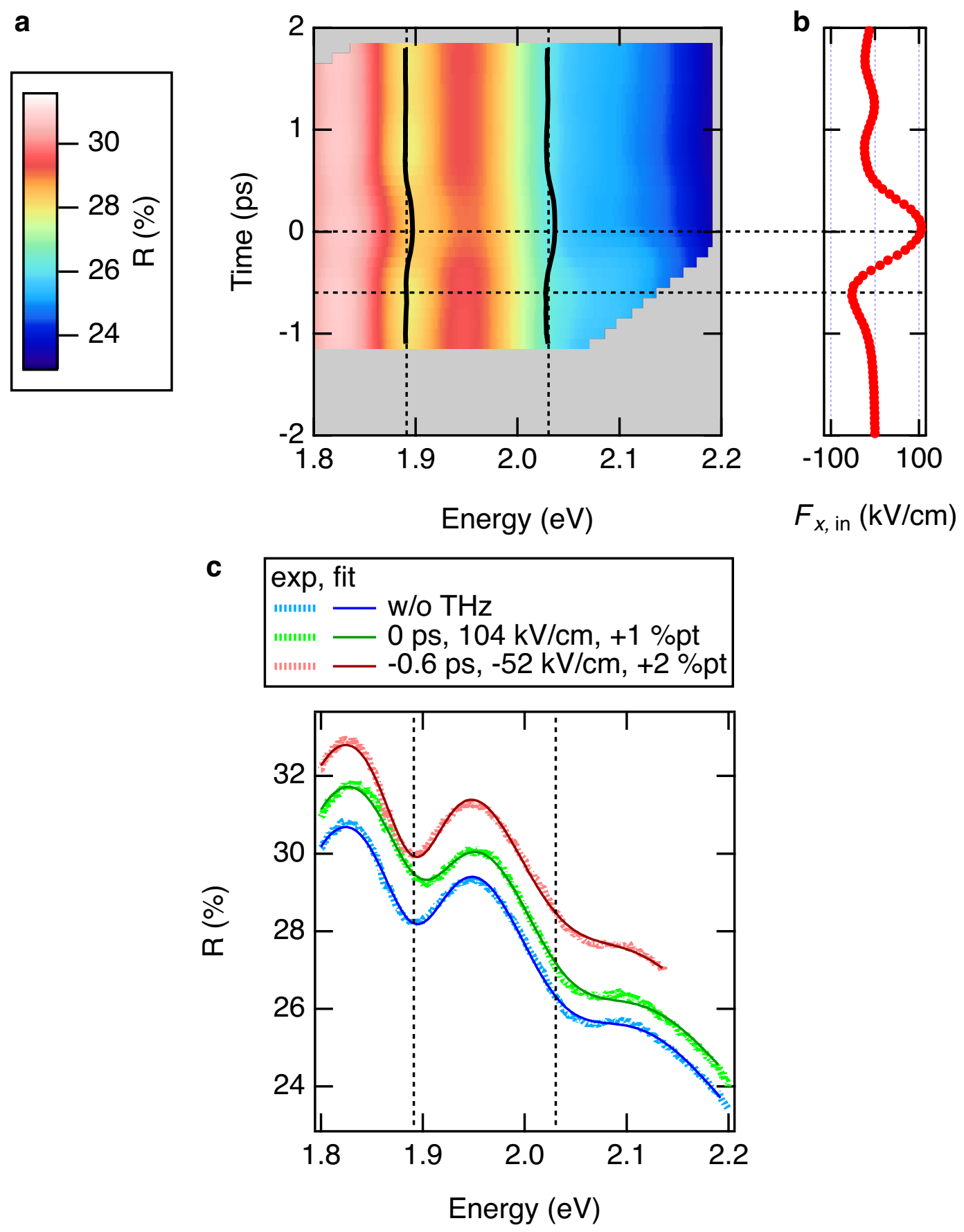

**Figure F2.** An example of a figure named "XXX_REtFit_and_EOS," that shows the fitting curves of the reflectance spectra. XXX is an identifier for each experiment. **a**, Transient reflectance spectra obtained via fitting. Two dotted vertical lines show the original position of A- and B-dip without the THz field. **b**, THz waveform incident to the antenna (same as Fig. F1c.) **c**, Reflectance spectra without THz pump (blue), under THz pump of 104 kV/cm at 0 ps time delay (light green, vertically offset by +1%pt for clarity), and under THz pump of -52 kV/cm at -0.6 ps time delay (dark red, vertically offset by +2%pt for clarity). Traces of dotted lines are experimental data, and traces of solid lines are fitting. The vertical dotted line shows the position of A and B dip without the THz field.

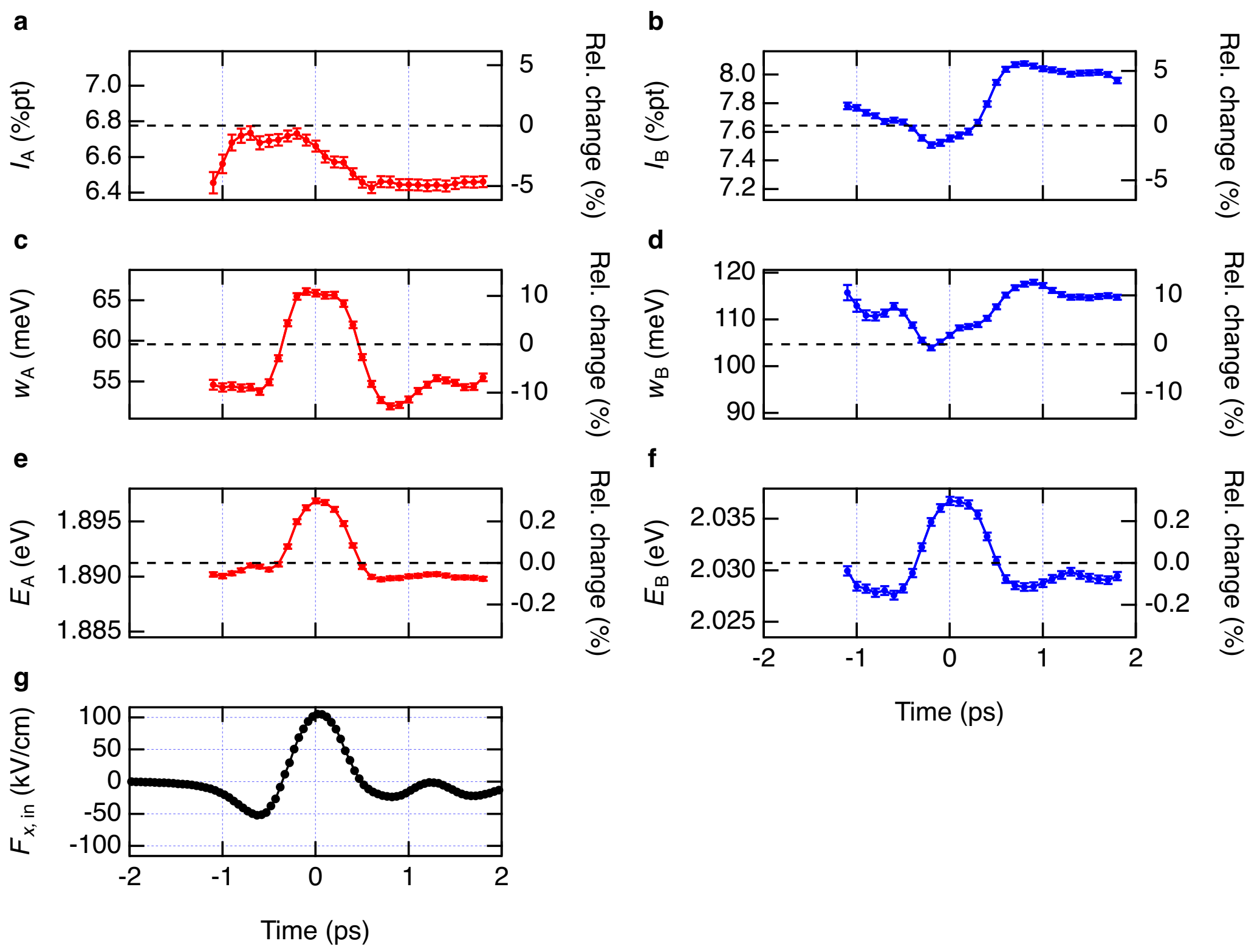

**Figure F3.** An example of a figure named "XXX_REtFit_ParamVsTime," that shows the temporal change of all the fitting parameters. XXX is an identifier for each experiment. **a-b**, Amplitude of A-dip (**a**) and B-dip (**b**.) **c-d**, Width of A-dip (**a**) and B-dip (**b**.) **e-f,** Position of A-dip (**e**) and B-dip (**f**.) **g**, THz waveform incident to the antenna measured with EO sampling, under which the response of **a-f** was observed. Each horizontal dashed line in **a-f** show the original value of the parameter. The right axis in **a-f** is the relative change to the original value. Error bars in **a-f** show standard deviation as fitting parameters.

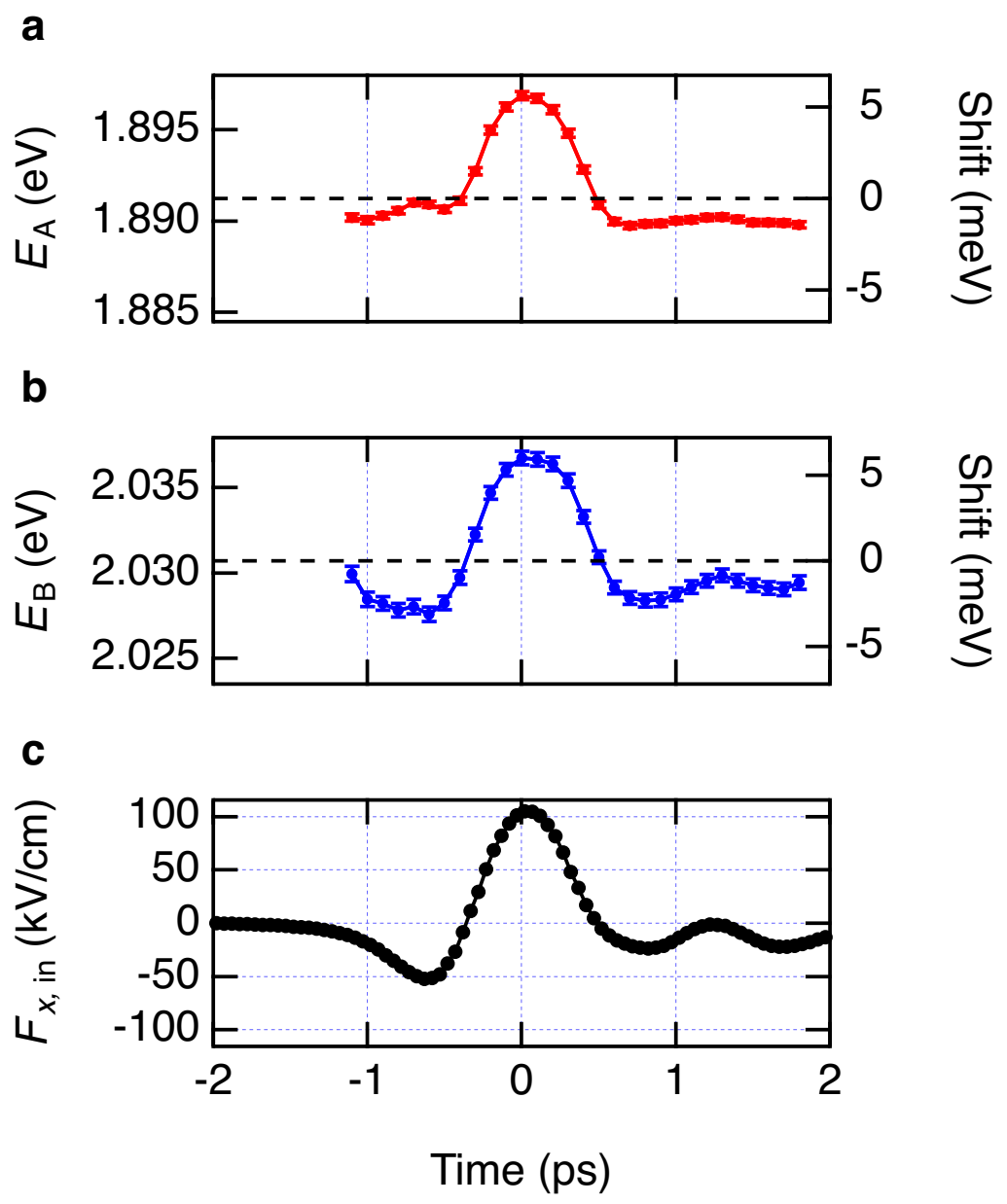

**Figure F4.** An example of a figure named "XXX_REtFit_ParamEshiftVsTime," that shows the temporal change of the peak position with the energy shift shown on the right axis. XXX is an identifier for each experiment. **a-b**, The time dependence of the position of A-dip (**a**) and B-dip (**b**). Horizontal dashed lines show original dip positions without the THz field. **c**, THz waveform incident to the antenna measured with EO sampling, under which the response of **a** and **b** was observed. Horizontal dashed lines show original dip positions. The right axis in **a** and **b** is the relative change to the original value. Error bars in **a** and **b** show standard deviation as fitting parameters.

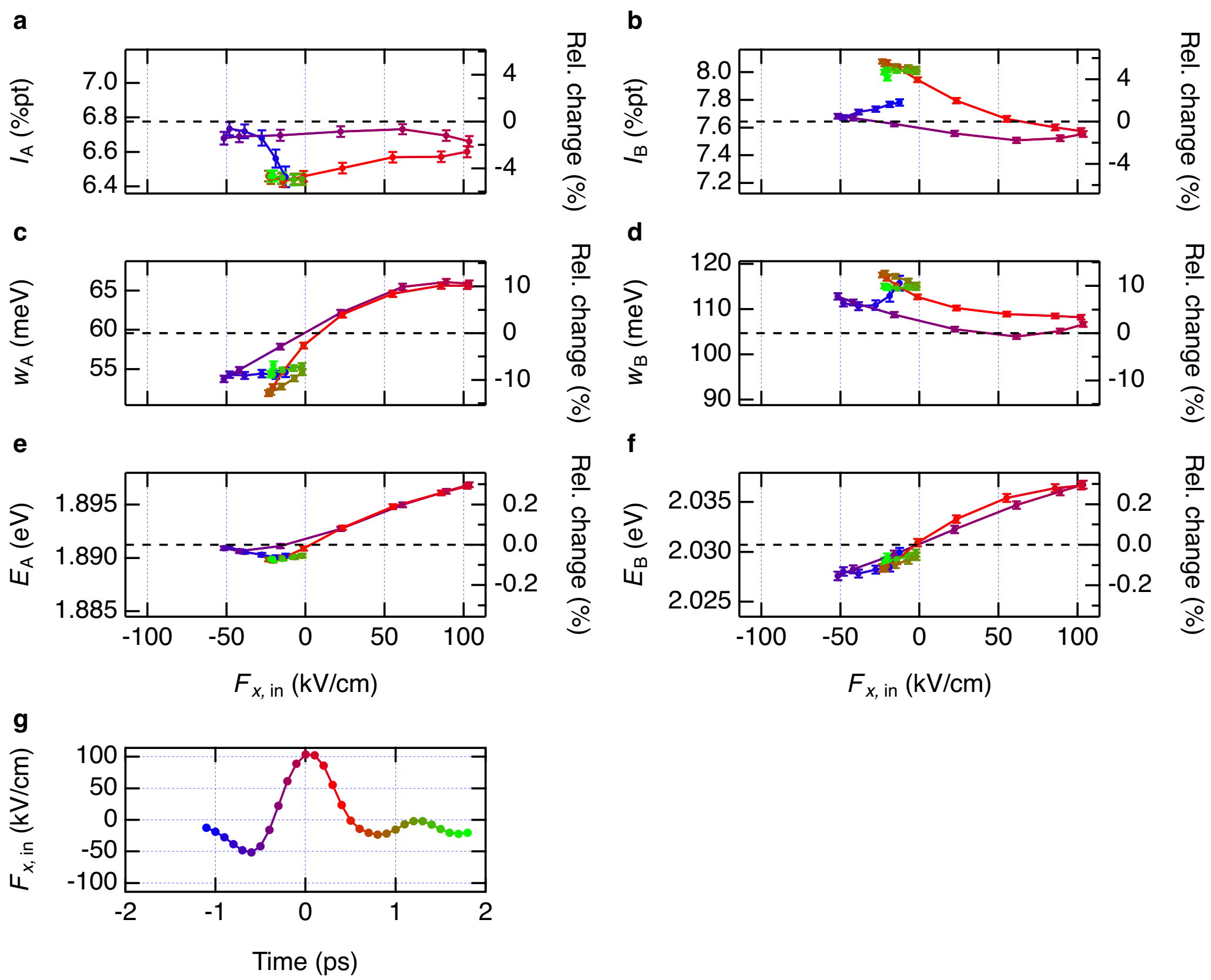

**Figure F5.** An example of a figure named "XXX_REtFit_ParamVsETHz," that shows the incident-field dependence of all the fitting parameters. XXX is an identifier for each experiment. **a-b**, Amplitude of A-dip (**a**) and B-dip (**b**.) **c-d**, Width of A-dip (**a**) and B-dip (**b**.) **e-f,** Position of A-dip (**e**) and B-dip (**f**.) **g**, THz waveform incident to the antenna under which the response of **a-f** was observed. Each point represents the THz feild for the timing at the timing of each TPOP data point in **a-f**, obtained by linearly interpolating the measured data shown in Fig. F3g. Each horizontal dashed line in **a-f** show the original value of the parameter. The right axis in **a-f** is the relative change to the original value. Error bars in **a-f** show standard deviation as fitting parameters.

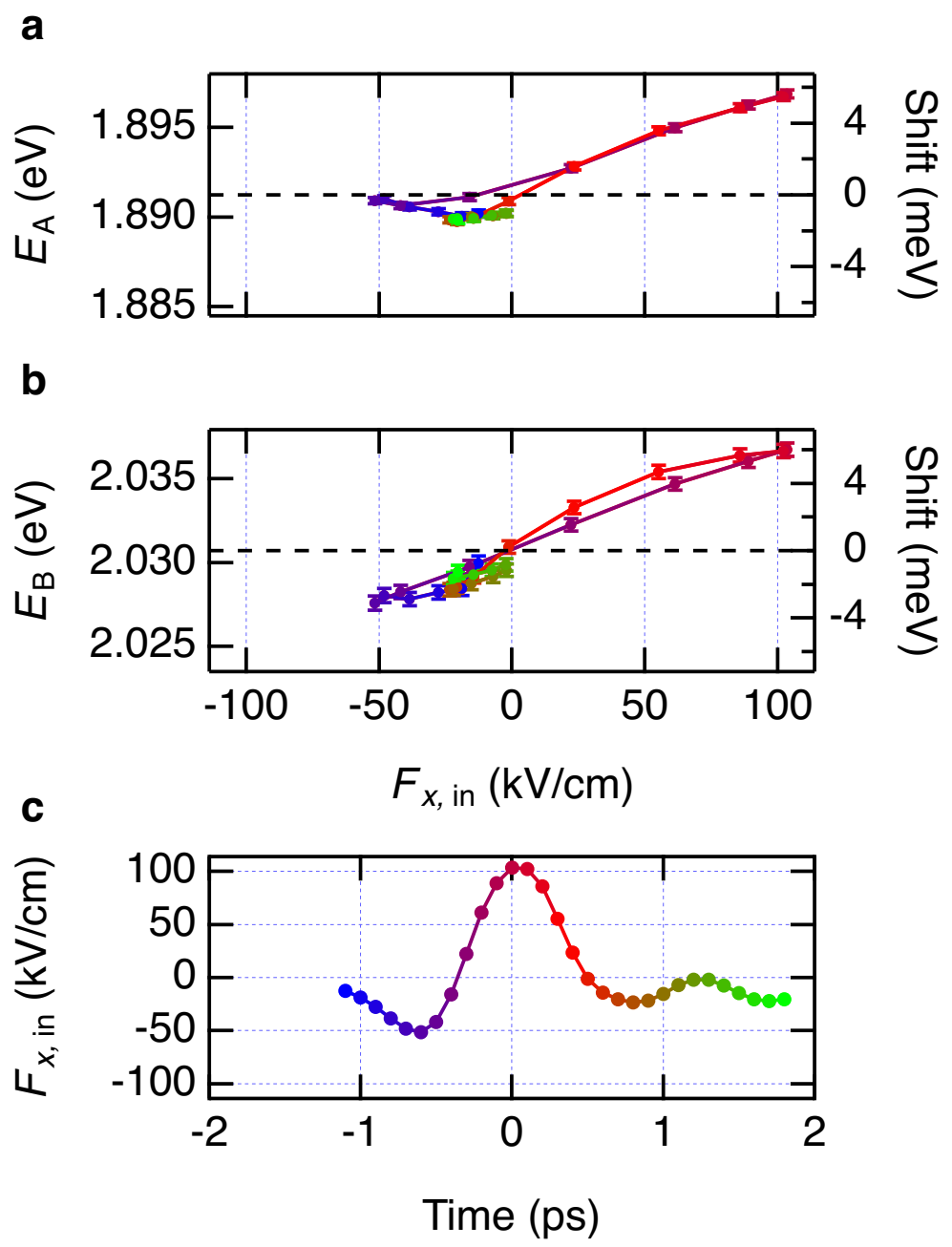

**Figure F6.** An example of a figure named "XXX_REtFit_ParamEshiftVsETHz," that shows the incident-field dependence of the peak position with the energy shift shown on the right axis. XXX is an identifier for each experiment. **a-b**, The time dependence of the position of A-dip (**a**) and B-dip (**b**). Horizontal dashed lines show original dip positions without the THz field. **c**, THz waveform incident to the antenna measured with EO sampling, under which the response of **a** and **b** was observed. Each point represents the THz feild for the timing at the timing of each TPOP data point in **a-f**, obtained by linearly interpolating the measured data shown in Fig. F3g. Horizontal dashed lines in **a** and **b** show original dip positions. The right axis in **a** and **b** is the relative change to the original value. Error bars in **a** and **b** show standard deviation as fitting parameters.

## Section G. Photoluminescence and reflectance measurement data

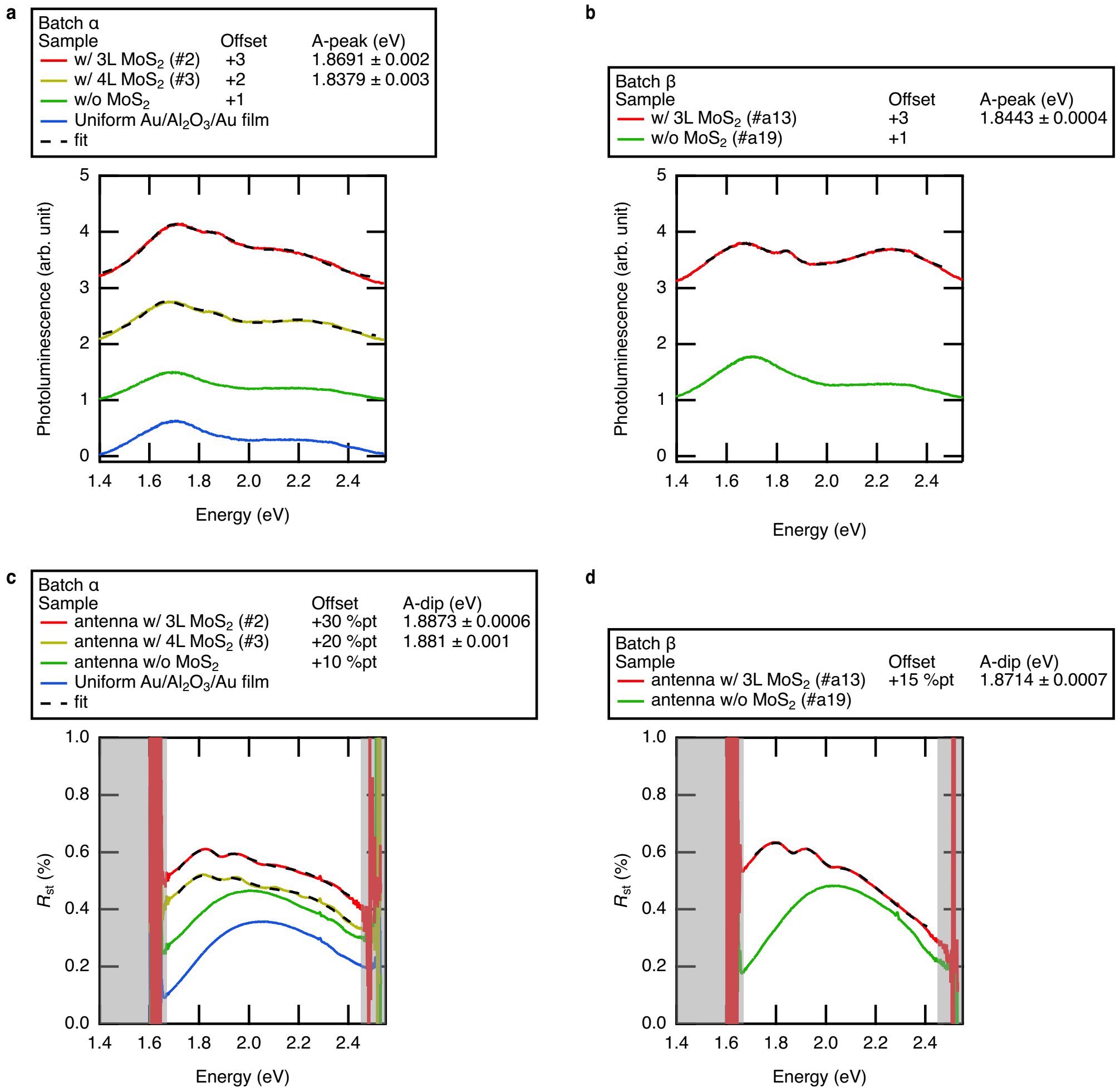

**Figure G1.** PL and reflectance spectra of the devices at the field-enhancement section. **a**, PL spectra of the devices from α-batch. **b**, PL spectra of the devices from β-batch. **c**, Reflectance spectra of the devices from α-batch. **d**, Reflectance spectra of the devices from β-batch. Traces in each graph are offset by the value indicated in the legend. The black dashed lines on the traces are fitting curves. Position of the A-peak and A-dip is indicated for the fitting result of each trace is also shown in the legend.

**Table G1**. Position of A-peak in PL spectrum and A-dip in reflectance spectrum for each device. Δ is the difference of the energy of A-dip to A-peak.

| Batch | Fabrication number | Layer num. | Name in the main text | A-peak, PL (eV) | A-dip, Refl. (eV) | Δ (meV) |
|---|---|---|---|---|---|---|
| α | #2 | 3 | Device I | 1.8691 ± 0.002 | 1.8873 ± 0.0006 | 18.2 |
| α | #3 | 4 | Device III | 1.8379 ± 0.003 | 1.881 ± 0.001 | 43.1 |
| β | #13 | 3 | Device II | 1.8443 ± 0.0004 | 1.8714 ± 0.0007 | 27.1 |

Figure G1 shows the photoluminescence (PL) and reflectance spectra of devices fabricated in this study at the field-enhancement section. We measured not only antennae with $MoS_2$ flakes but also an antenna without $MoS_2$ and an unstructured uniform film with the same layer structure to the antenna without $MoS_2$ of α-batch (8-nm Au/75-nm ALD-$Al_2O_3$/75-nm ALD-$Al_2O_3$/3-nm Cr/ 25-nm Au).

The PL spectra of antennae without $MoS_2$ and the uniform film has two broadband peaks, which is likely originated from photoluminescence of gold and an optical interference in the layer structure. In addition to these two peaks, each PL spectrum of antenna with $MoS_2$ has one more peak at 1.838 - 1.868 eV, which can be assigned as A-exciton luminescence [14]. The peak positions were derived by fitting to the PL spectra with three Lorentzian peaks.

The reflectance spectra of the antennae without $MoS_2$ and the uniform film have smooth curve originating from the optical interference in the layered structure, and that of antennae with $MoS_2$ have two additional dips corresponding to the A- and B-exciton absorption as discussed in Supplementary Section B. The dip positions were determined via fitting with two-Lorentzian-dip model shown in Fitting of Transient Reflectance Spectra in Method.

The PL peak in each device was red shifted to the reflectance dip, i.e., absorption peak, by 18 to 35 meV, as shown in Table G1. This is consistent to the Stokes shift in $MoS_2$, which can be on the order of several tens of meV and dependent on the doping level [13]. The positions of the A-peaks, position of A-dips, and Stokes shift were different sample to sample. This should be originated from the different number of layer [14], dielectric disorder [15], and so on.

## Section H. Optical measurement detail

### H1. THz pump-optical probe reflection microspectroscopy

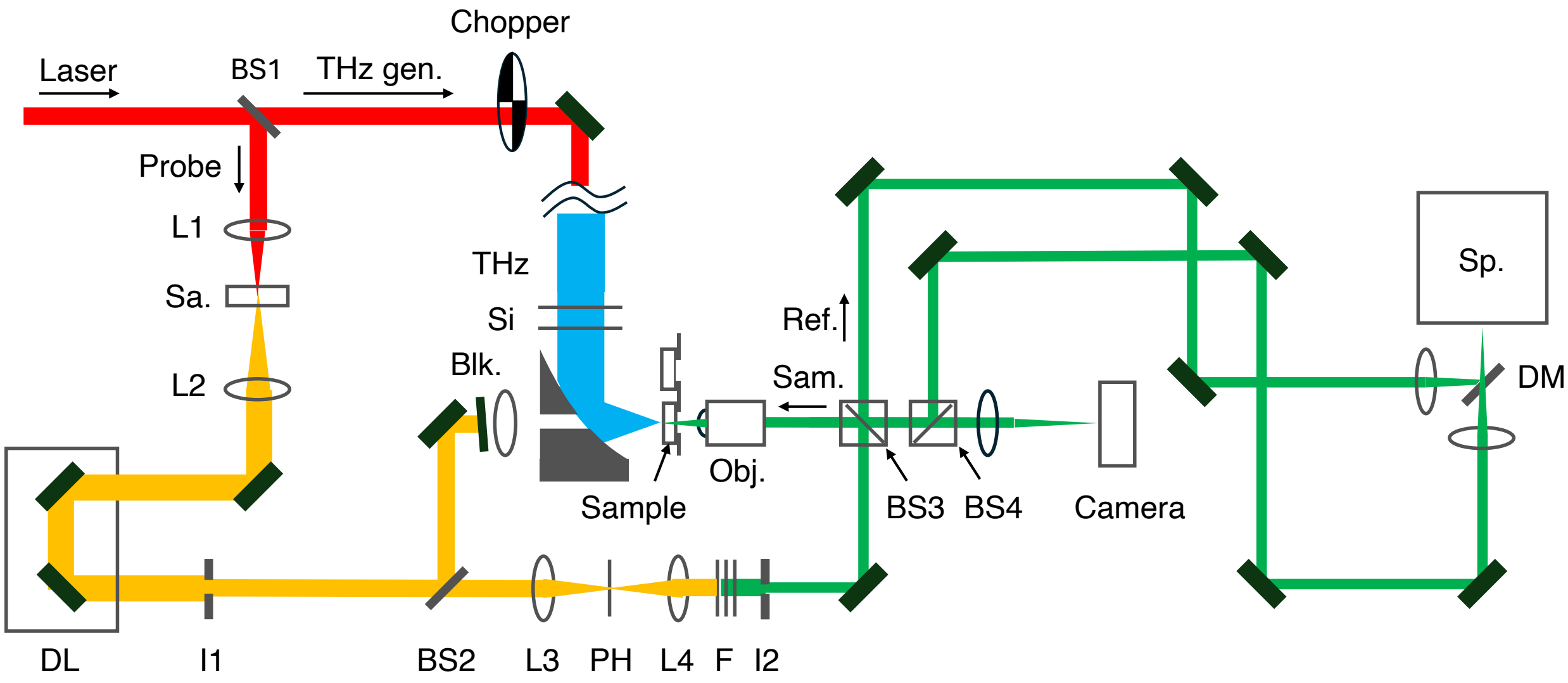

**Figure H1-1.** Schematic optical setup for THz pump-optical probe reflection microspectroscopy. The input laser (Solstice from Spectraphysics with center wavelength of 800 nm, bandwidth of 50 nm (FWHM), repetition rate of 1 kHz, pulse energy of 6 mJ) is split into THz-generation (THz gen.) path to probe-beam path (Probe) by a beam splitter (BS) 1. In the THz-generation path, the laser was converted to an intense THz pulse via tilted-pulse-front scheme using an $LiNbO_3$ crystal. The THz pulse was attenuated via silicon (Si) wafers depending on experimental conditions and focused onto a sample, i.e., a 2D-3D THz antenna device on a grass substrate. Multiple samples and EO crystal are mounted on a single sample holder on a three-axis stage to make the condition of TPOP and EO sampling measurements the same as possible. A chopper in the THz-generation path chops the THz pulse with 50 Hz. The probe beam is loosely focused on a 4-mm thick sapphire to generate a white light. The white-light probe beam passes a mechanical delay line (DL) to control the time delay between the THz pump pulse and optical probe pulse. The probe beam is split via BS2. The reflected beam at BS2 is blocked in this measurement, and the transmitted beam is used as a probe. Optical filters are used to make the spectrum of the probe beam flat in the measured spectral range and to attenuate the energy of the probe beam to 900 fJ/pulse. Iris H1, focusing onto a pinhole of 20-μm diameter (PH), and iris H2 is used for beam shaping of the probe beam. Then, the probe beam is split via BS3. The transmitted beam is a reference beam (Ref.) which enters the spectrometer (Sp.) after being reflected via D-shaped mirror (DM). The reflected beam at BS3 is a sample beam (Sam.) which is focused on the sample, the 2D-3D hybrid THz antenna device on the backside (left side in the figure) of the glass substrate, with a 10x objective lens (Obj.). The reflected sample beam enters BS4. The sample beam transmitted BS4 is focused onto a camera to monitor the probe beam spot on the sample. The sample beam reflected at BS4 enters the spectrometer after passing through the DM. The spectra of Ref. and Sam. beams are measured simultaneously at the spectrometer. The spectra with THz and without THz are measured to obtain the transient differential reflectance.

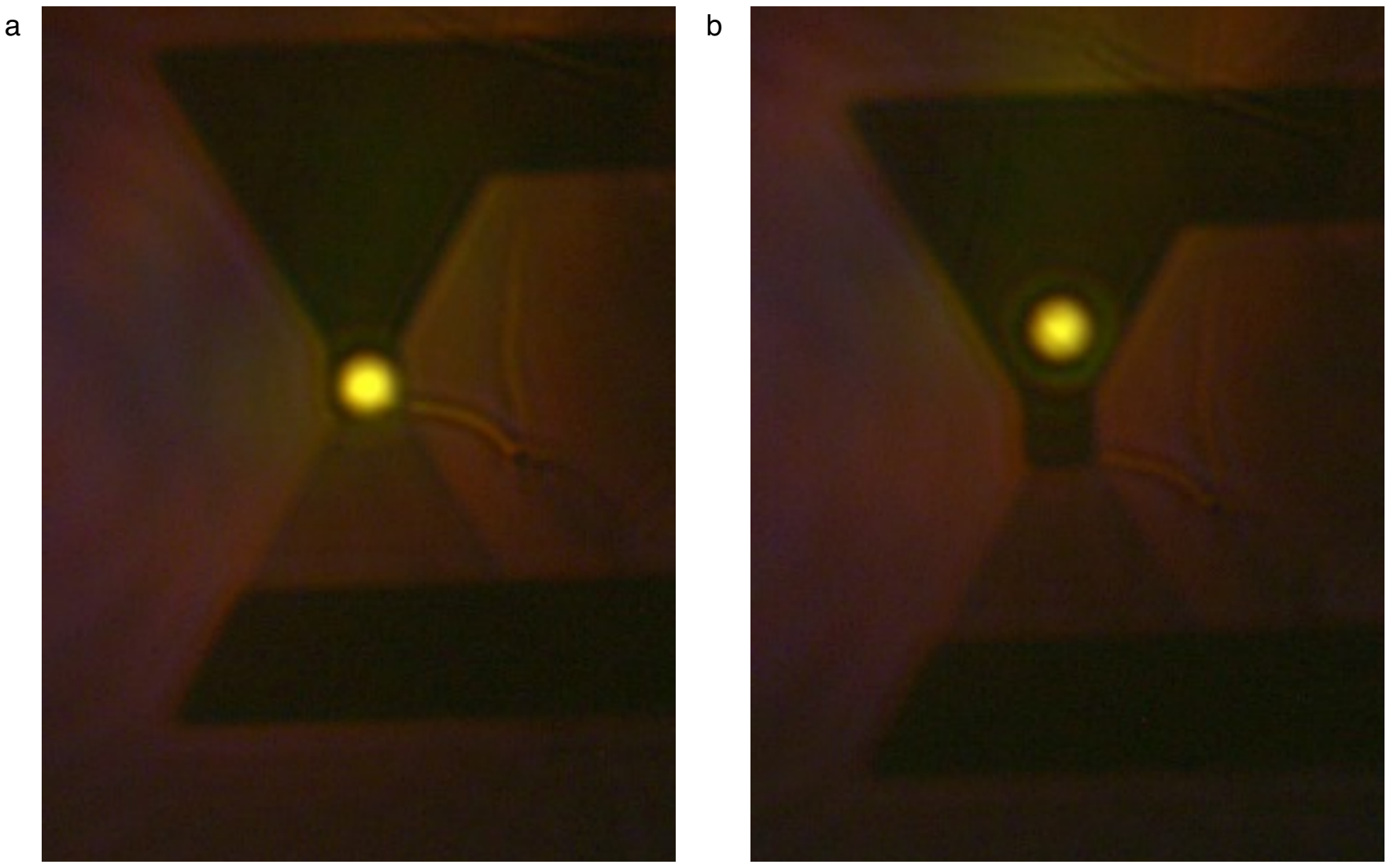

**Figure H1-2.** Microscopic image of the probe beam and a device in TPOP measurement taken via Camera in Fig. H1-1. The white light reflected at BS2 in Fig. H1-1 was unblocked and used to illuminate the whole device in this image. **a**, Image with the probe beam focused on the field-enhancement section. **b**, Image with the probe beam out of the field-enhancement section. The focused probe beam has a shape of Airy disk. By comparing the size of the Airy disk to that of the field-enhancement section of 10 μm × 10 μm, the radius of the dark ring of the Airy disk is estimated to be 5 μm.

Figure H1-1 shows the schematic setup of the TPOP measurement, and Fig. H1-2 shows the microscopic image of the probe beam and a device in the measurement.

The peak fluence of the probe beam at the sample was 4 μJ/cm$^2$. In a previous study of optical pump-probe spectroscopy of few-layer $MoS_2$ with a white-light pump covering the range of 1.68-2.21 eV, it has shown to be a one-photon absorption regime up to the pump energy of 900 μJ/cm$^2$ [16]. Therefore, our probe fluence was well in the linear regime.

In TPOP measurement, the probe spot was intentionally made to be comparable to the field-enhancement section to make the power of probe beam as high as possible without increasing the fluence. The higher probe power is necessary to perform the pump-probe time scan in a reasonable measurement time. The small clipping of the probe beam at the edge of field-enhancement section is negligible for this measurement, since the effect of clipping is cancelled for the transient differential reflectance $\Delta R/R$, which is a ratio of the reflectance change to the absolute reflectance.

**H2. Reflection microspectroscopy for static absolute reflectance measurement**

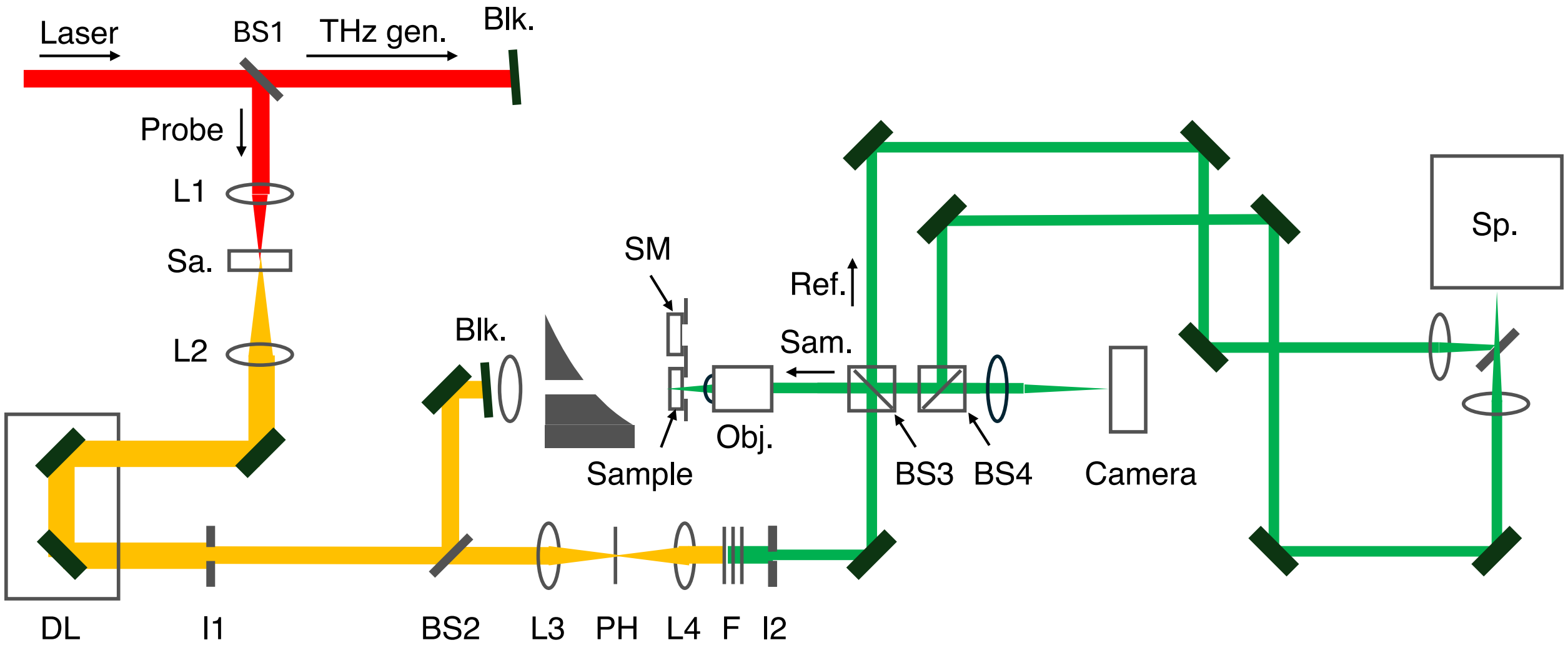

**Figure H2-1.** Schematic optical setup for reflection microspectroscopy. The same setup as TPOP measurement is used with the THz-generation path blocked. The antenna sample and a reference silver mirror (SM) with known reflectance spectrum are mounted on the same sample holder on a three-axis stage. The measurement result of SM is used to calibrate the transfer function of the optical system and to obtain the reflectance spectrum of the sample. In this measurement, a 20x objective lens was used to make the probe beam spot small enough for the measurement of static absolute reflectance.

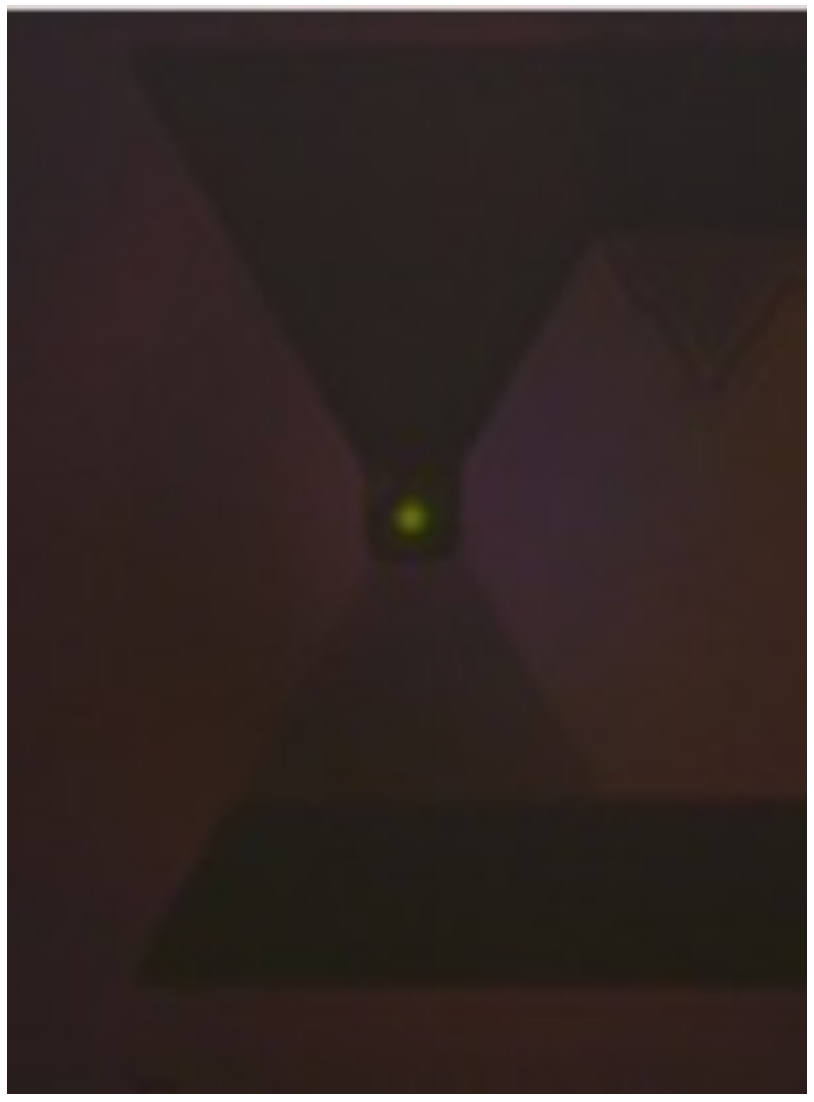

**Figure H2-2.** Microscopic image of the probe beam and a device in TPOP measurement taken via Camera in Fig. H2-1. The white light reflected at BS2 in Fig. H2-1 was unblocked and used to illuminate the whole device in this image. The probe beam is focused on the field-enhancement section of the device. By comparing the size of the Airy disk of the probe beam to that of the field-enhancement section of 10 μm × 10 μm, the radius of the dark ring of the Airy disk is estimated to be 2 μm.

Figure H2-1 shows the schematic setup of the reflection microscopy to measure the static absolute reflectance of the device, and Fig. H2-2 shows the microscopic image of the probe beam and a device in the measurement. The estimated peak fluence of the probe beam at the sample is 3 $\mu J/cm^2$, which is much smaller than the peak fluence needed to cause nonlinear processes. The probe spot was made smaller than the field-enhancement section to eliminate the effect of the beam clipping for the absolute reflectance measurement.

### H3. EO sampling

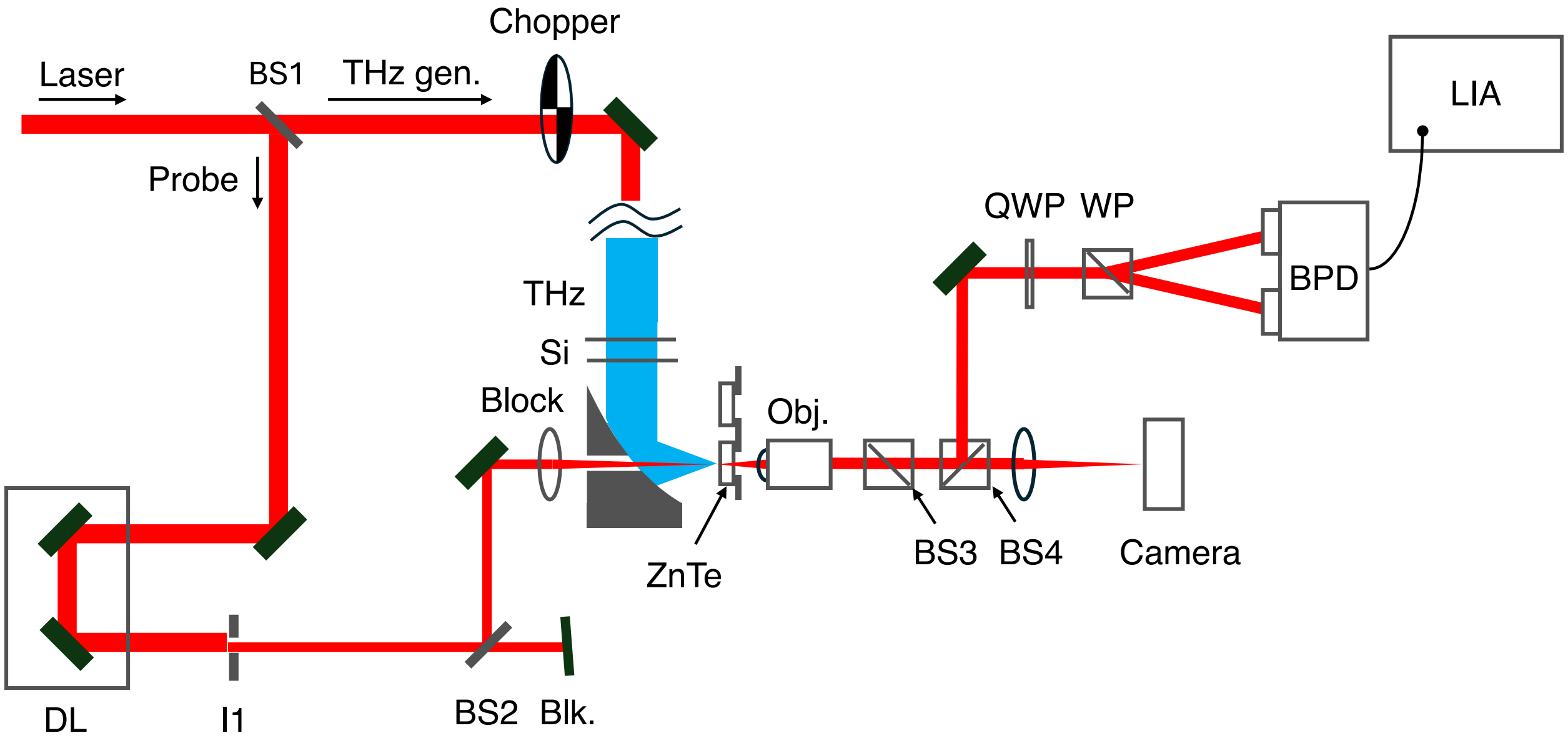

**Figure H3.** Schematic optical setup for EO sampling [6]. In this measurement, the probe beam of 800 nm is used without conversion to the white light. The probe beam reflected at BS2 is focused on a 1-mm Zinc telluride (ZnTe) crystal at the focus of the THz beam. The probe beam reflected at BS4 passes through quarter-wave plate (QWP) and Wollaston prism (WP) to resolve the polarization and is detected at the balanced photodiode (BPD). The THz-generation beam is chopped at 500 Hz, and the signal from BPD is measured with lock-in amplifier (LIA). The mechanical delay line (DL) is used to sweep the time delay between probe pulse and THz pulse. The ZnTe and the 2D-3D hybrid THz antenna devices for the TPOP measurement were mounted on the same sample holder to locate the ZnTe and the antenna at the same position in the THz-beam focus in the two measurements. The EO-sampling probe beam transmitted BS4 was monitored by a camera to adjust the focus spot of the transmission probe beam of EO sampling to the focus spot of the reflection probe beam of TPOP measurement. The position of antenna devices was adjusted to this reflection probe spot in the TPOP measurement.

Figure H3 shows the schematic setup for the EO sampling for the characterization of THz waveform.

**References**

[1] Neu, J. & Schmuttenmaer, C. A. Tutorial: An introduction to terahertz time domain spectroscopy (THz-TDS). *J. Appl. Phys.* **124**, 231101 (2018).

[2] Glover, R. E. & Tinkham, M. Conductivity of superconducting films for photon energies between 0.3 and40kTc. *Phys. Rev.* **108**, 243–256 (1957).

[3] Born, M., Born, L. N. L., Wolf, E., Born, M. A. & Bhatia, A. B. *Principles of Optics: Electromagnetic Theory of Propagation, Interference and Diffraction of Light*. (Cambridge University Press, 1999).

[4] Walther, M. *et al.* Terahertz conductivity of thin gold films at the metal-insulator percolation transition. *Phys. Rev. B Condens. Matter* **76**, 125408 (2007).

[5] Robertson, J. High dielectric constant oxides. *Eur. Phys. J. - Appl. phys.* **28**, 265–291 (2004).

[6] Nahata, A., Weling, A. S. & Heinz, T. F. A wideband coherent terahertz spectroscopy system using optical rectification and electro-optic sampling. Appl. Phys. Lett. 69, 2321–2323 (1996).

[7] R. Olmon, B. Slovick, T. Johnson, D. Shelton, S.-H. Oh, G. Boreman, and M. Raschke, "Optical dielectric function of gold," Phys. Rev B. 86(23),235147 (2012).

[8] Z. Mei, S. Deng, L. Li, X. Wen, H. Lu, and M. Li, "Dielectric function of sub-10 nanometer thick gold films," Appl. Phys. A 127(6), 437 (2021).

[9] E. Palik, ed., *Handbook of optical constants of solids*, 3rd ed. (Academic press, 1998).

[10] Y. Yu, Y. Yu, Y. Cai, W. Li, A. Gurarslan, H. Peelaers, D. Aspnes, C. Van de Walle, N. Nguyen, Y.-W. Zhang, and L. Yu, "Exciton-dominated dielectric function of atomically thin $MoS_2$ films," Sci. Rep. 5(1), 16996 (2015).

[11] D. Marques, J. Guggenheim, R. Ansari, E. Zhang, P. Beard, and P. Munro, "Modelling Fabry-Pérot etalons illuminated by focussed beams," Opt. Express 28(5), 7691-7706 (2020).

[12] Radisavljevic, B. & Kis, A. Mobility engineering and a metal-insulator transition in monolayer $MoS_2$. *Nat. Mater.* **12**, 815–820 (2013).

[13] Mak, K. F. *et al.* Tightly bound trions in monolayer $MoS_2$. *Nat. Mater.* **12**, 207–211 (2013).

[14] Splendiani, A. *et al.* Emerging photoluminescence in monolayer $MoS_2$. *Nano Lett.* **10**, 1271–1275 (2010).

[15] Raja, A. *et al.* Dielectric disorder in two-dimensional materials. *Nat. Nanotechnol.* **14**, 832–837 (2019).

[16] Nie, Z. *et al.* Ultrafast carrier thermalization and cooling dynamics in few-layer $MoS_2$. *ACS Nano* **8**, 10931–10940 (2014).